\documentclass[usenatbib]{mnras}

\usepackage{newtxtext,newtxmath}
\usepackage[T1]{fontenc}
\usepackage[utf8]{inputenc}
\usepackage{ae,aecompl}

\usepackage[usenames, dvipsnames]{color}
\usepackage{graphicx}	\usepackage{amsmath}	\usepackage{pifont}
\usepackage{multirow}
\usepackage{siunitx}
 
\newcommand{\cm}{\ensuremath{\, \mathrm{cm}}}

\newcommand{\km}{\ensuremath{\, \mathrm{km}}}

\newcommand{\Mpc}{\ensuremath{\, \mathrm{Mpc}}}

\newcommand{\cMpc}{\ensuremath{\, \mathrm{cMpc}}}

\newcommand{\Msun}{\ensuremath{\, \mathrm{M}_{\odot}}}
\newcommand{\Rsun}{\ensuremath{\, \mathrm{R}_{\odot}}}
\newcommand{\s}{\ensuremath{\, \mathrm{s}}}

\newcommand{\Myr}{\ensuremath{\, \mathrm{Myr}}}

\newcommand{\Hz}{\ensuremath{\, \mathrm{Hz}}}
\newcommand{\G}{\ensuremath{\, \mathrm{G}}}

\newcommand{\dex}{\ensuremath{\mathrm{\, dex}}}
\newcommand{\erg}{\ensuremath{\mathrm{\, erg}}}

\newcommand{\ie}{\emph{i.e.}\, }
\newcommand{\eg}{\emph{e.g.}\, }

\renewcommand{\d}{\mathrm{d}}
\renewcommand{\log}{\mathrm{log}_{10}}
\newcommand{\few}{\mathrm{few\, }}

\newcommand{\Mbh}{\ensuremath{M_\bullet}}
\newcommand{\mstar}{\ensuremath{m_\star}}
\newcommand{\rstar}{\ensuremath{r_\star}}

\renewcommand{\vv}[1][args]{\mathbf}

\newcommand{\hugo}[1]{#1}

\title[GW from TDEs and their EM counterpart]{Observable gravitational waves from tidal disruption events and their electromagnetic counterpart}

\author[H. Pfister et al.]{
Hugo Pfister,$^{1,2}$\thanks{Sophie and Tycho Brahe Fellow; pfisterastro@gmail.com}
Martina Toscani,$^{3}$
Thomas Hong Tsun Wong,$^{1}$
Jane Lixin Dai,$^{1}$\thanks{lixindai@hku.hk} \newauthor
Giuseppe Lodato$^{3}$ 
and Elena M. Rossi$^{4}$\\
$^{1}$Department of Physics, The University of Hong Kong, Pokfulam Road, Hong Kong, China\\
$^{2}$DARK, Niels Bohr Institute, University of Copenhagen, Jagtvej 128, 2200 København, Denmark\\
$^{3}$Dipartimento di Fisica, Universit\`a Degli Studi di Milano, Via Celoria, 16, Milano, 20133, Italy\\
$^{4}$Leiden Observatory, Leiden University, PO Box 9513, 2300 RA, Leiden, the Netherlands
}

\date{Accepted XXX. Received YYY; in original form ZZZ}

\pubyear{2021}

\begin{document}
\label{firstpage}
\pagerange{\pageref{firstpage}--\pageref{lastpage}}
\maketitle

\begin{abstract}
We estimate the rate of tidal disruption events (TDEs) that will be detectable with future gravitational wave detectors as well as the most probable properties of these events and their possible electromagnetic counterpart. To this purpose we combine standard gravitational waves and electromagnetic results with detailed rates estimates. We find that the \emph{Laser Interferometer Space Antenna} (LISA) should not detect any TDEs, unless black holes (BHs) are typically embedded by a young stellar population which, in this situation, could lead up to few 10 events during the duration of the mission. If there are gravitational wave observations, these events should also be observable in the X-ray or the optical/UV part of the electromagnetic spectrum, which may open up the multi-messenger era for TDEs. The generation of detectors following LISA will at least yearly observe $10^4$ TDEs at cosmological distances, allowing to do population studies and constrain the black hole mass function. In all cases, most probable events should be around black holes with a mass such that the Keplerian frequency at the Schwarzschild radius is similar to the optimal frequency of the detector and with a large penetration factor.
\end{abstract}


\section{Introduction}
\label{sec:Introduction}

If a star gets too close to a black hole (BH), the tidal force of the latter can disrupt the former. The stellar debris is then accreted by the BH, and this results in bright flares. These events are known as tidal disruption events \citep[TDEs, ][]{Hills_75, Lacy_82, Rees_88}, and they are now routinely detected by optical wide field surveys \citep[\eg ][]{VanVelzen_20, YSE_paper}.

The dynamics of the debris while falling onto the BH is a complex, non-linear problem that depends on orbital properties of the initial BH-star system (\eg eccentricity and pericenter of the star, see \citealt{LawSmith_20} and \citealt{2020ApJ...904..100R}), on the intrinsic properties of the star (\eg internal structure: \citealt{LKP09,Guillochon_13,2019MNRAS.487..981G,2019ApJ...882L..26G}; or rotation: \citealt{golightly19,sacchi19}), and on the intrinsic properties of the BH \citep[\eg  spin; ][]{Kesden_12}. Nonetheless, several recent hydrodynamic codes allow to perform simulations of this problem \citep[see review of][]{lodato20}, sometimes in a fully general relativistic context \citep{liptai19, Ryu_20,2020ApJ...904...99R}. Since the evolution of the rate at which debris are accreted by the BH can be directly measured in these simulations, the time evolution of the luminosity can be directly mapped to the initial parameters of the TDE, and this idea has been recently exploited to reverse-engineer parameters of observed TDE lightcurves \citep{Mockler_19, Ryu_20b}.

These extreme events involving massive BHs also result in the emission of gravitational waves, which also carry information about the nature of the system \citep{Kobayashi_04, Stone_20, Toscani_20}. Gravitational waves being independent from electromagnetic waves, observing events in both domains allows us to better understand them: this is a branch of multi-messenger astronomy which has recently started with the spectacular observation of a binary neutron star merger both in the electromagnetic and the gravitational spectrum \citep{LIGO_17b}. The future space based gravitational wave detectors \textit{Laser Interferometer Space Antenna} \citep[LISA,][]{LISA_Proposal}, TianQin \citep{Tianqin_proposal}, \textit{DECI-hertz inteferometer Gravitational wave Observatory} \citep[Decigo,][]{Decigo_proposal}, \textit{Advanced Laser Interferometer Antenna} \citep[ALIA,][]{Alia_proposal} and \textit{Big bang observatory} \citep[Bbo][]{BBO_proposal} will be designed to study supermassive BHs. As such, one can naturally wonder if we will enter in the multi-messenger astronomy era for TDEs, and what this will unveil about our understanding of BHs.

To address these questions, we estimate in this paper the expected rates that will be observed with these future gravitational wave detectors, what the properties of observed events will be, and if there will be any associated electromagnetic counterpart. We start by describing the conditions to observe a single event in \S\ref{sec:DetectionOfSingleEvents}; we then compute the rates for a global population of galaxies in \S\ref{sec:TDErates}; we finally give our results and conclusions in \S\ref{sec:results} and \S\ref{sec:conclusions}.

\section{Detection of single events}
\label{sec:DetectionOfSingleEvents}

In this Section, we assume that a TDE occurs at a given redshift~($z$): a star of mass and radius $m_\star$ and $r_\star$ plunges into a BH with mass $M_\bullet$. The nearly parabolic orbit of the star is such that, at pericenter ($r_p$), it gets closer to the BH than the tidal radius ($r_T=\rstar(\Mbh/\mstar)^{1/3}$) where tidal force overcomes the stellar self-gravity, resulting in a TDE {\citep[we do not consider partial TDEs in this study and assume stars are fully disrupted for $r_p \leq r_T$, see ][]{2020ApJ...904...99R}}. During its journey, the star-BH system emits gravitational waves that may be detected when they arrive on Earth. In \S\ref{sec:GWfromTDE} we describe the formalism to define if a TDE is observable through gravitational waves and derive the maximum redshift for such observations; in \S\ref{sec:EMfromTDE} we perform the same exercise for the electromagnetic counterpart with the goal of determining whether multi-messenger detectable events are likely or not.

\subsection{Gravitational waves}
\label{sec:GWfromTDE}

\subsubsection{Formalism}
\label{sec:Formalism}

{When a star orbiting a BH arrives at pericenter, the time dependent mass quadrupole moment tensor of the system star-BH results in a burst of gravitational waves whose characteristic strain and frequency can be estimated as} \citep{Kobayashi_04, Stone_19}:
\begin{eqnarray}
h_\mathrm{GW} &=& \frac{2\G^2 m_\star M_\bullet}{\chi(z) c^4 r_p} \label{eq:hGW_1}\\
\nonumber &=&10^{-22} \times \\
&& \beta \left( \frac{\chi(z)}{16 \Mpc} \right)^{-1} \left(\frac{r_\star}{\Rsun}\right)^{-1} \left(\frac{m_\star}{\Msun}\right)^{4/3} \left(\frac{\Mbh}{10^6\Msun}\right)^{2/3} \nonumber\\
f_\mathrm{GW} &=& \left(\frac{\G \Mbh}{4\pi^2 r_p^3}\right)^{1/2} \label{eq:fGW_1}\\
\nonumber &=&10^{-4} \Hz \times \beta^{3/2} \left(\frac{m_\star}{\Msun}\right)^{1/2} \left(\frac{r_\star}{\Rsun}\right)^{-3/2} \label{eq:fGW_2} \, ,
\end{eqnarray}
where $\chi(z)$ is the comoving distance \citep[we assume a $\Lambda$CDM cosmology with \emph{Planck} parameters; ][]{Planck_15}; $\beta=r_T/r_p$ is the penetration factor {(recall that in this study we have $\beta \geq 1$ as we only consider full disruption of stars below the tidal radius)}; $c$ and G are respectively the speed of light and gravitational constant; and keeping in mind that the frequency is redshifted while travelling to Earth so that the observed frequency is:
\begin{eqnarray}
f_\mathrm{obs} =  \frac{f_\mathrm{GW}}{1+z} \label{eq:fobs}\, .
\end{eqnarray}

For highly penetrating orbits, the star is swallowed whole resulting in a direct plunge. While this may result in the emission of observable gravitational waves, the debris would not form a luminous accretion disk, and there would not be any electromagnetic counterpart. \hugo{In the simple ``Newtonian picture with a BH of size the Schwarzschild radius of the BH'' ${r_\mathrm{Sch}=2\G\Mbh/c^2}$, a star on parabolic orbit will directly plunge for ${r_p \leq r_\mathrm{Sch}}$. In the more realistic relativistic picture, the orbit of the star will follow the geodesic, and it is not straightforward to know what should be the initial pericenter of the parabolic orbit such that the star penetrates the BH. For simplicity, we only keep orbits with:}
\begin{eqnarray}
\beta \leq \beta_\mathrm{max}&=& \frac{r_T}{\kappa r_\mathrm{Sch}} \label{eq:beta_max}\\ 
&=&\frac{\rstar c^2}{2 \kappa G\mstar^{1/3}\Mbh^{2/3}} \\
\nonumber &=&  12 \left(\frac{\kappa}{2}\right)^{-1} \left(\frac{\mstar}{\Msun}\right)^{-1/3} \left(\frac{\rstar}{\Rsun}\right) \left(\frac{\Mbh}{10^6\Msun}\right)^{-2/3} \, ,
\end{eqnarray}
where $\kappa$ indicates some critical radius (in units of $r_\mathrm{Sch}$) for direct plunge. \hugo{When it comes to the number of events per year, a \textit{larger} $\kappa$ naturally results in \textit{less} TDEs, and vice-versa. We tried with $\kappa=1$ and $\kappa=2$ and found the results to be changed by a factor of $\sim 2$ only. However, since \cite{Kesden_12} has shown that $\kappa \sim 2$ \textit{nearly reproduced the correct relativistic} rates (see their \S~IV.A.), we will use $\kappa=2$ ($r_p\leq 2 r_\mathrm{Sch}$) throughout the paper. We stress again that these are purely dynamical considerations, and that some of these events may actually not be observable. For instance, \cite{2020ApJ...904...68K} suggest that the rate of detectable events with $r_p\leq 7r_\mathrm{Sch}$ could actually be lower than that of direct captures.}

In order to reduce the dimensionality of the study, we assume that the mass and radius of stars are broken power-law related \citep[{$r_\star\propto m_\star^\theta$}; ][]{Kippenhahn_90}:
\begin{eqnarray}
\frac{r_\star}{\Rsun} = \left\{
    \begin{aligned}
        &\left(\frac{m_\star}{\Msun}\right)^{0.8} &&\mathrm{\, if\, }m_\star \leq \Msun\\
        &\left(\frac{m_\star}{\Msun}\right)^{0.57} &&\mathrm{\, if\, }m_\star \geq \Msun \, .
     \label{eq:rstarVSmstar}
    \end{aligned}
    \right.
\end{eqnarray}
{While this relation is technically valid for stars with $m_\star~\lesssim~60~\Msun$, for some models (see \S\ref{sec:StellarMassFunction}), we will extrapolate up to $100 \Msun$ in order to mimic a young stellar population.}

With all this, for a TDE involving a star with mass $m_\star$ on an orbit with a penetration factor $\beta$ around a BH with mass $\Mbh$ occuring at redshift $z$, we are now able to estimate the strain ({$h_\mathrm{GW}\propto \beta \chi^{-1} \mstar^{4/3-\theta} \Mbh^{2/3}$, where $4/3-\theta>0$)}) and frequency ({$f_\mathrm{obs}\propto \beta^{3/2} \mstar^{(1-3\theta)/2}(1+z)^{-1}$, where $(1-3\theta)/2<0$}) of gravitational wave when they arrive on Earth.

{As a remark, during TDEs, there can also be other mechanisms resulting in the emission of gravitational waves, \eg pulsation of the star due to the tides \citep{2009ApJ...705..844G, Stone_19} or instabilities once the accretion disk is formed \citep{Toscani_19}. We do not consider these processes in this work, such that we finally obtain lower estimates of the strain. We note however that these other processes are usually negligible \citep{ Stone_19}.}

\subsubsection{Maximum redshift for detection}
\label{sec:MaximumRedshiftForDetection}

In order to know if the event is detectable, we must compare the strain to the sensitivity of the detector. We define TDEs which signal is at least a factor $ S/N_\mathrm{lim}$ larger than the characteristic amplitude noise of the detector at the observed frequency $h_\mathrm{det}(f_\mathrm{obs})$ \citep{Maggiore_08,Colpi_17}, as ``detected'' events.  In other words the strain of the signal has to be above the sensitivity curve in Fig.~\ref{fig:SensitivityDet}, this yields:
\begin{eqnarray}
\frac{h_\mathrm{GW}}{h_\mathrm{det}(f_\mathrm{obs})} = S/N  \geq S/N_\mathrm{lim}  \label{eq:SNR}\, .\end{eqnarray}

\begin{figure}
\includegraphics[width=\columnwidth]{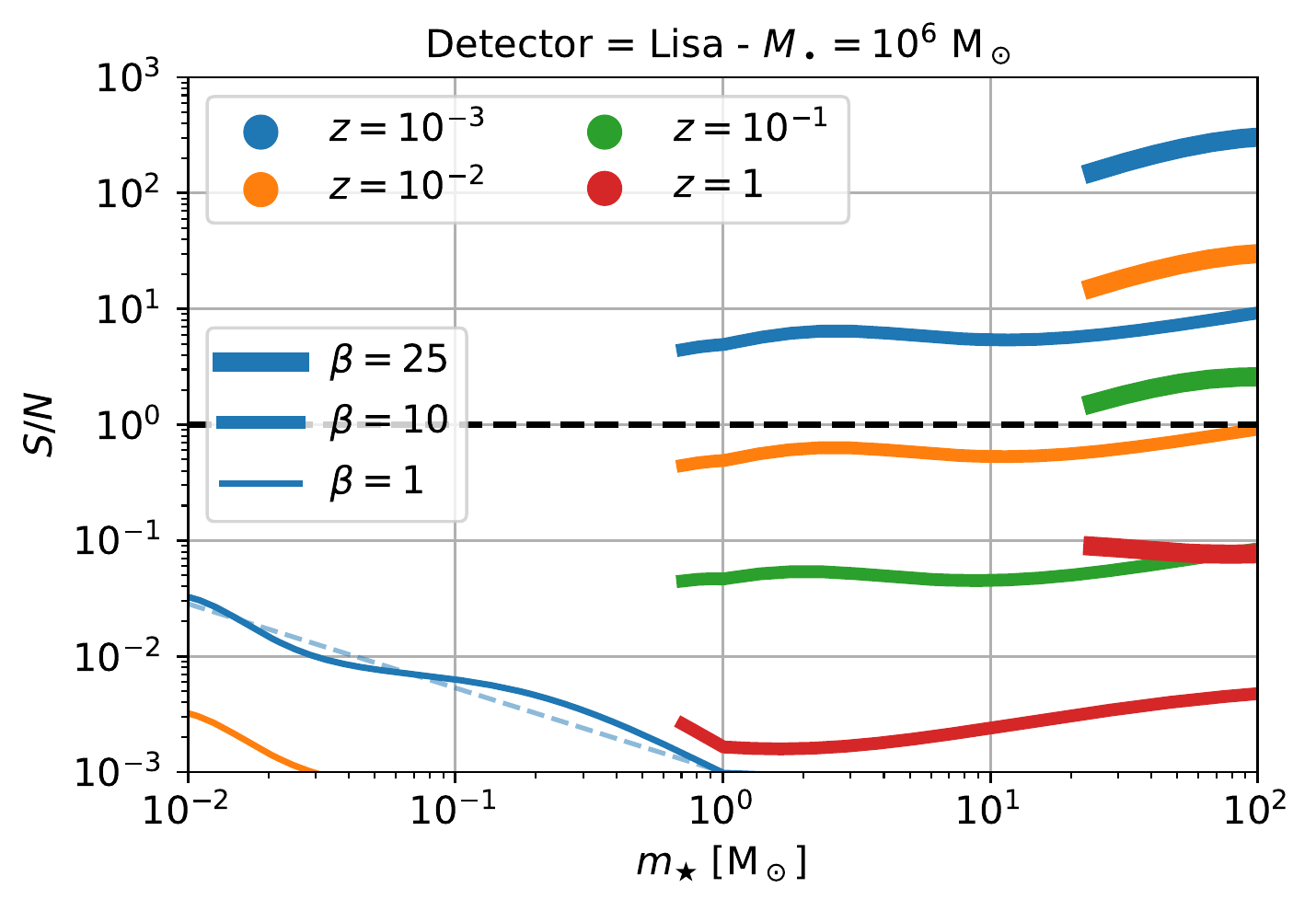}
\caption{Signal to noise ratio of a LISA observation of a TDE as a function of the mass of the disrupted stars for a $10^6\Msun$ BH. We show the results when the penetration factor is changed (thicker lines referring to larger $\beta$) and when the TDE occurs at a different redshift, hence at a different comoving distance (shown as different colors). For larger $\beta$, the curves are truncated at small masses, when stars penetrate twice the Schwarzschild radius of the BH. The horizontal black dashed line indicates a $S/N=1$ which can be considered as an (optimistic) limit for detection, and the thin light blue dashed line guide the eye to indicates $S/N\propto m_\star^{-0.7}$ as predicted by Eq.~\eqref{eq:SNVSmstar}. Gravitational waves from $\beta=1$ TDEs (thin lines) will not be observed by LISA; gravitational waves from TDEs of Sun-like stars will be observed up to $z\sim10^{-2}$ for moderately $\beta=10$ penetrating orbits (orange medium lines); at $z\sim0.1$, only gravitational waves from extreme events ($\beta=25$ and $m_\star \gtrsim 20\Msun$; green thick line) can be observed.}
\label{fig:SNVSmstar}
\end{figure}

To give an example, we show in Fig.\ref{fig:SNVSmstar} the value of $S/N$ for an observation with LISA\footnote{See sensitivity curve of different detectors in Fig.~\ref{fig:SensitivityDet}.} of gravitational waves emitted by a TDE of a star with mass $m_\star$, for different $\beta$ (thickness of the lines) and $z$ (colors) around a $10^6\Msun$ BH. We also show the (optimistic) limit for detection $S/N_\mathrm{lim}=1$ with an horizontal black dashed line. Orbits penetrating twice the Schwarzschild radius ($\kappa=2$ in Eq.~\eqref{eq:beta_max}) are excluded, and this is why the curves at high $\beta$ are truncated at low mass. We find that $\beta=1$ orbits (thin line), \ie orbits that barely penetrate the TDE radius, cannot be detected with LISA; moderately penetrating orbits with $\beta=10$ (medium lines) can be detected up to $z\sim0.01$ (orange line) for $\gtrsim 1\Msun$ stars; and $\beta=25$ extremely penetrating orbits (thick lines) may be observed up to $z\sim 0.1$ (green line) for massive $\gtrsim 20\Msun$ stars. 

It can be somewhat surprising that sometimes, for instance when $\beta=1$ (thin lines), $S/N$ \textit{decreases} with \textit{increasing} \mstar while Eq.~\eqref{eq:hGW_1} predicts that the strain \textit{increases} with \textit{increasing} \mstar, \ie larger stellar mass results in more ``violent'' event. The reason is that $f_\mathrm{GW}$ (hence $f_\mathrm{obs}$) also varies (Eq.~\eqref{eq:fGW_1}), and that the characteristic amplitude noise of LISA is not flat (see Fig.~\ref{fig:SensitivityDet}). {To be more quantitative, we start with Eq.~\eqref{eq:SNR} in which we substitute $h_{\rm GW}$ and $f_{\rm obs}$ by their expressions from Eqs. \eqref{eq:hGW_1} and \eqref{eq:fobs}:
\begin{equation}
S/N = \frac{\pi^{2/3} 2^{5/3} \G^{5/3}}{c^4} \times
m_\star \Mbh^{2/3} \frac{1}{\chi(z)} \frac{f_\mathrm{GW}^{2/3}}{h_\mathrm{det}\left( \frac{f_\mathrm{GW} }{ 1+z}\right)}\, . \label{eq:zmax_start}
\end{equation}
If we further use  for $h_\mathrm{det}$ a broken power-law 
($h_\mathrm{det}(f) = h_\mathrm{opt} \left( {f}/{f_\mathrm{opt}} \right)^{-a} \left( \left( {f}/{f_\mathrm{opt}} \right)^{c} +1  \right)^{{(b + a)}/{c}}$, where $f_\mathrm{opt}$ is the optimal frequency of the detector, see Appendix~\ref{sec:SensitivityOfGWDetectors}), we find:
\begin{eqnarray}
S/N \propto  \left\{
    \begin{aligned}
        & m_\star^{1+(1-3\theta)(\frac{1}{3}+\frac{a}{2})} &&\mathrm{\, if\, } f_\mathrm{GW} \ll f_\mathrm{opt}\\
        & m_\star^{1+(1-3\theta)(\frac{1}{3}-\frac{b}{2})} &&\mathrm{\, if\, } f_\mathrm{GW} \gg f_\mathrm{opt} 
        \label{eq:SNVSmstar}
        \, .
    \end{aligned}
    \right.
\end{eqnarray}
At this point, it is worth noting that, for $\beta=1$, $f_\mathrm{GW} \lesssim 10^{-3}\Hz$ for $ m_\star \gtrsim 0.01 \Msun$ such that for the different detectors (see Table~\ref{tab:fdet}), $\beta=1$ TDEs are always in the regime $f_\mathrm{GW} \ll f_\mathrm{opt}$. For the particular case of LISA ($a=1.8$) shown for $m_\star \leq \Msun$ ($\theta=0.8$ in Eq.~\eqref{eq:rstarVSmstar}) in Fig.~\ref{fig:SNVSmstar}, we find $S/N \propto m_\star^{-0.7}$ (thin light blue dashed line), in excellent agreement with the numerical estimate. Conversely, the highest possible $f_\mathrm{GW}$ is obtained for $\beta=\beta_{\max}$:
\begin{eqnarray}
f_\mathrm{GW,\, max} &=& \left( \frac{c^6}{32 \kappa ^3 \pi^2 \G^2 \Mbh^2} \right)^{1/2} \\
\nonumber &\sim& 4\times 10^{-3}\Hz\,  \left(\frac{\kappa}{2}\right)^{-3/2} \left(\frac{\Mbh}{10^6\Msun}\right)^{-1} \, ,
\end{eqnarray}
which only gives $f_\mathrm{GW,\, max} \gg f_\mathrm{opt} $ for $\lesssim 10^4 \Msun$ BHs not considered in this study.}

\begin{figure}
\includegraphics[width=\columnwidth]{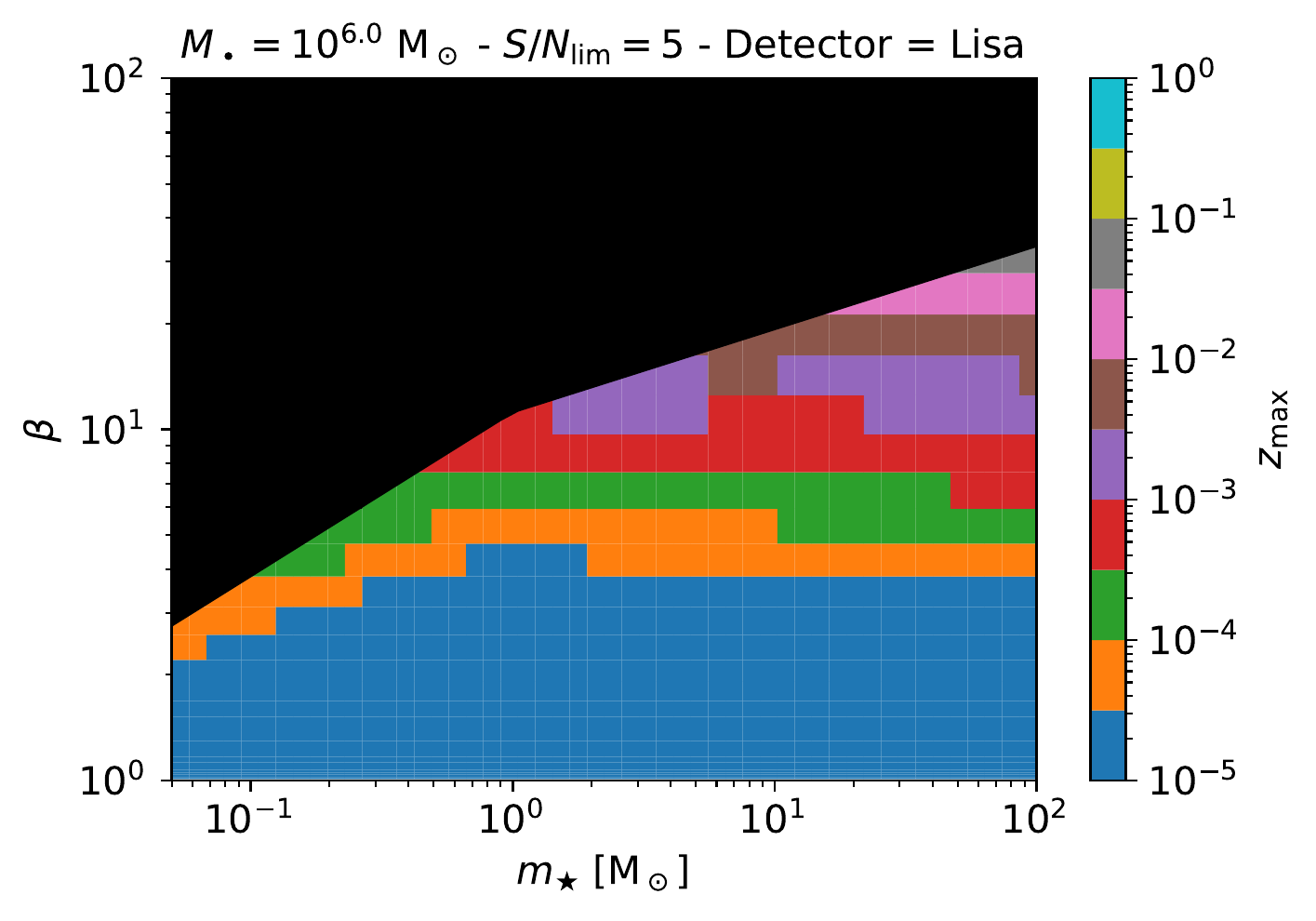}
\caption{Maximum redshift for observation with LISA of a TDE of a star with mass $m_\star$ on an orbit with penetration factor $\beta$ around a $10^6\Msun$ BH assuming detection if $S/N\geq5$. The black region indicates when the orbit plunges directly toward the BH ($r_T/\beta < 2 r_\mathrm{Sch}$). Conclusions are identical to that of Fig.~\ref{fig:SNVSmstar}.}
\label{fig:zmaxVSbetaANDmstar_1e6}
\end{figure}

{For a given detector ($h_{\rm det}$ known) and fixed $S/N=S/N_{{\lim}}$, \Mbh, \mstar\, and $\beta$, we can solve Eq.~\eqref{eq:zmax_start} to obtain $z_{\max}$. In Fig.~\ref{fig:zmaxVSbetaANDmstar_1e6} we show, for the particular case of LISA, and for $S/N_{{\lim}} =5$ and $M_\bullet=10^6\Msun$, the value of $z_\mathrm{max}$ as a function of the two parameters left, $m_\star$ and $\beta$. Here we recognize that, overall, a good rule of thumb is that TDEs produced by massive stars on penetrating orbits can be detected up to higher redshift, but the details ultimately depend on the complex shape of the sensitivity curve of the gravitational wave detector (LISA in this case).}

\begin{figure*}
\includegraphics[width=\columnwidth]{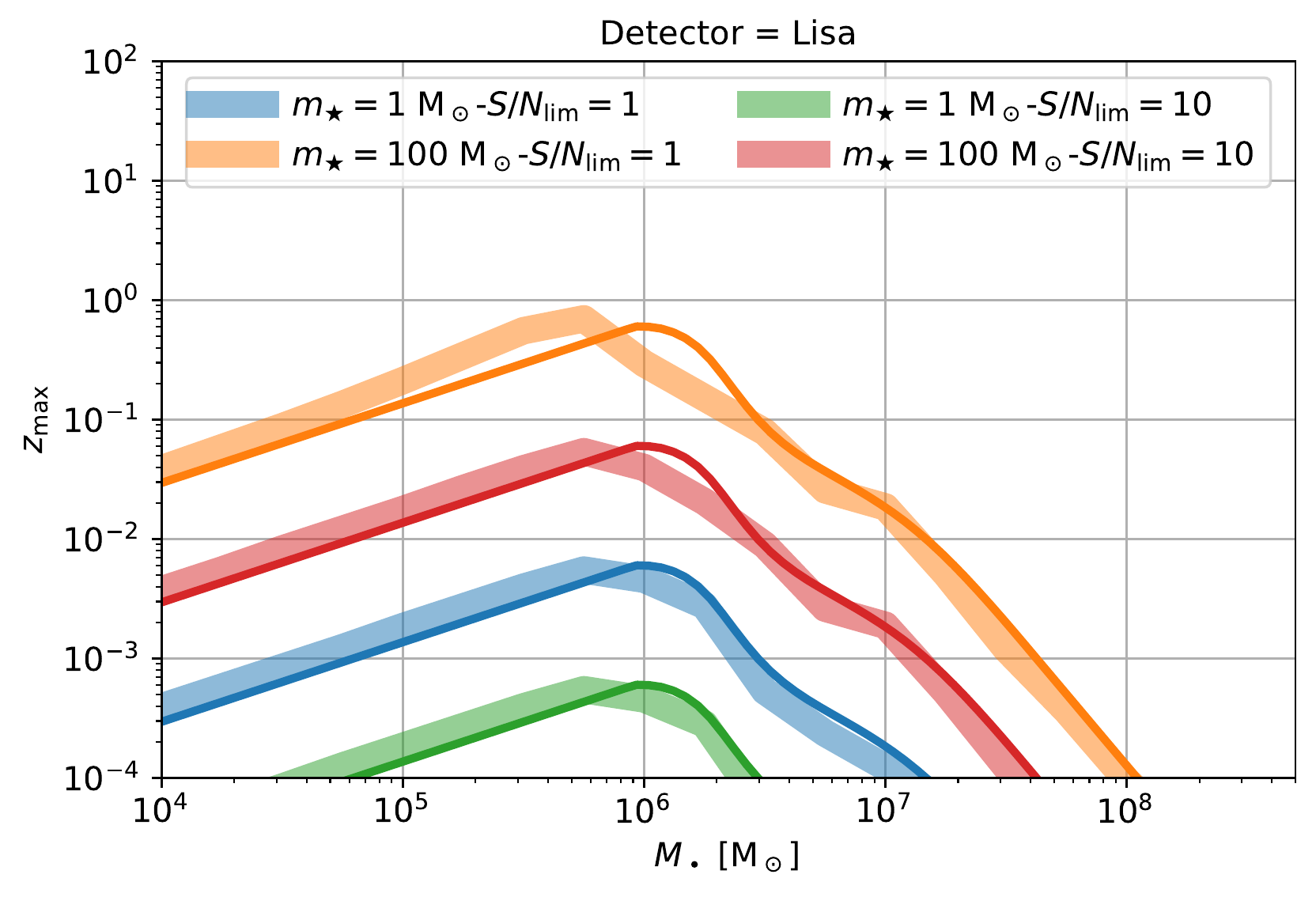}\hfill
\includegraphics[width=\columnwidth]{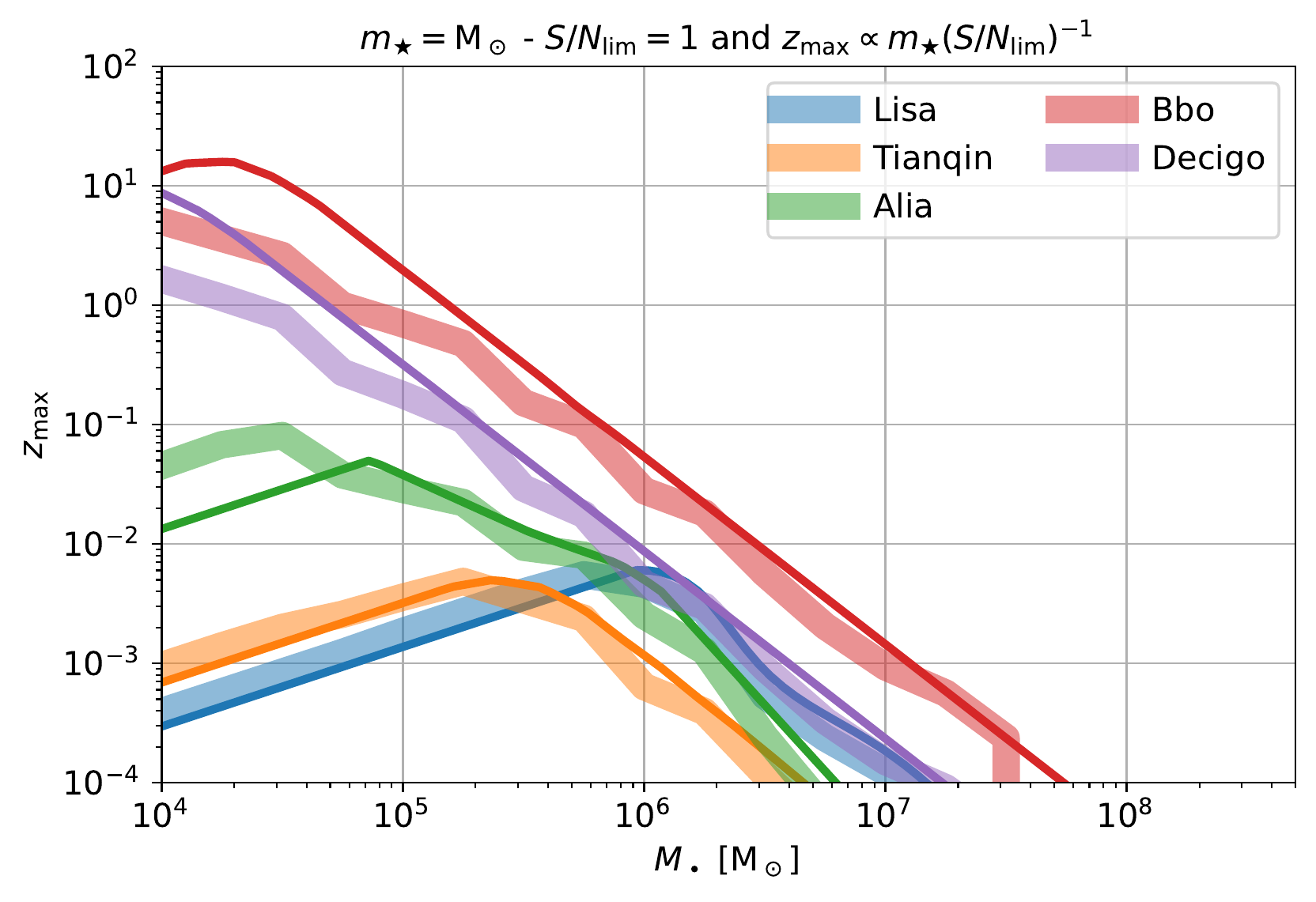}
\caption{\textbf{Left:} Maximum redshift at which a TDE can be detected with LISA as a function of the mass of the BH around which the TDE occurs. We explore different stellar masses and thresholds for the signal to noise ratio (colors) and show the results of a numerical search (thick light lines) and from Eq.~\eqref{eq:zmax} (thin dark lines). LISA is optimal for a detection around a $10^6\Msun$ BH. \textbf{Right:} Same, but for $\mstar=1\Msun$ and $S/N_\mathrm{lim}=1$ and different detectors (colors). The generation of gravitational waves detectors following LISA will be able to detect TDEs from intermediate mass BHs up to cosmological redshifts.}
\label{fig:zmaxVSMBH}
\end{figure*}

{In order to understand the dependency of $z_{\max}$ with \Mbh, we fix \mstar\, and $S/N_{\lim}$, and we solve Eq.~\eqref{eq:zmax_start} across all the possible $\beta$. We show the results in Fig.~\ref{fig:zmaxVSMBH} (light thick lines), where we explore different stellar masses $m_\star$ and $S/N_{{\lim}}$ for LISA (left panel), as well as for future gravitational waves detectors but with fixed $m_\star=\Msun$ and $S/N_{{\lim}}=1$ (right panel). Here, the exact shape of the sensitivity curve appears even more clearly: LISA is optimal around $10^6\Msun$ BHs.}

All this can be understood as follows. Starting with Eq.~\eqref{eq:zmax_start} and using that, for $z\ll1$, we can approximate $\chi\sim~c z / H_0$ ($H_0$ is the Hubble constant), we have: 
\begin{equation}
z \underset{z \ll 1}{\sim}\frac{\pi^{2/3} 2^{5/3} \G^{5/3} H_0}{c^5} \times \label{eq:zmax_long} 
\frac{m_\star \Mbh^{2/3}}{S/N} \frac{f_\mathrm{GW}^{2/3}}{h_\mathrm{det}\left( f_\mathrm{GW} \right)}.
\end{equation}
If $f_\mathrm{GW,\, max} \geq f_\mathrm{opt}$, then there exists $\beta$ for which $h_\mathrm{det}\left( f_\mathrm{GW} \right)=h_\mathrm{opt}$, and as $h_\mathrm{det}$ is a steep function of $f$, this is where the maximal $z$ is obtained. In the other situation, if $f_\mathrm{GW,\, max} \leq f_\mathrm{opt}$, then, given the U-shape of $h_\mathrm{det}$, the maximal $z$ is obtained for $f_\mathrm{GW,\, max}$.
Overall, Eq.~\eqref{eq:zmax_long} giving the maximum redshift can be wrapped up as:
\begin{eqnarray}
z_\mathrm{max} &=& 0.01 \left(\frac{m_\star}{\Msun}\right)^{} \left(\frac{\Mbh}{10^6\Msun}\right)^{2/3}  \times \label{eq:zmax} \\
&& S/N_{{\lim}}^{-1}
\left(\frac{f_\star}{10^{-2}\Hz}\right)^{2/3} \left(\frac{h_\mathrm{det}(f_\star)}{10^{-21}}\right)^{-1},   \nonumber
\end{eqnarray}
where
\begin{equation}
f_\star = \mathrm{min}\left( 4\times 10^{-3}\Hz\, \left(\frac{\kappa}{2}\right)^{-3/2} \left( \frac{\Mbh}{10^6\Msun} \right)^{-1} , f_\mathrm{opt}\right) \, .
\end{equation}
We conclude that $z_\mathrm{max}\propto m_\star  S/N_{{\lim}}^{-1}$ is always true: more massive stars can be detected farther away. However, for BHs with ${f_\mathrm{GW,\, max} \geq f_\mathrm{opt}}$ (BHs lighter than $M_{\bullet,\,{\rm opt}}$, see Table~\ref{tab:fdet}), we have $z_\mathrm{max}\propto~M^{2/3}_\bullet$; and for ${f_\mathrm{GW,\, max} < f_\mathrm{opt}}$ (``massive'' BHs) $z_\mathrm{max}$ decreases faster than $M^{-1/3}_\bullet$: even though the signals are stronger for more massive BHs, there exists an optimal BH mass for detection given by ${f_\mathrm{GW,\, max}(M_\bullet)=f_\mathrm{det}}$. For LISA, this optimal mass is $M_{\bullet,\,{\rm opt}}\sim10^6\Msun$, and this is why $z_{\max}$ peaks at this value in Fig.~\ref{fig:zmaxVSMBH}~(left). We also show  in Fig.~\ref{fig:zmaxVSMBH} the results of Eq.~\eqref{eq:zmax} (thin lines), apart when $z_\mathrm{max}\sim1$ and our hypothesis $z\ll1$ is not correct anymore, the numerical estimate and Eq.~\eqref{eq:zmax} are in excellent agreement.

Overall, LISA can realistically detect TDEs only up to $z~\sim~0.01-0.1$ (depending on $\mstar$ and $S/N_{\lim}$), but the next generation detectors will be able to detect TDEs around intermediate mass BHs up to cosmological redshifts. In all cases, detectors are most sensible to BHs for which the \hugo{Keplerian frequency around the critical radius for direct plunge ($\kappa \times r_{\rm Sch}$)} is the same as the optimal frequency of the detector, about $10^6\Msun$ for LISA (see Table~\ref{tab:fdet}).

\begin{table}
    \centering
    \begin{tabular}{|l|c|c|c|}
    \hline
         \textbf{Detector} & $f_\mathrm{opt}$ & $h_\mathrm{det}(f_\mathrm{opt})$ & $M_{\bullet,{\rm opt}}$ \\
         & Hz & $10^{-21}$ & \Msun \\
         \hline
LISA    & $6\times 10^{-3}$& 0.2    & $7 \times 10^5 $\\
        Tianqin & $0.02$           &  7     &$2 \times 10^5 $\\
        Alia    & $0.08$           &  0.02  &$5 \times 10^4 $\\
        Bbo     & $0.3$            &  0.01  &$1 \times 10^4 $\\
        Decigo  & $0.4$            &  0.04  &$1 \times 10^4 $ \\
         \hline
    \end{tabular}
    \caption{Different detectors considered in this study. We indicate the optimal frequency ($f_{\rm opt}$), strain ($h_{\rm det}(f_{\rm opt})$) and BH mass ($M_{\bullet,\,{\rm opt}}$) of these detectors for detections of TDEs.}
    \label{tab:fdet}
\end{table}

\subsection{Electromagnetic counterpart}
\label{sec:EMfromTDE}

TDEs are very luminous electromagnetic sources bright both in the X-ray \citep[\eg][]{Saxton_20} and in the optical \citep[\eg][]{VanVelzen_20} part of the spectrum. With the aim of exploring possible TDEs observed as multi-messenger sources, we estimate, using observationally motivated models, the electromagnetic luminosity of TDEs. We begin with the optical emission in \S\ref{sec:OpticalEmission} and then move to the X-ray emission in \S\ref{sec:XrayEmission}.

\subsubsection{Optical emission}
\label{sec:OpticalEmission}
\paragraph*{Formalism: }
The origin of the optical emission is still debated: while some groups believe it is caused by shocks during the circularization process \citep[\eg][]{lodato12,Piran_15, Shiokawa_15}, others believe it is sourced by the accretion luminosity reprocessed in the expanding outflow \citep[\eg][]{Lodato_11, Roth_18, Dai_18}. Discussion about the origin of the optical component is beyond the scope of this paper, and we instead adopt an observational-driven approach. We {assume that the optical emission of TDEs can be well modeled with a black-body with temperature $T_{\rm opt}=3\times 10^4\,\text{K}$ (mean temperature from Table 3 of \citealt{Wevers_17} but see also \citealt{2020SSRv..216..124V})}, and a luminosity equals to the Eddington limit:
\begin{align}
L_{\rm Edd} = 3\times 10^4 \text{L}_{\odot}\left(\frac{M_{\bullet}}{M_{\odot}}\right) \, ,
\end{align} 
where \text{L}$_{\odot}$ is the luminosity of the Sun. It should be noted that not all observed TDEs emit up to Eddington luminosity \citep[\eg][]{Hung_17}. Furthermore, in our vast parameter space, the Eddington luminosity can sometime exceed the luminosity at peak (\eg \citealt{Evans_89, Stone_13}; Eq.~(24) of \citealt{Stone_16a}), for this reason the final adopted luminosity is:
\begin{eqnarray}
L=L_{\rm Edd}\times{\rm min}\left(1; 133 \left(\frac{M_\bullet}{10^6\Msun}\right)^{-3/2} \left(\frac{m_\star}{\Msun}\right)^{2}\left(\frac{r_\star}{\Rsun}\right)^{-3/2}\right) \, , \label{eq:L_adopted}
\end{eqnarray}
where we adopted a constant radiative efficiency of 0.1.

This allows to write the spectral flux density \citep[appropriately redshifted as in the Appendix~A of ][]{Roth_20}:
\begin{align}
 F_{\rm opt}(\lambda)&= \label{eq:flux}\\
 &\frac{(1+z)^2}{\chi(z)^2}\times\frac{2hc^2}{\lambda^5}\times\frac{R^2_{\rm opt}}{\exp\left(hc(1+z)/\lambda k_B T_{\rm opt}\right)-1}, \nonumber
\end{align}
where $\lambda$ is wavelength in the observer frame; $h$, $k_B$ and $\sigma_S$ are respectively the Planck, the Boltzmann and the Stefan constants; and we defined the typical black-body radius of emission ($R_{\rm opt}$) from the Stefan-Boltzmann law:
\begin{eqnarray}
 R_{\rm opt} = \sqrt{\frac{L}{4 \pi \sigma_S T^4_{\rm opt}}}\, .
\end{eqnarray}
From all this, the $G$ magnitude associated to the flux is
\begin{align}
    M_G = -2.5 \log\Bigg(9\times\left(\frac{\lambda_{\rm G}}{{\si\angstrom}}\right)^2\times \frac{F_{\rm opt}(\lambda_{\rm G})}{\text{erg}\,\text{cm}^{-2}\text{s}^{-1}{\si\angstrom}^{-1}}\Bigg),
\label{eq:Def_G}
\end{align}
where $\lambda_{\rm G}\sim 464\,\text{nm}$ is the central wavelength of the $G$ band. 

Note that, for $L=L_{\rm Edd}$ (most cases in our parameter space), we can approximate Eq.~\eqref{eq:Def_G} for small $z$ similarly to \S\ref{sec:MaximumRedshiftForDetection} to obtain the following simple expression:
\begin{align}
    M_G \sim 28 -2.5\log\left( \frac{\Mbh}{10^6\Msun} \right) + 5\log(z) \, .
\label{eq:G_lowz}
\end{align}
Under the observationally motivated assumption $L=L_{\rm Edd}$, the magnitude of a TDE solely depends on the mass of the BH and the cosmological distance.

\paragraph*{Multimessenger TDEs:}
For a TDE with given properties, we can now estimate what the maximum redshift is at which it can be observed with a gravitational wave detector (Eq.~\eqref{eq:zmax}) as well as the $G$ magnitude of this event (Eq.~\eqref{eq:Def_G}). In order to address which events can be detected both in the electromagnetic and the gravitational spectrum, we can estimate the $G$ magnitude in the most pessimistic scenario ($M_{G, {\rm lim}}$): when the TDE occurs at $z_{\rm max}$.

\begin{figure}
    \centering
\includegraphics[width=\columnwidth]{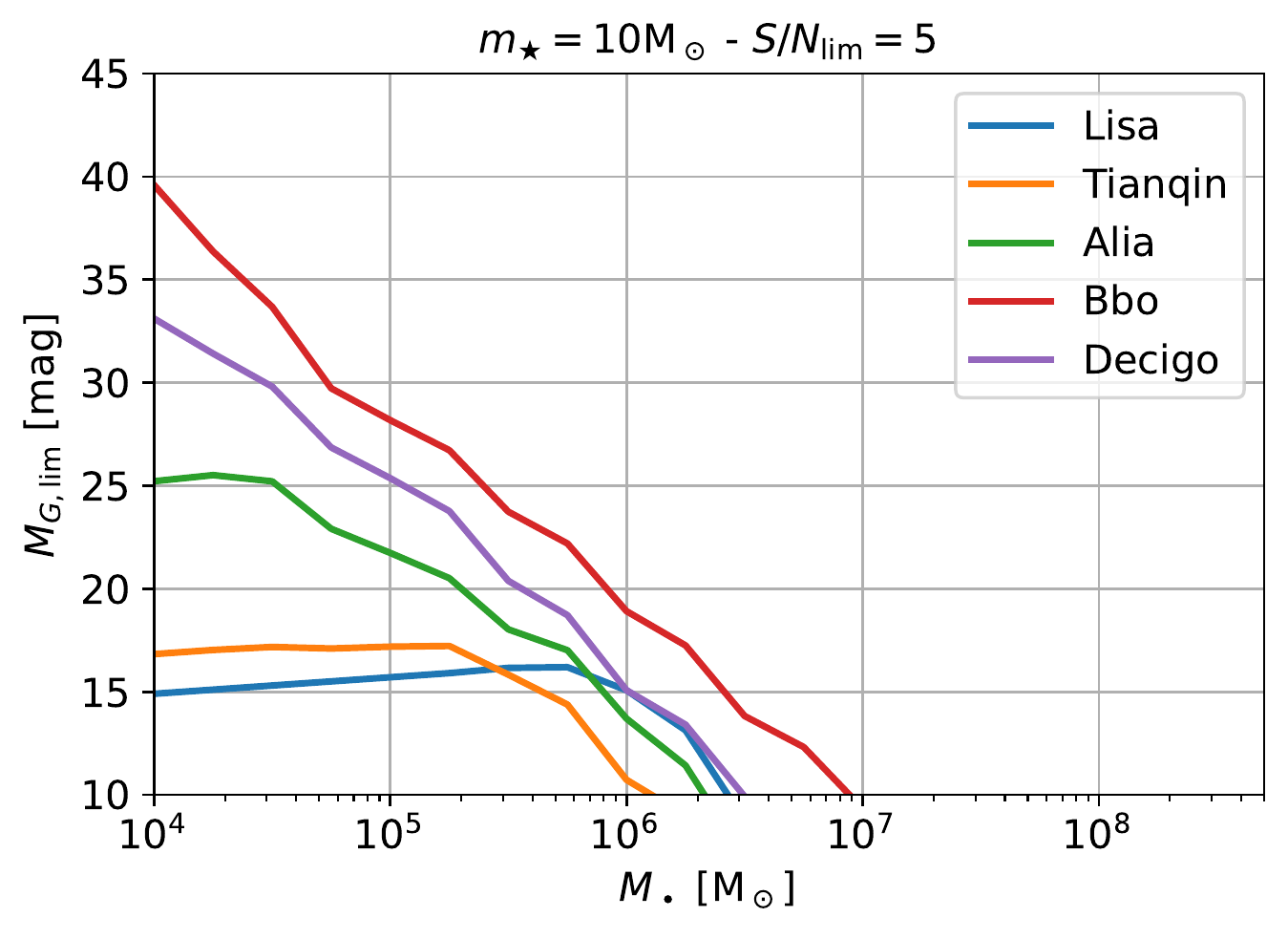}
\caption{G magnitude as a function of $M_\bullet$ for a TDE of a $m_\star=10$ star occurring at the maximum observable redshift with different gravitational wave detectors.}
\label{fig:g_band}
\end{figure}

We show in Fig.~\ref{fig:g_band} $M_{G, {\rm lim}}$ as a function of $\Mbh$ for TDEs of $10\Msun$ stars, and assuming events can be gravitationally detected with $S/N_{\lim} =5$. This can easily be generalized to any $\mstar$ and $S/N_{\lim}$, combining the approximate expressions for $M_G$ (Eq.~\eqref{eq:G_lowz}) and $z_{\max}$ (Eq.~\eqref{eq:zmax}) one finds:
\begin{eqnarray}
M_{G, {\rm lim}} &=& 18+5\log\left( \frac{m_\star}{\Msun} \right)-5\log\left( S/N_{\lim} \right)+ \label{eq:Glim_all}\\
&& \frac{5}{6}\log\left( \frac{M_\bullet}{10^6\Msun} \right)+5\log\left( \left(\frac{f_\star}{10^{-2}\Hz}\right)^{2/3}\left(\frac{h_{\rm det}(f_\star)}{10^{-21}}\right)^{-1} \right) \, . \nonumber
\end{eqnarray}
Any gravitational detection with the two upcoming interferometers LISA and Tianqin should be associated with a maximum $M_{G, {\rm lim}}$ magnitude $\lesssim 20$, even detectable with current wide field facilities \citep[YSE and ZTF $M_G$ limit is 21.5, see][]{YSE_paper}. As a consequence, if there is a detection of a TDE through gravitational waves, there should be a detection of its electromagnetic counterpart. However, future generation detectors may see gravitational waves that are not observed in the optical counterparts: there will be orphans. Since for BHs lighter than the optimal mass for gravitational waves detection ($M_{\bullet,\,{\rm opt}}$, see Table ~\ref{tab:fdet}), $M_{G, {\rm lim}}$ \textit{increases} with the mass of the BH (as $5/6 \times \log\Mbh$, see Eq.~\eqref{eq:Glim_all}), while it \textit{decreases} for more massive BHs (faster than $-5/2\times\log\Mbh$ given the expression of $f_\star$ in this regime), these orphans are most likely to be powered by BHs with a mass around $M_{\bullet,\,{\rm opt}}$.

\subsubsection{X-ray emission}
\label{sec:XrayEmission}

\paragraph*{Formalism:} The origin of the X-ray emission is thought to be associated to the inner parts of an accretion disc \citep[\eg][]{Ulmer_99, Lodato_11}. We adopt a similar method as in \S\ref{sec:OpticalEmission}, but instead of fixing the temperature, we fix the black-body radius at the circularization radius of the stellar debris:
\begin{align}
    R_{\rm x}=\frac{2r_{\rm T}}{\beta} \, ,
\end{align}
and, similarly to the optical case \citep{2016Sci...351...62V}, we assume the luminosity equals the Eddington limit capped at the luminosity at peak (Eq.~\eqref{eq:L_adopted}). This allows to estimate the spectral flux density (Eq.~\eqref{eq:flux}) which can be integrated in the 0.3-10 keV band in order to obtain the X-ray flux.

\paragraph*{Multimessenger TDEs:}

Similarly to \S\ref{sec:OpticalEmission}, we estimate the X-ray flux in the pessimistic regime of a TDE occurring at the maximum redshift at which it can be observed with a gravitational wave detector. Note that, contrary to the optical emission, this flux depends on the mass of the star not only through $z_{\max}$, but also through $R_X$. Note also that the value of $\beta$ matters, and to estimate the flux we take $\beta$ for which $f_{\rm GW}=f_\star$, \ie such that the event is observable at $z_{\max}$.

\begin{figure}
    \centering
    \includegraphics[width=\columnwidth]{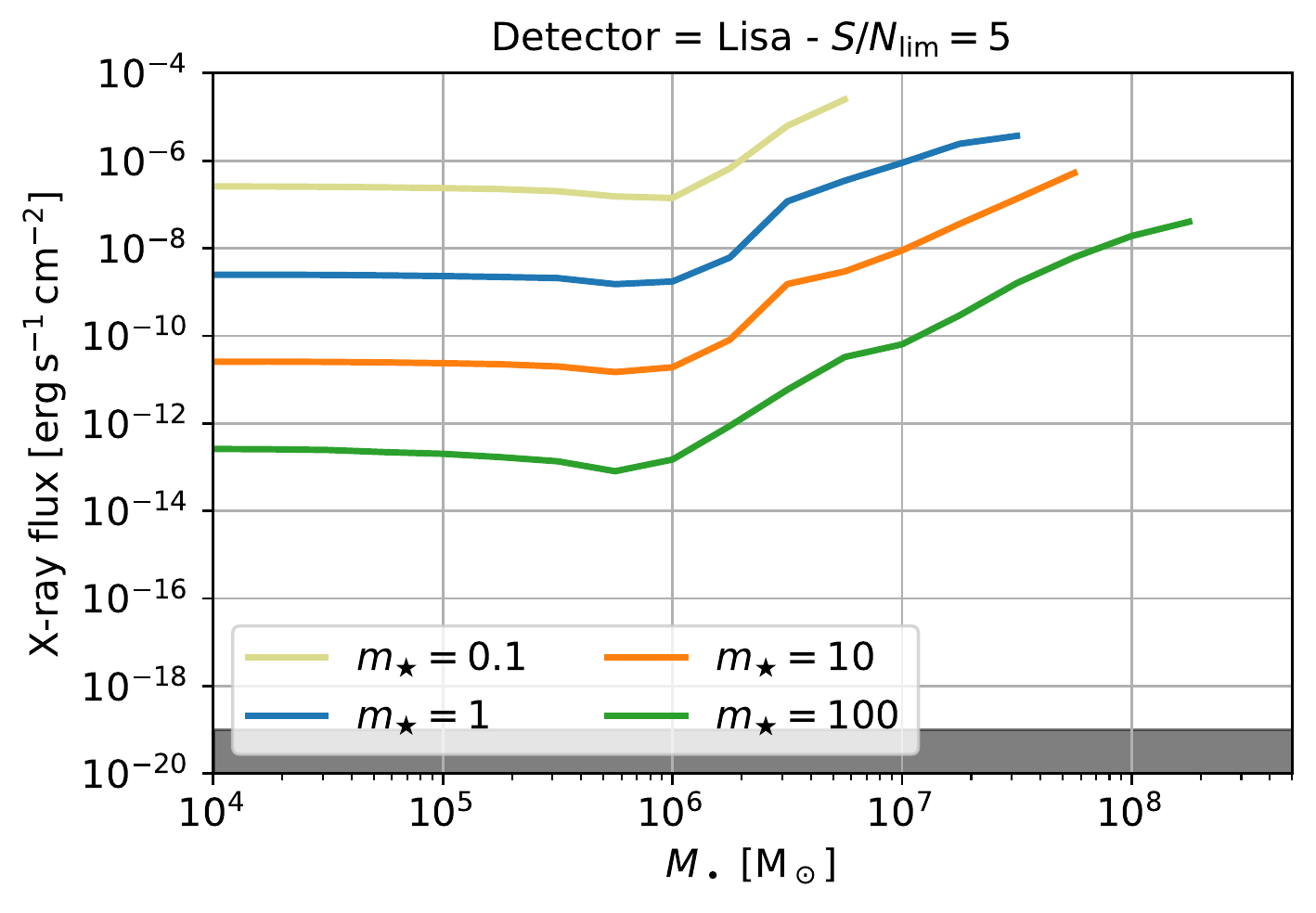} \\
    \includegraphics[width=\columnwidth]{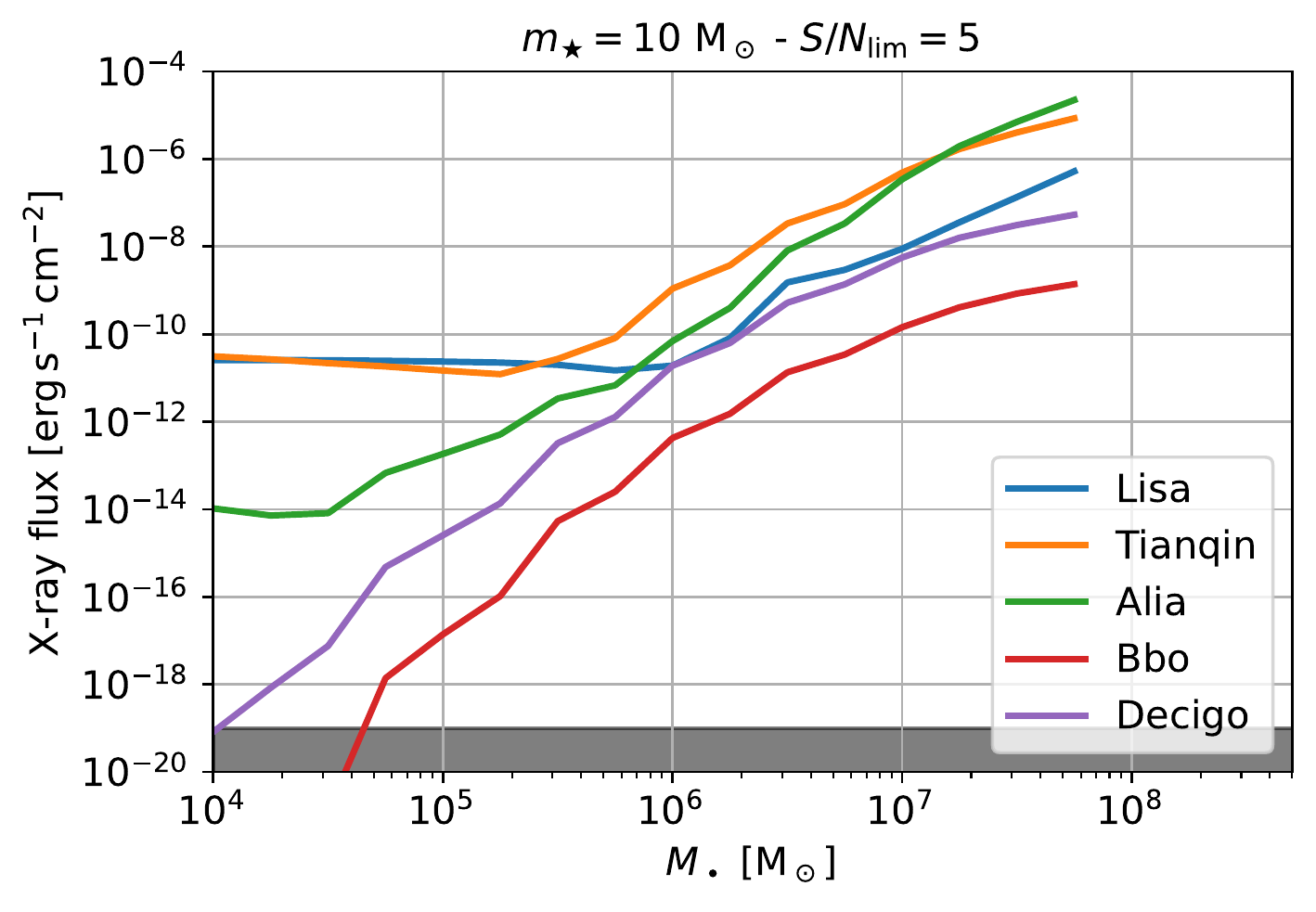}
    \caption{X-ray flux as a function of $M_\bullet$. \textbf{Top:} Case of LISA, we vary the total luminosity (varying the fraction of the Eddington limit, line style) and the mass of the disrupted star (color, low mass stars curves are truncated for large BH masses, when stars penetrate the Scwhazschild radius before being disrupted). \textbf{Bottom:} Case of future detectors (colors). In both cases, the Lynx limit is indicated in black.}
    \label{fig:LISA_xray}
\end{figure}

We show in Fig.~\ref{fig:LISA_xray} the X-ray flux as a function of $M_{\bullet}$. We also mark the optimistic flux limit of Lynx \citep[$10^{-19} \erg \cm^{-2} \s^{-1}$; ][]{lynx_18}, but similar conclusions can be reached for eROSITA \citep[flux limit of $10^{-14} \erg \cm^{-2} \s^{-1}$;][]{Merloni_12}. In the upper plot we focus on the case of LISA varying $m_\star$ (colors). We find that most TDEs that LISA may reveal through gravitational waves should be detectable in the X-ray. Similarly to the optical counterpart, we find that, in general, if we consider more massive stars, the X-ray counterpart is fainter as the maximum redshift is larger. In the lower plot we show the results for future gravitational wave detectors, with the same conclusions.

\section{TDE rates}
\label{sec:TDErates}

As shown in \S\ref{sec:DetectionOfSingleEvents}, the detectability of a TDE, both in gravitational and electromagnetic waves, depends on the mass of the BH as well as the properties of the disrupted star (mass and radius) and on the pericenter of  its orbit. In order to obtain the rate of observable events, we compute the total rate as a function of these parameters in one single galaxy in \S\ref{sec:SingleGalaxy}, and we then generalize to a population of galaxies in \S\ref{sec:PopulationOfGalaxies}.

\subsection{Single galaxy}
\label{sec:SingleGalaxy}

In this section, for a given galaxy with a central BH, we derive the TDE rate at which stars with a given mass and pericenter penetrate the tidal radius and are disrupted. A summary of the loss cone dynamics is given in \S\ref{sec:OrbitalParametersOfTDEs}; as the rate depends on the structure of the galaxy (\eg density profile), we describe our assumptions in \S\ref{sec:StellarPropertiesAroundTheBH}; and we compare our results with previous studies in \S\ref{sec:Summary}.

\subsubsection{Loss cone dynamics}
\label{sec:OrbitalParametersOfTDEs}

For a spherically symmetric bath formed by a monochromatic population of stars $m_\mathrm{bg}$, we classically \citep{BT_87, Merrit_book, Stone_16a} define the absolute value of the specific energy, $E$, the specific angular momentum $J$, the specific angular momentum of a circular orbit at a given energy, $J_c(E)$, their ratio $R=J^2/J^2_c$, the radial period, $P(E)$\footnote{In general, $P$ depends on $R$ except for the isochrone potential \citep{BT_87}. However, its dependency is usually weak and therefore neglected.}, and the mean-$R$ distribution function $f(E)$.

The flux of stars with a given energy and impact parameter that diffuse through 2-body interactions within an angular momentum limit $R_{\rm LC}$ is given by \citep[Eq.~(4.90-4.94) of][and variable change]{Strubbe_11}:
\begin{eqnarray}
\frac{\d^2 \Gamma}{\d E \d \ln \beta} &=& \frac{8 \pi^2 \G \Mbh r_T}{\beta} \frac{f(E)}{1+q^{-1}\xi\ln(R^{-1}_\mathrm{LC})} \times \label{eq:d2Gammad2Edbeta} \\
&& \left[1-2\sum_{m=1}^\infty \frac{e^{-\alpha^2_m q/4}}{\alpha_m} \frac{J_0\left(\alpha_m \beta^{-1/2}\right)}{J_1(\alpha_m)} \right] \nonumber \\
\xi &=& 1 - 4  \sum_{m=1}^\infty \frac{e^{-\alpha^2_m q/4}}{\alpha^2_m} \\
q(E, m_\star, m_\mathrm{bg}) &=& \frac{P \overline{\mu}(E, m_\star, m_\mathrm{bg})}{R_\mathrm{LC}}\, ,
\end{eqnarray}
where $q$ is the loss cone filling factor corresponding to the relative change of $R$ to $R_\mathrm{LC}$ per orbit; $\overline{\mu}(E,m_\mathrm{bg},m_\star)$ is the orbit averaged diffusion coefficient in $R$ corresponding to the inverse of the typical time to change $R$ by order unity (Eq.~(18) of \citealt{Merrit_book}, Eq.~(11) of \citealt{Stone_16a}); $J_k$ are the Bessel functions of first kind of order $k$; and $\alpha_m$ is the m-th zero of $J_0$. 

We define the TDE rate as the rate at which stars diffuse within the angular momentum limit such that they are are doomed to enter the disruption zone on a time scale of the orbital period, that is:
\begin{eqnarray}
R_\mathrm{LC}(E, m_\star) \sim \frac{4 E r_T}{\G \Mbh} \ll 1 \label{eq:RLC}\, ,
\end{eqnarray}
where we have used that $E \ll \G \Mbh /r_T$, as most stars arrive on eccentric orbit with semi-major axis much larger than $r_T$.

If we consider now a stellar population described by a mass function $\phi$, then, in principle, \textit{(i)} for a given test mass $\mstar$, the diffusion caused by a multi-species stellar background should differ from the simple monochromatic case; \textit{(ii)} different test particles with different masses should diffuse differently. However, by a happy coincidence \citep[][summary in Appendix \ref{sec:FPEqWithAStellarPopulation}]{Magorrian_99}, for a stellar population, the resulting rate happens to be the same as if the distribution was made by a monochromatic bath with mass $m_\mathrm{bg}=\langle m_\star^2 \rangle^{1/2}$, where $\langle m_\star^2 \rangle = \int m^2\phi(m)\d m$ is the root mean square of the mass of a star. Obviously, this is not true anymore when $\phi$ depends on the position: for instance if there is mass segregation that bring more massive objects to the center, or close enough to the BH so that more massive stars are disrupted. We neglect these processes in what follows. Consequently, the rate at which test particles diffuse within some angular momentum limit can be obtained using Eq.~\eqref{eq:d2Gammad2Edbeta} with $m_\mathrm{bg}=\langle m_\star^2 \rangle^{1/2}$. 

In order to obtain the rate for a given mass, one has to consider that only a fraction $\phi(m_\star)\d m_\star$ of test particles have the mass $m_\star$, and that the boundary for disruption ($R_\mathrm{LC}$) depends on $m_\star$, this yields:
\begin{eqnarray}
\frac{\d^2 \Gamma}{\d \ln m_\star \d \ln \beta} &=& \frac{8 \pi^2 \G \Mbh r_T}{\beta} \phi(m_\star)m_\star \times \label{eq:d2Gammadlogedlogrp} \\
&& \int_{0}^{\G\Mbh / r_T} \mathcal{G}(E,\beta,m_\star) \d E \nonumber \\
\mathcal{G}(E,\beta,m_\star) &=& \frac{f(E)}{1+q^{-1}\xi\ln(R^{-1}_\mathrm{LC})} \times \\
&&\left[1-2\sum_{m=1}^\infty \frac{e^{-\alpha^2_m q/4}}{\alpha_m} \frac{J_0\left(\alpha_m \beta^{-1/2}\right)}{J_1(\alpha_m)} \right]  \nonumber \\
\xi &=& 1 - 4  \sum_{m=1}^\infty \frac{e^{-\alpha^2_m q/4}}{\alpha^2_m} \\
q(E, m_\star) &=& \frac{P(E) \overline{\mu}(E, \langle m^2_\star \rangle^{1/2})}{R_\mathrm{LC}} \, ,
\end{eqnarray}
{where the upper bound in the integral comes from that orbits with higher $E$ are within the tidal radius. We note that this upper bound on $E$ is in apparent contradiction with Eq.~\eqref{eq:RLC}, but we have checked that the integral is independent of the upper value chosen, \ie  the rate is not dominated by the energy range near $\G \Mbh / r_T$.}

From Eq.~\eqref{eq:d2Gammadlogedlogrp} we can obtain the total TDE rate ($\Gamma$) around a particular BH as:
\begin{eqnarray}
\Gamma = \int_{m_\star=m_{\star,\min}}^{m_{\star,\max}} \int_{\beta=0}^{\beta_{\max}} \frac{\d^2 \Gamma}{\d \ln m_\star \d \ln \beta} \d \ln m_\star \d \ln \beta\, , \label{eq:GammaVSMBH}
\end{eqnarray}
where $m_{\star,\min}$ and $m_{\star,\max}$ are the boundaries of the stellar mass function (see following Section).

\subsubsection{Stellar properties around the BH}
\label{sec:StellarPropertiesAroundTheBH}

\paragraph*{Stellar density function:}

In general, the estimate of ${f}$ and $q$ requires numerical integration \citep[using for instance using \textsc{PhaseFlow};][]{Vasiliev_17,Pfister_19b,Pfister_20b,2021MNRAS.500.4628P} or approximations \citep{Wang_04, Pfister_20c}. However, we further assume that the stellar density profile surrounding the BH is a power law:
\begin{eqnarray}
\rho(r) &=& \rho_0\left( \frac{r}{r_\mathrm{inf}} \right)^{-\alpha} \label{eq:rho}
\end{eqnarray}
where $r_\mathrm{inf}$ is the influence radius of the BH, corresponding to the radius encompassing a stellar mass equal to that of the BH (which also defines $\rho_0$). In this situation, ${f}$ and $q$ can be obtained analytically within $r_\mathrm{inf}$ \citep{Magorrian_99, Strubbe_11, Merrit_book, Stone_16a}, when the potential is dominated by the BH, as:
\begin{eqnarray}
{f}(E) &=& (2\pi\sigma^2_\mathrm{inf})^{-3/2}\frac{\rho_0}{\langle m_\star \rangle} \frac{\gamma(\alpha+1)}{\gamma(\alpha-1/2)} \left( \frac{E}{\sigma^2_\mathrm{inf}} \right)^{\alpha-3/2} \\
q(E,m_\star) &=& \nu \left( \frac{E}{\sigma^2_\mathrm{inf}} \right)^{\alpha-4} \\
\nu(m_\star) &=& \frac{8\sqrt{\pi}}{3}(3-\alpha) \frac{\gamma(\alpha+1)}{\gamma(\alpha-1/2)} \left[ \frac{5}{32(\alpha-\frac{1}{2})} + \right.\\
&&\left.  \frac{3I_B(\frac{1}{2},\alpha)-I_B(\frac{3}{2},\alpha)}{4\pi} \right] \left( \frac{\G \langle m_\star^2 \rangle}{\sigma^2_\mathrm{inf} \langle m_\star \rangle r_T(m_\star)} \right) \ln\Lambda \nonumber
\end{eqnarray}
where $\sigma_\mathrm{inf}=(\G \Mbh/r_\mathrm{inf})^{1/2}$ is the velocity dispersion at $r_\mathrm{inf}$; $\langle m_\star \rangle=\int m\phi(m)\d m$ is the mean stellar mass; $\ln~\Lambda~=~\ln(0.4\Mbh/\langle m_\star \rangle)$ is the Coulomb logarithm \citep{Spitzer_71}; $\gamma$ is the Euler Gamma function;\footnote{$\gamma(z)=\int_0^\infty t^{z-1}e^{-t}\d t$} and we define $I_B$ as:
\begin{eqnarray}
I_B(\frac{n}{2},\alpha) &=& \int_0^1t^{-\frac{n+1}{2}}(1-t)^{3-\alpha} B(t,\frac{n}{2},\alpha-\frac{1}{2}) \d t \, ,
\end{eqnarray}
where $B$ the incomplete Euler Beta function.\footnote{$B(x,a,b)=\int_0^x t^{a-1}(1-t)^{b-1} \d t$}

In order to reduce the dimensionality of the problem, we assume that the density profile is an isothermal sphere ($\alpha=2$ in Eq.~\eqref{eq:rho}):
\begin{eqnarray}
\rho(r) &=& \frac{\sigma^2}{2\pi \G r^2} \\
r_\mathrm{inf} &=& \frac{\G \Mbh}{2 \sigma^2} \\
\sigma_\mathrm{inf} &=& \sqrt{2}\sigma \, ,
\end{eqnarray}
where $\sigma$ is the velocity dispersion of the galaxy, and is such that the BH lies on the $\Mbh-\sigma$ relation \citep[][]{2001ASPC..249..335M}:
\begin{eqnarray}
\frac{\sigma}{\km \s^{-1}} = 68 \left( \frac{M_\bullet}{10^6\Msun} \right)^{0.22} \, .
\end{eqnarray}

{The assumptions of an isothermal sphere lying on the $\Mbh-\sigma$ relation is clearly a simplification of reality, as galaxies exhibit different shapes \citep[\eg ][]{Lauer_07} and are not uniquely defined by their BH mass \citep[there is scatter in the relation, \eg ][]{Kormendy_13}. One possibility to overcome this issue would be to use a mock catalog \citep[\eg ][]{Pfister_20b,2020ApJ...900..191C} but \textit{(i)} this is beyond the scope of this study which only aims at providing trends and orders of magnitude on the gravitationally observed TDE rates; and \textit{(ii)} these mock catalogs are constructed from real observations for which the structure within the influence radius (the relevant region for TDE rates estimates) is usually poorly resolved for BHs with $\Mbh \lesssim 10^6 \Msun$ \citep{Pechetti_19, SanchezJanssen_19}. This said, we note that the isothermal sphere lying on the $\Mbh-\sigma$ has been widely used in Astronomy \citep[\eg ][]{Volonteri_03, 2020ApJ...904...16B}, including TDE studies for which it has shown to reproduce well observations \citep{Wang_04, Kochanek_16b}. We also note that the use of the $\Mbh-\sigma$ relation of \cite{2001ASPC..249..335M} among the different observationally found \citep[\eg ][]{Kormendy_13} has little effects on the rate, as shown by \S 3.2 of \cite{Kochanek_16b}.}

\paragraph*{Stellar mass function:}
\label{sec:StellarMassFunction}

In order to take into account that stars can have a variety of masses, which will produce differences in the strain, we assume the stellar population follows the usual Kroupa stellar mass function \citep{Kroupa_01}:
\begin{eqnarray}
\phi(m_\star) = \phi_0
    \left\{
      \begin{aligned}
      & \left( \frac{m_\star}{0.5 \Msun} \right)^{-1.3} &&\mathrm{\, for\, } m_\mathrm{\star,\, min} \leq m_\star \leq 0.5 \Msun\\
      & \left( \frac{m_\star}{0.5 \Msun} \right)^{-2.3} &&\mathrm{\, for\, }0.5 \Msun \leq m_\star \leq m_\mathrm{\star,\, max} &\\
      &0 &&\mathrm{\, else} \, ,
      \end{aligned}
    \right.
\end{eqnarray}
where $\phi_0$ is such that $\int \phi(m)\d m =1$. Our fiducial models is for $(m_\mathrm{\star,\, min}, m_\mathrm{\star,\, max})=(0.08,10)\Msun$ that is the stellar population is old enough so that massive stars have gone through supernovae \citep[similarly to ][]{Magorrian_99, Stone_16a}, but we explore populations with $m_\mathrm{\star,\, max}$ ranging from $2.5\Msun$ to $100\Msun$. For comparison with previous studies, we also consider the monochromatic Solar population $\phi(m_\star)= \delta(m_\star-\Msun)$, where $\delta$ is the delta Dirac function.

{The use of the Kroupa initial mass function among others \citep{1955ApJ...121..161S, 2003PASP..115..763C} is an arbitrary choice, but \cite{Stone_16a} have shown that the TDE rates depend more on the boundaries ($m_\mathrm{\star,\, min}$ and $m_\mathrm{\star,\, max}$) than on the mass function chosen. While it would be interesting to also vary the initial mass function, it is unfortunately impossible to explore all the possibilities in a finite and comprehensible paper.}\\

For a given BH with mass $\Mbh$, we now have a unique stellar density profile and distribution function ($f$). If we further assume a stellar mass function ($\phi$), we can estimate all the different terms in Eq.~\eqref{eq:d2Gammadlogedlogrp} to obtain the TDE rate for a given stellar mass (\mstar) and impact parameter ($\beta$).

\begin{figure}
\includegraphics[width=\columnwidth]{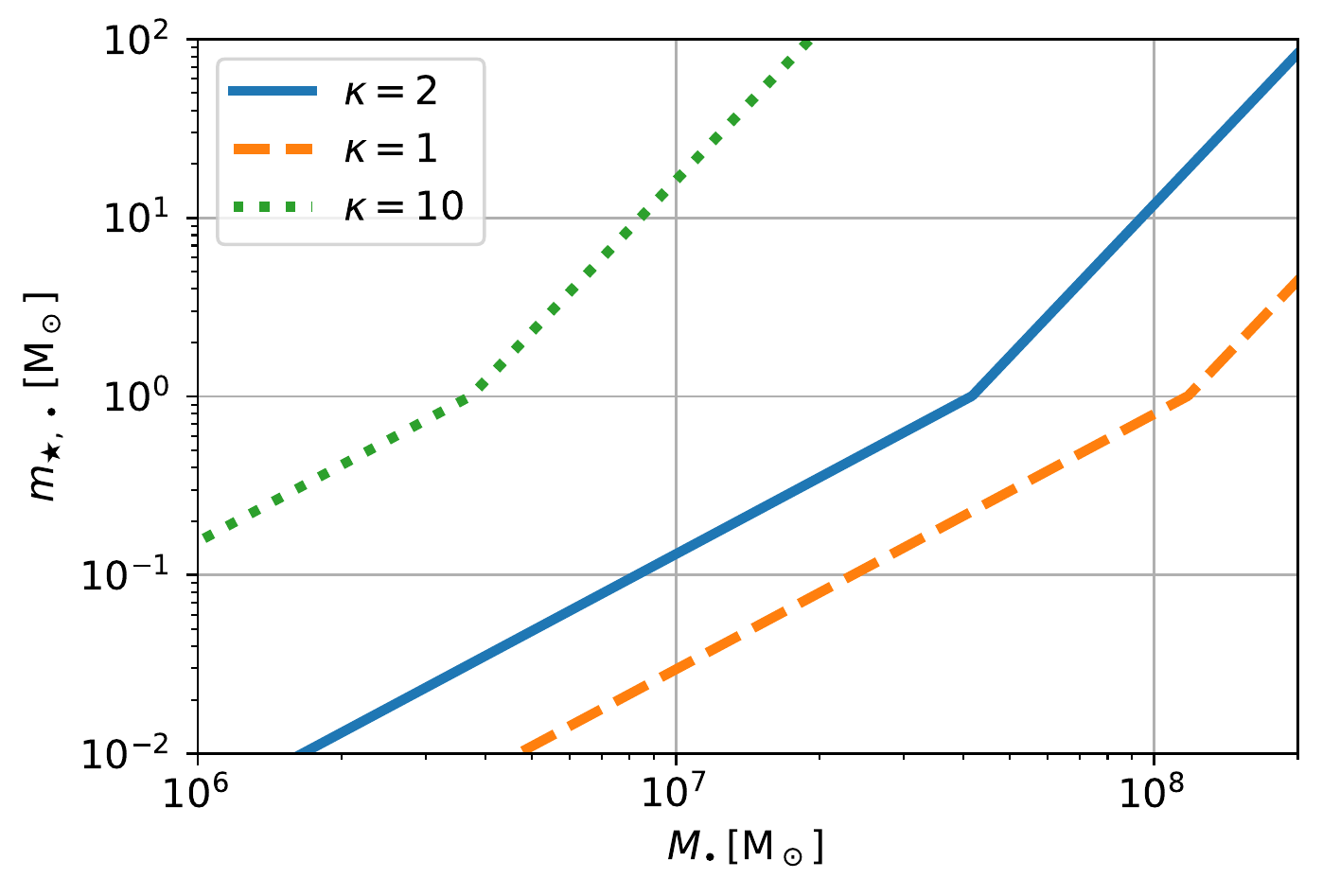}
\caption{{Minimum stellar mass that can produce a TDE as a function of the mass of the BH, for different $\kappa$ (the critical radius for direct plunge, see \S\ref{sec:Formalism}). For the $\kappa=2$ case, BHs with $\Mbh \lesssim  10^7\Msun$ can disrupt all stars; for more massive BHs, low-mass stars are gradually removed such that only massive stars can be disrupted; and for BHs with a mass above $\sim 2 \times 10^8\Msun$, even most massive stars with $\mstar \sim 100\Msun$ are swallowed whole.}}
\label{fig:mstarBHVSMBH}
\end{figure}

\begin{figure}
\includegraphics[width=\columnwidth]{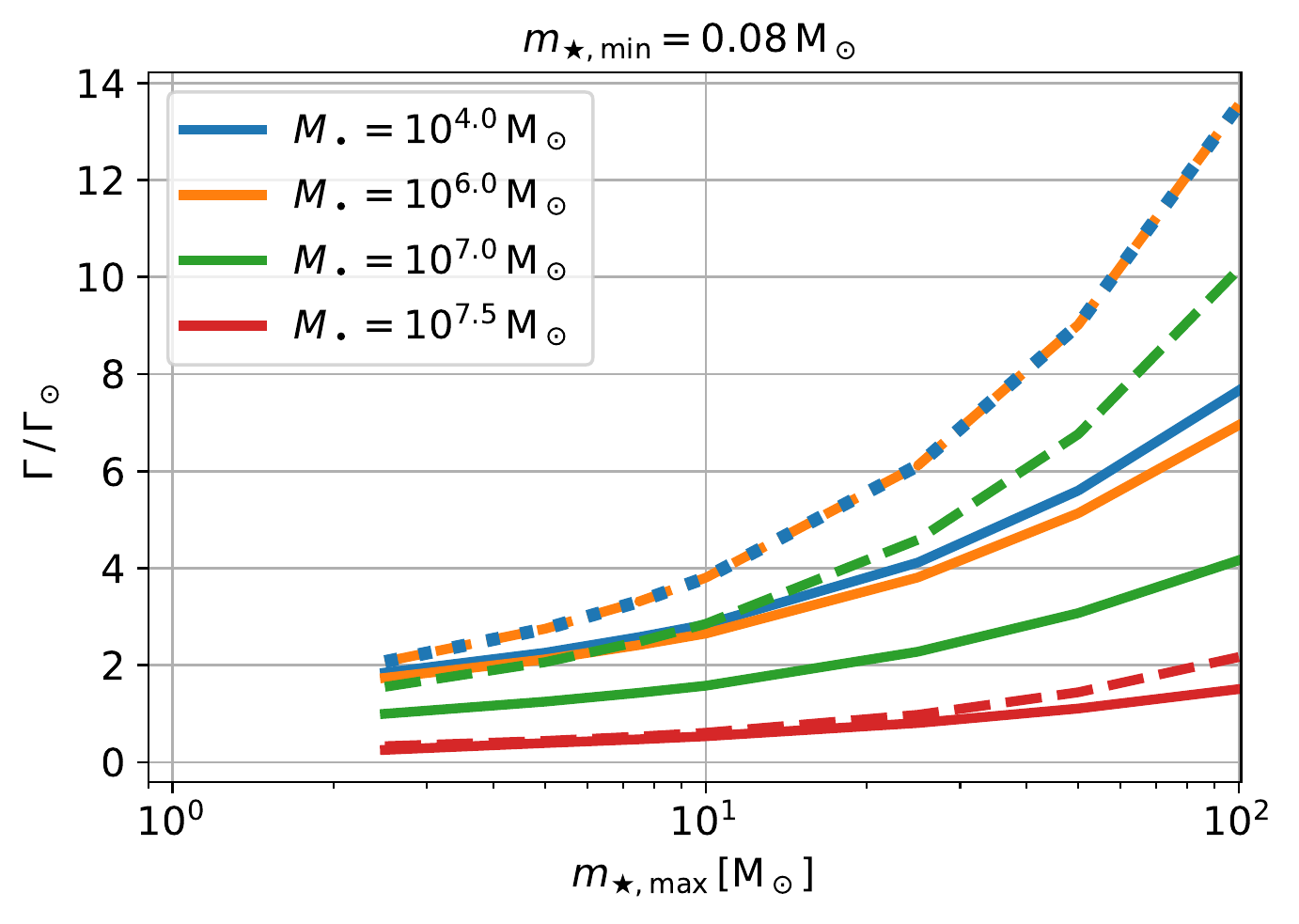}
\caption{{Ratio, for different BH masses (colors), of the TDE rate for a realistic stellar population over the TDE rate for a monochromatic solar pulation. The higher the BH mass the better the matching between our model (solid line, Eq.~\eqref{eq:GammaVSMBH}) and the empty loss cone model (dashed line, Eq.~\eqref{eq:GammaELC}): in the low mass end, the rate is not dominated by the empty loss cone regime.}}
\label{fig:GammaVSmstarmax}
\end{figure}

\begin{figure}
\includegraphics[width=\columnwidth]{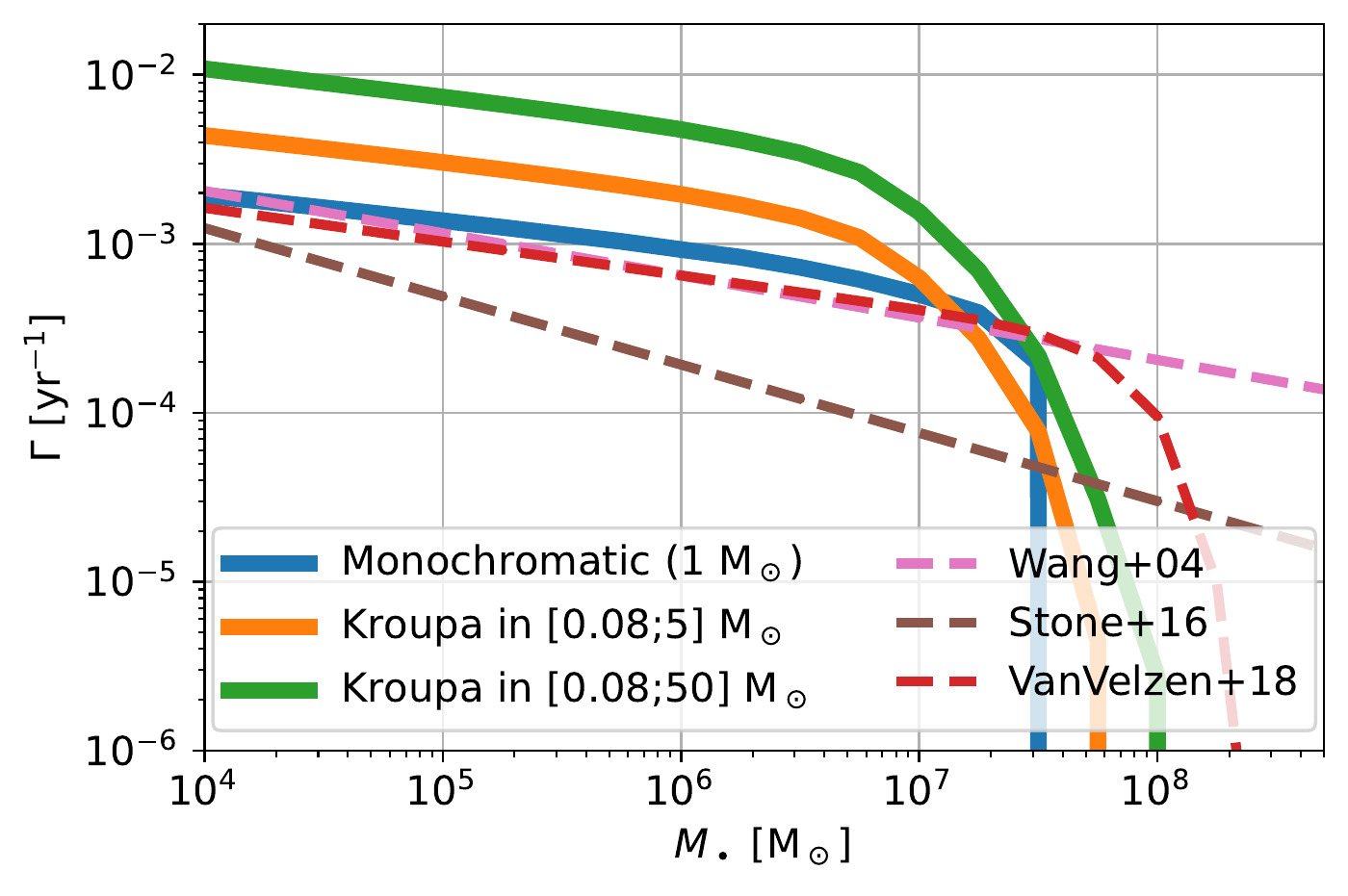}
\caption{TDE rate as a function as the mass of the central BH. We show the results from our study (thick lines) and from previous works \citep[dashed lines; ][]{Wang_04, Stone_16a, VanVelzen_18}. Rates are increased if we change the stellar mass function \citep[at first order it scales with $\langle m_\star^2 \rangle/\langle m_\star \rangle^2$; ][]{Magorrian_99,Stone_16a}, but our results for a monochromatic Sun-like population (blue line) are in excellent agreement with previous studies. When more massive stars are included, TDEs can occur around more massive BHs which explains why the drop shifts toward heavier BH masses when the stellar mass function is extended.}
\label{fig:GammaVSmBH}
\end{figure}

\begin{figure*}
\includegraphics[width=\columnwidth]{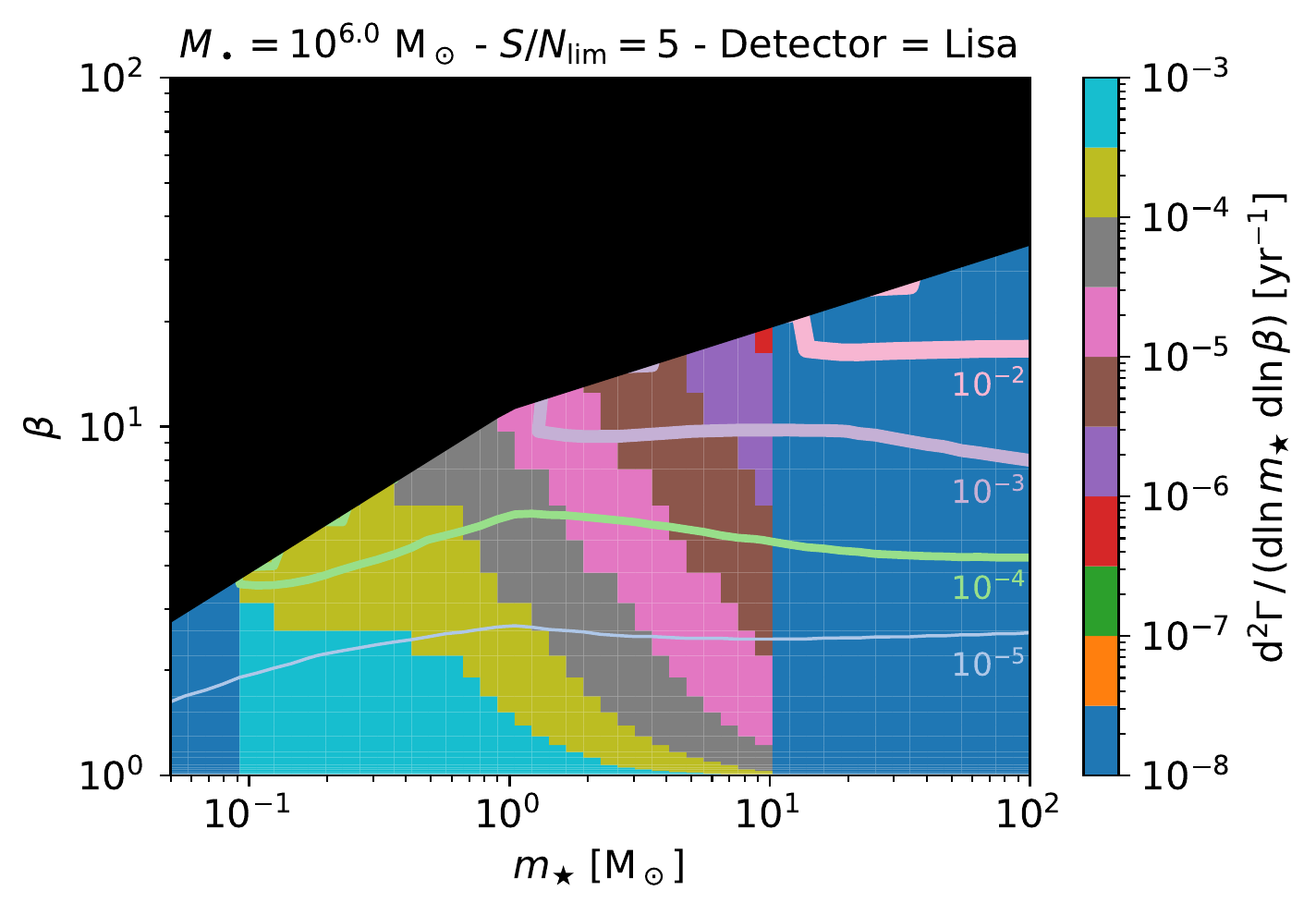} \hfill 
\includegraphics[width=\columnwidth]{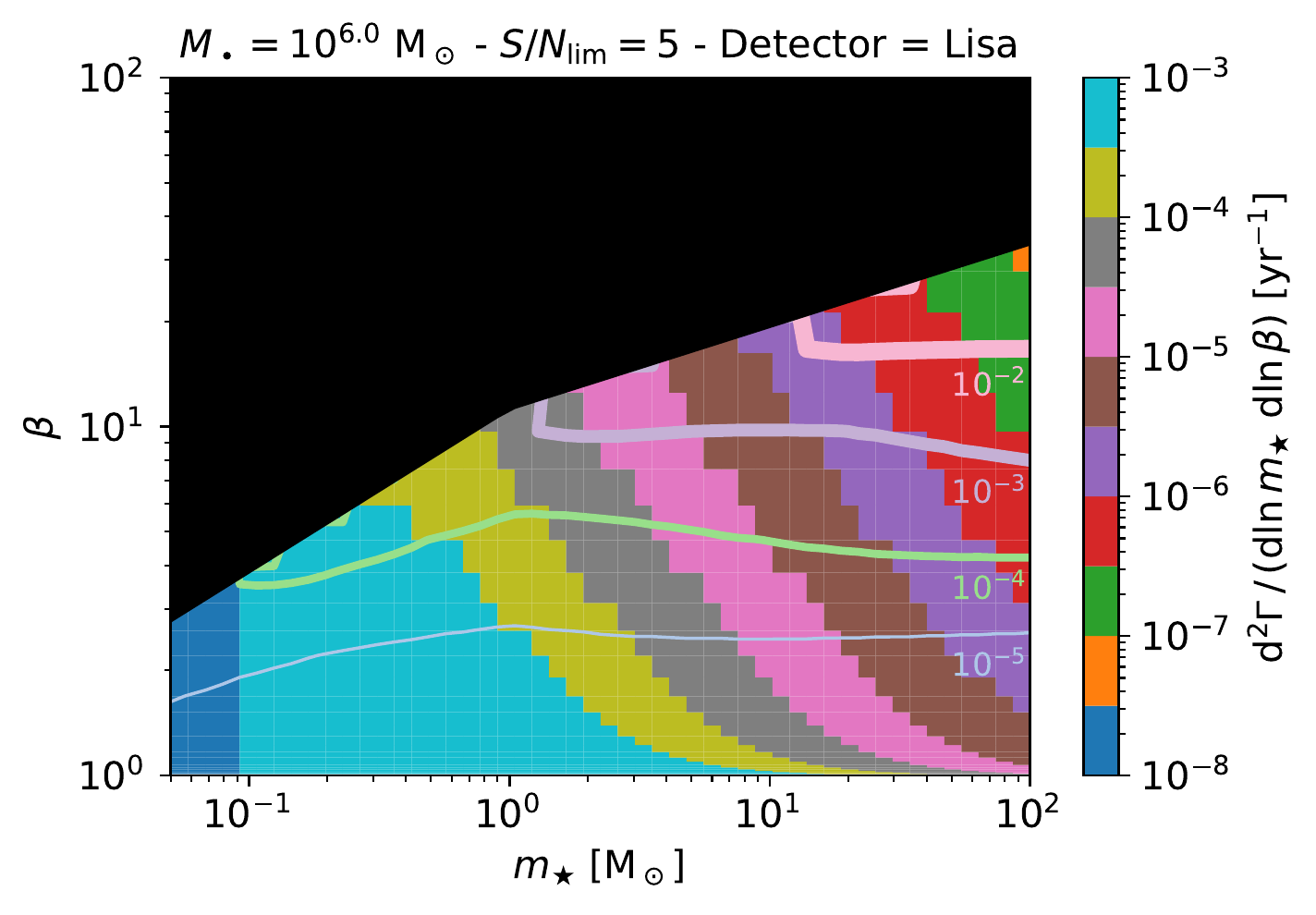}
\caption{Differential TDE rate as a function of the stellar mass and the impact parameter. We show the particular case a of $10^6\Msun$ BH and a Kroupa stellar mass function in $[0.08;10]\Msun$ (\textbf{left}) and in $[0.08;100]\Msun$ (\textbf{right}). We overplot the maximum redshift at which these events can be detected with LISA with $S/N_{\lim}=5$ (color lines). Most TDEs have $m_\star \lesssim 1\Msun$ and $\beta \lesssim 2$ but can be observed up to only $z\sim10^{-5}$ (which is cosmologically irrelevant and make them impossible to observe with gravitational waves). Rarer events events can be observed to larger distances yielding a larger observable volume.}
\label{fig:Example_GammaVSbetaANDrp}
\end{figure*}

\subsubsection{Summary}
\label{sec:Summary}

\paragraph*{Analytical considerations:} As noted by \cite{Magorrian_99} and \cite{Stone_16a}, if the TDE rate is dominated by the ``empty loss cone'' regime (or ``diffusive'' regime with $q \ll 1$), then the total rate scales as $\Gamma \propto \langle m_\star^2 \rangle / \langle m_\star \rangle^2$ ($\langle m_\star^2 \rangle/\langle m_\star \rangle$ from $\overline{\mu}$ and $1/\langle m_\star \rangle$ from $f$)). As we also take into consideration that low-mass stars can be swallowed whole, we have to remove the fraction $f_\bullet$ of stars with $m_\star \leq m_{\star, \bullet}$, where $m_{\star, \bullet}$ is solution to $\beta_{\max}=1$ (Eq.~\eqref{eq:beta_max}). In the end the total rate scales as:
\begin{eqnarray}
\frac{\Gamma}{\Gamma_\odot} = f_\bullet \frac{ \langle m_\star^2 \rangle }{ \langle m_\star \rangle^2} \, , \label{eq:GammaELC}
\end{eqnarray}
where $f_\bullet = \int_{m_{\star, \bullet}}^{m_{\star, \max}} \phi(u)\d u$, and $\Gamma_\odot$ is the TDE rate for a monochromatic solar population.

{We show in Fig.~\ref{fig:mstarBHVSMBH} $m_{\star, \bullet}$ as a function of the BH mass. Given our minimum stellar mass considered of 0.08 \Msun, BHs with $\Mbh \lesssim  10^7\Msun$ can disrupt all stars; for more massive BHs, low-mass stars are gradually removed such that only massive stars can be disrupted; and for BHs with a mass above $\sim 2 \times 10^8\Msun$, even most massive stars with $\mstar \sim 100\Msun$ are swallowed whole. As a consequence, rather independently on the maximum boundary $m_{\star,\, \max}$, the fraction of stars that can be disrupted, is $f_\bullet=1$ for $\Mbh \lesssim 10^7 \Msun$, and gradually drops to 0 for $\Mbh \gtrsim 2\times 10^8 \Msun$ passing by 0.77 and 0.16 for $\Mbh=10^7$ and $10^{7.5}\Msun$.}

{We show in Fig.~\ref{fig:GammaVSmstarmax}, for different BH masses (colors), the ratio of the TDE rate with respect to the one of a monochromatic solar population. This ratio is shown as a function of the maximum stellar mass $m_{\star,\, \max}$ in the stellar mass function. The results are presented for our model obtained using Eq.~\eqref{eq:GammaVSMBH} (solid lines) as well as in the empty loss cone regime using Eq.~\eqref{eq:GammaELC} (dashed line). The scaling with the empty loss cone is overall quite good. Yet, in the low mass BH regime ($10^{4}-10^5\Msun$, blue and orange), there is a mismatch between the empty loss cone ratio and the ``real'' ratio. In the higher mass BH regime ($10^{7.5}\Msun$, red), the agreement is better. This is in agreement with \cite{Stone_16a} who finds that the fraction of TDEs in the empty loss cone regime dominates in the high mass BH end, but not for low mass BH.}

\paragraph*{Comparison with previous results:} We show in Fig.~\ref{fig:GammaVSmBH} the TDE rate $\Gamma$ (Eq.~\eqref{eq:GammaVSMBH}) as a function of $\Mbh$ for our model (solid lines) and from previous studies (dashed lines). For our model, we perform the exercise for the three different stellar mass functions (shown as different colors). Our results with a monochromatic Sun-like population (solid blue) are in excellent agreement with those from \cite{Wang_04} (dashed pink): this was expected as both models use similar assumptions, but this is a nice test to confirm the validity of our model, of our numerical implementation and of our results. As discussed in the paragraph above, and similarly to \cite{Magorrian_99} or \cite{Stone_16a}, we find an enhancement of the rate when we extend the stellar mass function. Finally, we also note the TDE rates sharply drop to 0 at $\Mbh \sim 10^8\Msun$ for a monochromatic Sun-like population, and smoothly decreases starting from few $10^7\Msun$ for a more realistic stellar population. This is due to that low mass stars (see Fig.~\ref{fig:mstarBHVSMBH}) are gradually removed when we shift toward more massive BHs \citep[see also Fig.~4 of][]{Kochanek_16b}.

\paragraph*{Fraction of observable TDEs with gravitational waves:} Since we are  confident with our rate calculation, we look at the differential rate, as estimated using Eq.~\eqref{eq:d2Gammadlogedlogrp}, as a function of the impact parameter $\beta$ and the mass of the disrupted star \mstar. We show in Fig.~\ref{fig:Example_GammaVSbetaANDrp} the example of a $10^6\Msun$ BH surrounded by two fiducial $[0.08;10]\Msun$ and $[0.08;100]\Msun$ Kroupa stellar mass function, and we overplot the maximum redshift at which these TDEs can be observed with LISA (see also Fig.~\ref{fig:zmaxVSbetaANDmstar_1e6}). For the $[0.08;10]\Msun$ case (left), we find that most TDEs have $\beta \lesssim 2$ which reflects that we typically have $\partial_\beta \Gamma \propto 1/\beta^2$ \citep[][or see Eq.~\eqref{eq:d2Gammadlogedlogrp}; note however that this is not entirely true as this neglects the dependency of $\mathcal{G}$ with $\beta$]{Stone_16a, Kochanek_16b}. Most TDEs are also powered by low mass stars with $m_\star \lesssim~1\Msun$, which reflects the fact that the stellar mass function is bottom-heavy. These most numerous TDEs (cyan region) can unfortunately only be detected in our galaxy ($z_{\max} \sim10^{-5}$). Events with $\mstar \gtrsim 6\Msun$ and $\beta \gtrsim 10$ (purple region) are typically 100-1000 times rarer, but their gravitational waves also carry more energy and can be detected up to $z\sim~10^{-2}$, yielding a much larger volume. For the $[0.08;100]\Msun$ case (right), we find similar conclusions, even if at same $\mstar$ and $\beta$, rates are different. Overall, the rate of observed gravitationally detected TDEs will be a competition between their rarity, and the volume within which they can be seen.

\subsection{Population of galaxies}
\label{sec:PopulationOfGalaxies}

In the previous Section we have obtained the TDE rate for a single galaxy. In order to compute the total observable rate, one needs the BH mass function ($\Phi_\bullet$, giving the volumetric number of BHs within a certain mass range) to sum the contributions of all BHs. We describe our choice in \S\ref{sec:BHMassFunction} and finally wrap up everything in \S\ref{sec:Summary}.

\subsubsection{BH mass function}
\label{sec:BHMassFunction}

We adopt here two different models for the BH mass function.

Our first model ($\Phi_{\bullet,1}$) assumes that all galaxies host a central BH and that the mass of the BH can be inferred from the mass of the galaxy. In particular, we assume that the redshift-dependent galaxy mass function can be expressed as:
\begin{eqnarray}
\Phi_{\rm gal}(M_{\rm gal},z) \d M_{\rm gal}&=& e^{-M_{\rm gal}/M_\star} \frac{\d M_{\rm gal}}{M_\star}\times \\
&& \left[ \Phi^\star_1 \left( \frac{M_{\rm gal}}{M_\star} \right)^{\alpha_1} + \Phi^\star_2 \left( \frac{M_{\rm gal}}{M_\star} \right)^{\alpha_2} \right] \, ,\nonumber
\end{eqnarray}
where $\Phi^\star_1$, $\Phi^\star_2$, $M_\star$, $\alpha_1$ and $\alpha_2$ depend on redshift and are obtained fitting the ``total sample'' galaxy mass function of the COSMOS field \citep[see Table~1 of][]{Davidzon_17}. Using in addition the BH mass - galaxy mass relation from \cite{Reines_15}:
\begin{eqnarray}
\log\left( \frac{\Mbh}{\Msun} \right) = 7.45 + 1.05 \times \log \left( \frac{M_{\rm gal}}{10^{11} \Msun} \right) \, ,
\end{eqnarray}
we can express the BH mass function as:
\begin{eqnarray}
\Phi_{\bullet,1}(\Mbh, z) = \Phi_{\rm gal}(M_{\rm gal},z) \frac{\d M_{\rm gal}}{\d \Mbh} \, . \label{eq:BHMF_model1}
\end{eqnarray}

Our second model ($\Phi_{\bullet,2}$) is simply the BH mass function from \cite{Gallo_19}\footnote{We report here the correct equation as there is a typo in the original paper, private communications with A.~Sesana.}\footnote{We use $\cMpc$ for comoving Mpc.}:
\begin{eqnarray}
&&\log \frac{\Phi_{\bullet,2}}{\cMpc^{-3}\Msun^{-1}} = \\
&&-9.82 - 1.10 \times \log \left( \frac{\Mbh}{10^7 \Msun} \right) - \left( \frac{\Mbh}{128 \times 10^7\Msun} \right)^{1/\ln(10)} \, . \nonumber
\end{eqnarray}

While the first model seems more realistic, as it depends on redshift, it assumes that \textit{(i)} all galaxies host a central BH; and that \textit{(ii)} the mass of this central BH correlates with the mass of the galaxy. While these are reasonable assumptions in the high galaxy~mass/BH~mass end \citep[$\gtrsim 10^{10}\Msun / 10^6\Msun$; ][]{Kormendy_13}, in the dwarf/intermediate mass BH regime the occupation fraction may be less such that some dwarfs do not host any BHs in their center \citep{Tremmel_15, Pfister_17, Pfister_19a}; and the scaling relations between galaxies and BHs may break down \citep{Greene_19}. Furthermore, both \cite{Davidzon_17} and \cite{Reines_15} use a sample of $\gtrsim 10^{9.5}\Msun$ ($10^6\Msun$) galaxies (BHs) that we extrapolate to lower masses. Our second model from \cite{Gallo_19} takes into account that the occupation fraction may not be unity through the entire mass spectrum and explores BHs with masses down to $10^4\Msun$, but is only valid in the local Universe.

We report in Fig.~\ref{fig:BHMF} the BH mass functions used in this work. 

\subsubsection{Summary}
\label{sec:CompleteEquation}

We now have all the ingredients to estimate the rate of TDEs observable with gravitational waves ($\dot{N}$). 

Combining all the Sections above, the rate at given $z$ and BH with mass $\Mbh$ of a star with mass $\mstar$ on an orbit with penetration factor $\beta$ is:
\begin{eqnarray}
\frac{\d^4 \Gamma}{\d z\, \d\Mbh\, \d \mstar\, \d \beta} &=& \frac{\d^2 \Gamma}{\d \mstar\, \d \beta} \frac{1}{1+z} \label{eq:MasterEqGlobal} \\
&& \Phi_\bullet \times 4\pi c \frac{\chi^2(z)}{ H(z)} \nonumber \, ,
\end{eqnarray}
where the first line is simply the differential rate for a single BH measured in the observer frame, and the second line expresses the number of BH at given mass and redshift ($H$ is the Hubble parameter).

In the end, the final equation of interest for us is:
\begin{eqnarray}
\frac{\d^4 \dot{N}}{\d z\, \d\Mbh\, \d \mstar\, \d \beta} &=& \label{eq:MasterEqObserved} \\
\frac{\d^4 \Gamma}{\d z\, \d\Mbh\, \d \mstar\, \d \beta} &\times& \Psi(z,\Mbh,\mstar,\beta,S/N_{\lim}) \nonumber \, , 
\end{eqnarray}
where $\Psi$ is either $0$ or $1$ depending on whether the particular event can be detected:
\begin{eqnarray}
 \Psi = 
    \left\{
      \begin{aligned}
      & 0 &&\mathrm{\, if\, } z \geq z_{\max}\\
      & 1 &&\mathrm{\, else}  \, ,
      \end{aligned}
    \right.
\end{eqnarray}
and $z_{\max}$ can approximately be obtained through Eq.~\eqref{eq:zmax} (see \S\ref{sec:MaximumRedshiftForDetection}).

\subsection{{Caveats}}
\label{sec:Caveats}

Our method is fully analytical, this has several advantages and downsides. On the one hand, this allows us to study a variety of models (\eg different detectors and abilities to extract physical signals from the noise, different maximum stellar masses surrounding BHs, or different BH mass functions) and to understand what are the physical relevant parameters for detection of gravitational waves from TDEs. On the other hand, the simplicity of the method comes at the price of many assumptions due to our still incomplete understanding of the physics (\eg maximum penetration factor, mass-to-radius relation of massive stars), or due to that incorporating such physics would add an extra layer of complexity beyond the scope of this paper (\eg isothermal sphere lying on the $\Mbh-\sigma$ relation). As such, our predictions should be regarded only as guidelines and order of magnitude estimates. Yet, we believe that our results provide insight on the feasibility of gravitationally detected TDEs.

\section{{Total rates of gravitationally observed TDEs}}
\label{sec:results}

In this Section we compute the number of TDEs emitting gravitational waves we can detect and their properties. We first focus in \S\ref{sec:ExpectedNumberOfObservationsAndDistributions} on one particular model and exemplify what different detectors can tell about this model, in \S\ref{sec:TheNeedForDetectorImprovement} we detail what these future observations can tell us about the underlying properties of TDEs.

\subsection{Typical numbers of observations and distributions}
\label{sec:ExpectedNumberOfObservationsAndDistributions}

\begin{figure}
\includegraphics[width=\columnwidth]{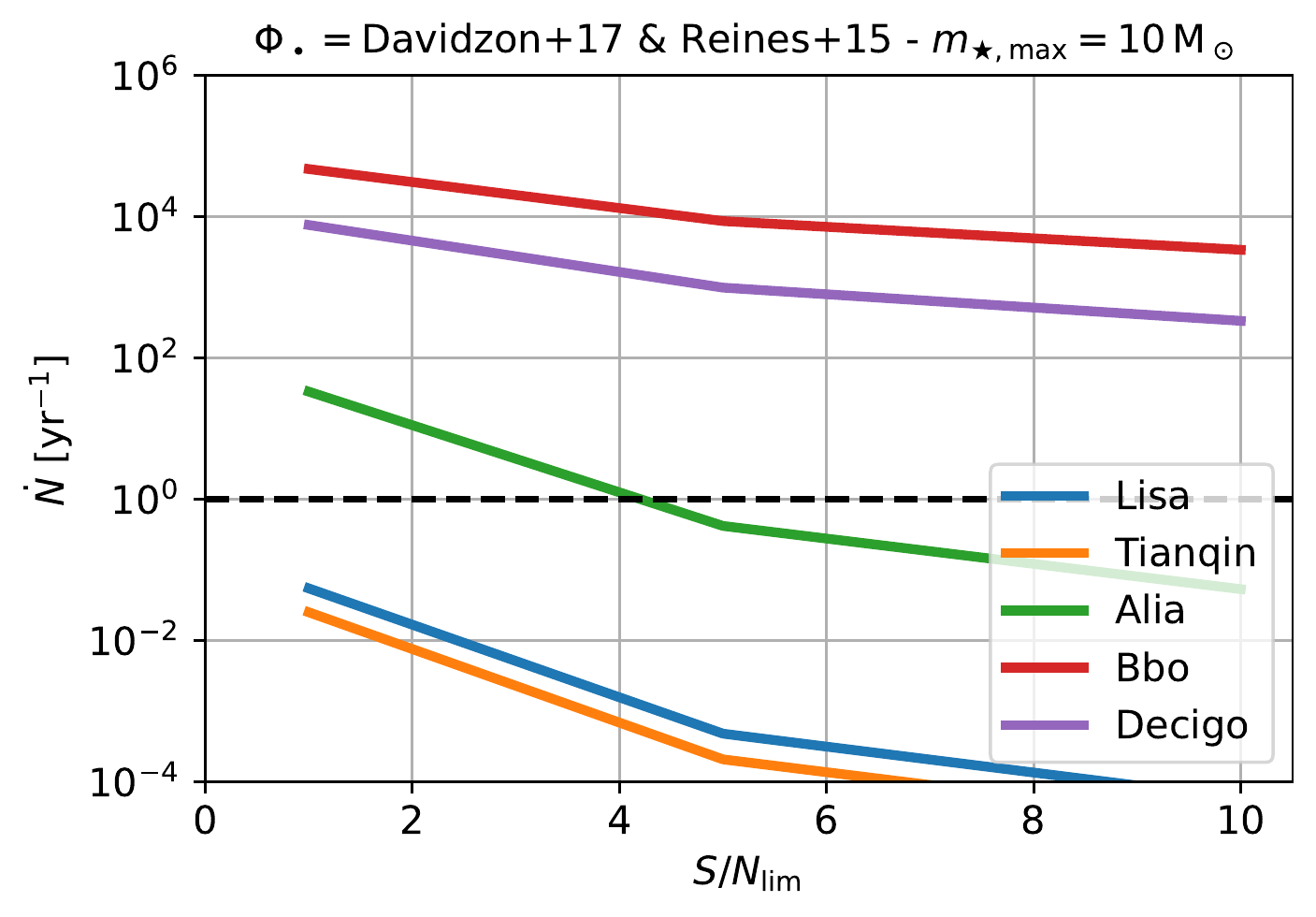}
\caption{Observable TDE rate with gravitational waves as a function of our ability to detect events from the noise for different gravitational wave detectors. While a detection with LISA and Tianqin are unlikely, the second generation of space-based gravitational wave detectors will observe gravitational waves of TDEs.}
\label{fig:ObsAll}
\end{figure}

In this Section we focus on one model: the Kroupa stellar mass varies between $[0.08;10]\Msun$ and the BH population is obtained from the galaxy mass function and BH-galaxy mass scaling relations ($\Phi_{\bullet,\, 1}$, Eq.~\eqref{eq:BHMF_model1}).  We chose this model as ``fiducial'' because $m_{\star,\,\max}=10\Msun$ corresponds to a relatively old stellar population \citep[$10-100\Myr$, ][]{Choi_16} consistent with previous works \citep{Stone_16a}; and because this model for the BH population depends on redshift, which is necessary as Bbo and Decigo can detect TDEs up to $z \gtrsim 1-10$ (see Fig.~\ref{fig:zmaxVSMBH}).

We show in Fig.~\ref{fig:ObsAll} the observable rate of TDEs $(\int \d^4 \dot{N})$ as a function of the $S/N_{\lim}$ threshold for detection, or in other words the number of detection in a year as a function of our (in)ability to detect events in the noise. We show the rates for different detectors (colors) and we also mark the critical rate of 1 event per year (horizontal black dashed line). We find that it is unlikely that LISA and TianQin will detect TDEs. Future generations of gravitational wave detectors (Bbo and Decigo) should however detect tens of thousands of events per year. In both cases we note that the observed rate is extremely sensitive to our ability to detect signal from the noise.

\begin{figure*}
\includegraphics[width=\columnwidth]{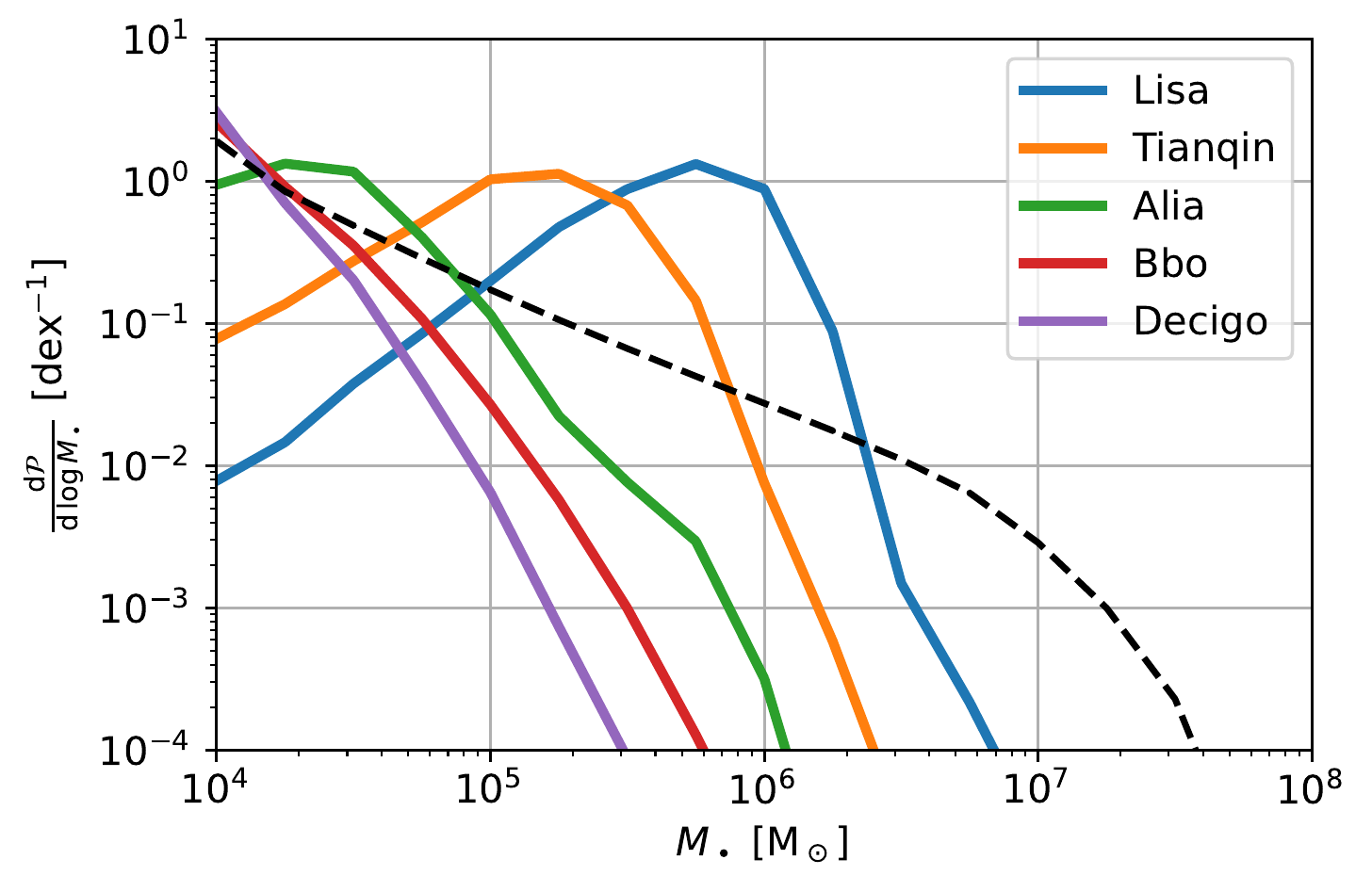} \hfill 
\includegraphics[width=\columnwidth]{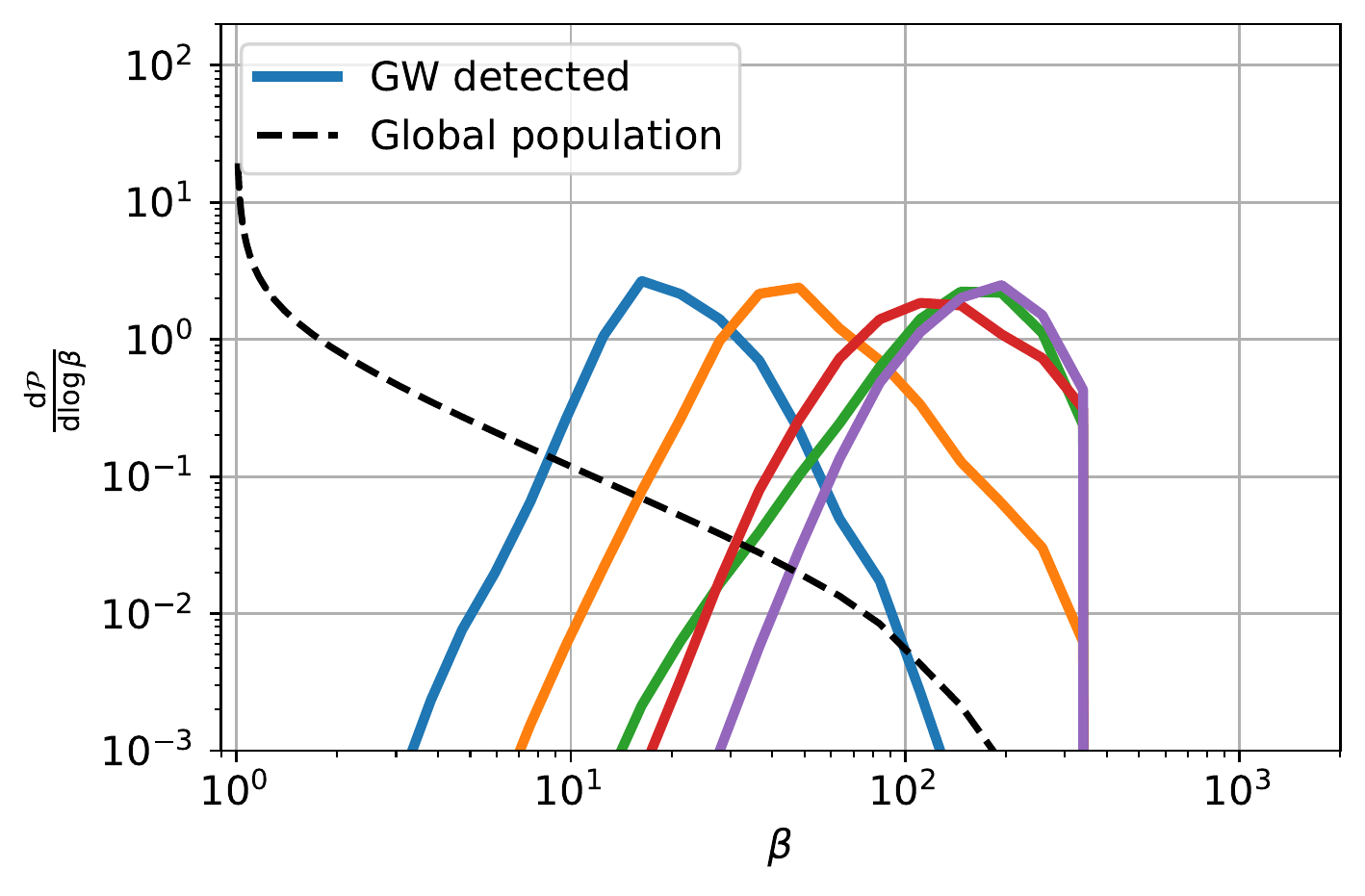} \\
\includegraphics[width=\columnwidth]{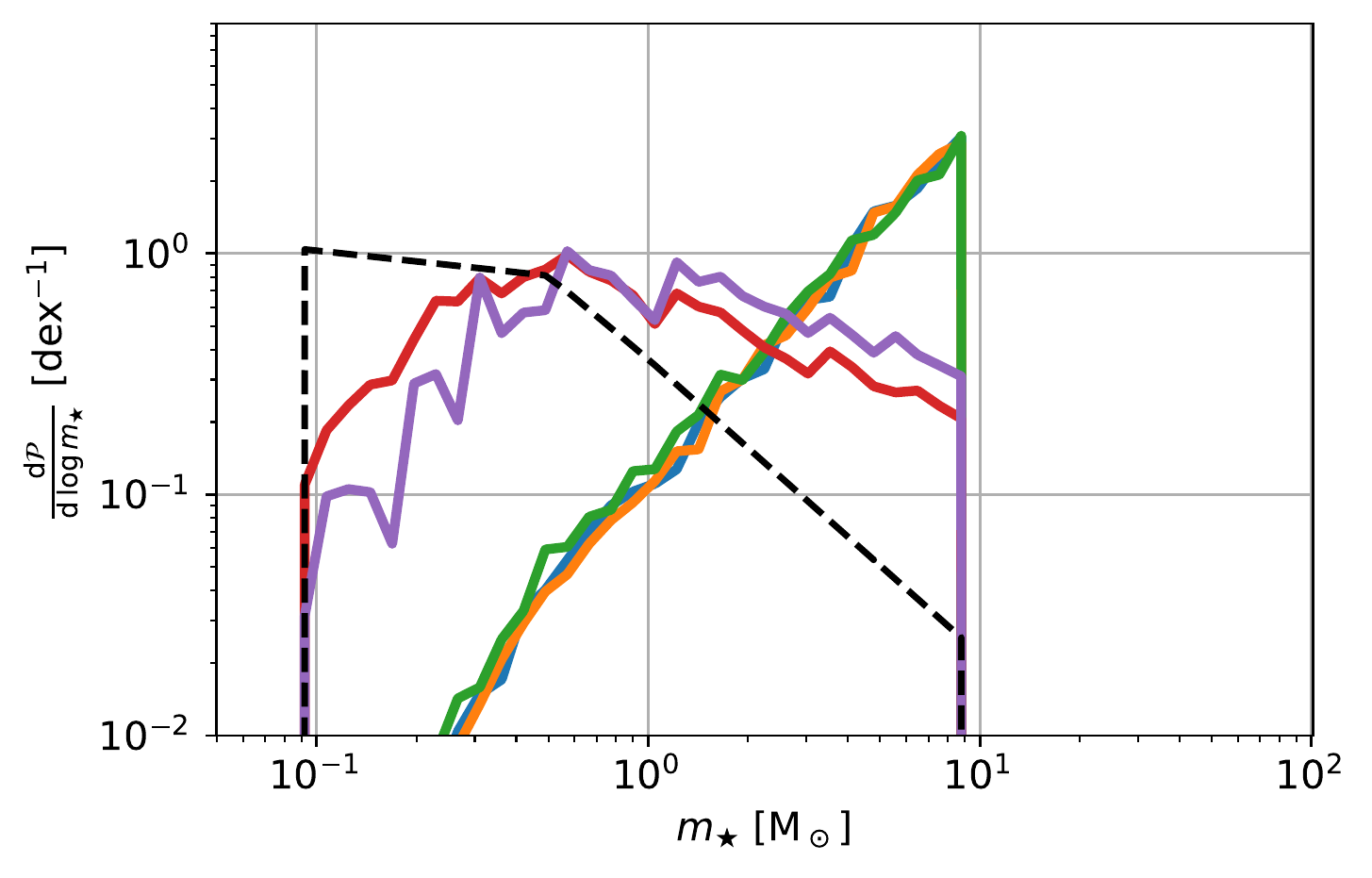}  \hfill
\includegraphics[width=\columnwidth]{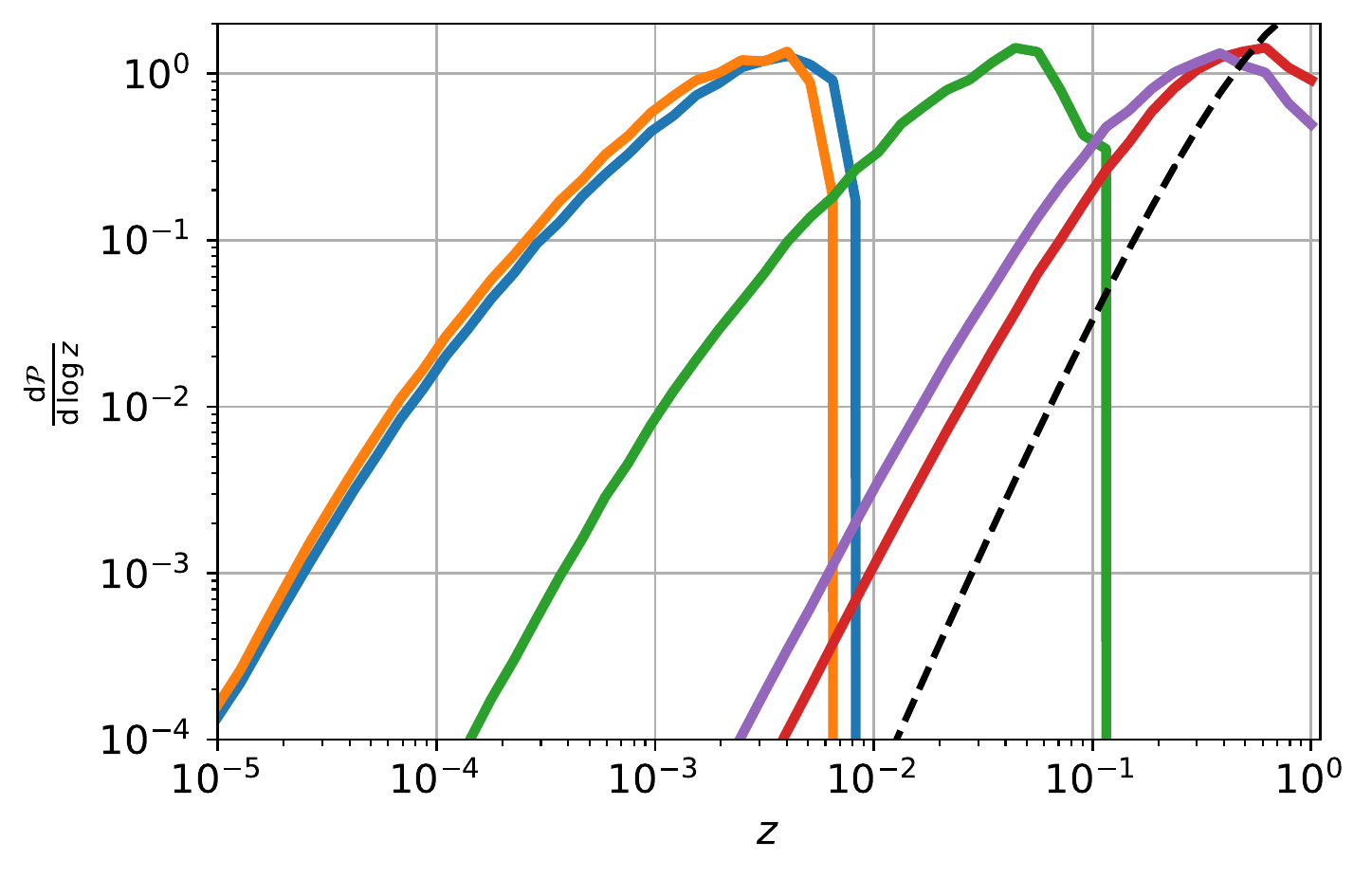}
\caption{Distribution of TDEs if we were able to detect all of them (``global population'', dashed lines), and distributions of ``GW detected'' TDEs (solid lines) for different gravitational wave detectors (colors).}
\label{fig:PropObs}
\end{figure*}

In order to know what the typical parameters of these possible detections will be, we show in Fig.~\ref{fig:PropObs} the distribution functions $\d^4 \mathcal{P}=\d^4\Gamma/\dot{\Gamma}$ ($\d^4 \mathcal{P}=\d^4\dot{N}/\dot{N}$) marginalized onto the different relevant parameters ($\Mbh,\, \beta,\, \mstar,\, z$) for the global (gravitationally observed) population of TDEs.

We begin with the global populations of TDEs (black dashed lines in Fig.~\ref{fig:PropObs}, obtained with $\Gamma$), that is the intrisic distribution of TDEs for our model that may or may not be observed with electromagnetic or gravitational waves.

The distribution with the BH mass (upper left) decreases by $\sim 2 \dex$ in the range $10^4-10^6\Msun$ which reflects the fact that both the BH mass function (see Fig.~\ref{fig:BHMF}) and the TDE rate at fixed BH mass (see Fig.~\ref{fig:GammaVSmBH}) decrease by $\sim 1 \dex$. The distribution with the penetration factor (upper right) peaks at $\beta \sim 1$, then scales as $1/\beta^2$ on a wide range of values, and finally smoothly decreases to reach 0 at $\beta~=\beta_{\max}$. This is in agreement with previous results \citep{Stone_16a, Kochanek_16}, suggesting a combination of $\delta$-dirac and $1/\beta^2$ distributions respectively in the empty and full loss cone, but we find here an exact continuous distribution which will be detailed more in depth in Wong et al. (in prep). The distribution with the stellar mass (lower left) is similar to that of the underlying stellar mass function chosen, which reflects that stars diffuse similarly as diffusion coefficients are independent of the stellar mass. Finally, the distribution with (the log of) redshift (lower right) scales as $z^3$ which reflects that Eq.~\eqref{eq:MasterEqGlobal} scales as $\partial_z \Gamma \propto z^2$. In summary, this analysis shows that, albeit inexact, it is reasonable to overall write the global differential rate as:
\begin{eqnarray}
&&\frac{\d^4 \Gamma}{\d z\, \d\Mbh\, \d \mstar\, \d \beta}\sim  \label{eq:PartialRateProduct}\\
&& \frac{\Gamma(M_\bullet)}{(1+z)} \times \frac{\beta_{\max} }{\beta^2 (\beta_{\max} -1)} \times \phi \times \Phi_\bullet \times 4\pi c \frac{\chi^2(z)}{ H(z)} \nonumber \, .
\end{eqnarray}
To our knowledge, this is the first time such a demonstration is obtained, although similar forms of this equation have been used in previous works \citep[\eg ][]{Kochanek_16b, Toscani_20}. When expressed as this, it clearly appears that if we know the distribution of TDEs, then we have a viable way to probe the BH mass function, the stellar mass function or even cosmological parameters through the $H$ and $\chi$ dependency.

However, we do not have direct access to the distribution of TDEs, but to the distribution of gravitationally observed TDEs. This is why we now move the population of TDEs observed with different detectors (solid lines in Fig.~\ref{fig:PropObs}, obtained with $\dot{N}$). We recall that we may not even observe TDEs with LISA or TianQin ($S/N_{\lim}=5$ in this case) such that these distributions really makes sense for future generation detectors. Nonetheless, they can be useful to obtain the most probable events, in the situation in which, by chance, we have a detection with LISA or TianQin. 

The distribution with the BH mass (upper left) differs greatly from the one of the intrinsic populations. This reflects that, as discussed in \S\ref{sec:MaximumRedshiftForDetection}, detectors are particularly sensible to BHs for which the \hugo{Keplerian frequency around the critical radius for direct plunge ($\kappa \times r_{\rm Sch}$)} is the same as the optimal frequency of the detector
: similarly to Fig.~\ref{fig:zmaxVSMBH} (right) the peak is at $M_{\bullet, {\rm opt}}$ (see Table~\ref{tab:fdet}). The distribution with the penetration factor (upper right) typically exhibits a peak at $\beta \gtrsim 20$, \eg $\beta \sim 20$ for LISA and $\beta \sim 250$ for Decigo. This reflects that, for an average population of stars, say with $\langle \mstar \rangle \sim 1\Msun$, typically disrupted around BHs with mass $\Mbh \sim M_{\bullet,\,{\rm opt}}$, detectors are particularly sensible to events with $\beta \sim \beta_{\max}$ (Eq.~\eqref{eq:beta_max}): that is $\beta \sim 15$ for LISA, and $\beta \sim 250$ for Decigo. The distribution with the stellar mass (lower left) is rather difficult to interpret as, on the one hand, low mass stars are more numerous; but on the other hand, high mass stars can be detected to higher redshift (Eq.~\eqref{eq:zmax}) and can be disrupted across the entire BH mass range ($\beta_{\max}$ scales positively with $m_\star$). In the end, the distribution for most sensitive detectors, which will be able to detect most TDEs (Bbo and Decigo), is similar to that of the underlying stellar mass function, while the distribution for less sensitive detectors (LISA, TianQin and Alia) is skewed toward high mass stars. Finally, the distribution with redshift (lower right), similarly to that of the global rate, scales as $z^3$ at ``low'' redshift; subsequent evolution is a competition between volume which makes events more and more numerous and distance which makes them less and less detectable, it results in gradual flattening of the distribution where it reaches its maximum ($z \sim \few 10^{-3}$ for LISA and $z \sim \few 10^{-1}$ for Decigo) and then decreases. This reflects that most detected TDEs will be of stars with a mass $m_{\star,\,{\rm opt}}$ (10\Msun for LISA and 1\Msun for Decigo) around $M_{\bullet,\,{\rm opt}}$ BHs at $z_{\max}$ (Fig.~\ref{fig:zmaxVSMBH}) for volume effects, which yields $\sim \few 10^{-3}$ for LISA and $\sim \few 10^{-1}$ for Decigo.

In summary, we will unlikely detect TDEs with gravitational waves during the LISA and TianQin missions, but next generations detectors will observe hundreds to tens of thousands of events per year. We also derive a complete $\beta$-distribution encompassing, but consistent with, both the full and empty loss cone regimes (black dashed line in the upper right panel of Fig.~\ref{fig:PropObs}). Finally we show how the underlying and the observed distributions of TDEs are affected by the different detectors, allowing to predict the properties of the most probable observed events. For LISA, although detections are unlikely, most probable TDE detection will be disruptions of $\sim10\Msun$ stars on $\beta\sim20$ orbits around a $\sim 5\times10^5\Msun$ BHs at $z\sim0.005$. For Decigo, most probable TDE detection will be disruptions of $\sim1\Msun$ stars on $\beta\sim200$ orbits around a $\sim 10^4\Msun$ BHs at $z\sim0.5$. This results in a $G$ magnitude of respectively 15 and 29, or X-ray flux of $10^{-11}\,\text{erg cm}^{-2}\text{ s}^{-1}$ and $10^{-16}\,\text{erg cm}^{-2}\text{ s}^{-1}$. As a consequence, these TDEs observed with gravitational waves will also be observed by facilities in the electromagnetic spectrum like Lynx \citep{lynx_18}, Athena \citep{Nandra_13} or the LSST \citep{Ivezic_19}. The information encoded in the gravitational wave signal \citep[$\Mbh$, $\beta$, $\mstar$, $z$ but also informations on the internal structure of the star, or the spin of the BH, see][]{Stone_19} combined with those of the electromagnetic signal \citep[which are already used to undersand TDEs, \eg ][]{Mockler_19} will open the multi-messenger era for TDEs and unveil new physics currently not well constrained \citep[\eg][]{Roth_20b, Bonnerot_21, Dai_21}. 

In this Section, we focused on the detection rate of several detectors assuming one fiducial model. In the following Section, we focus on the case of LISA (in construction detector) and Decigo (next generation detector) and discuss different models.

\subsection{The need for detector improvement}
\label{sec:TheNeedForDetectorImprovement}

\begin{figure*}
\includegraphics[width=\columnwidth]{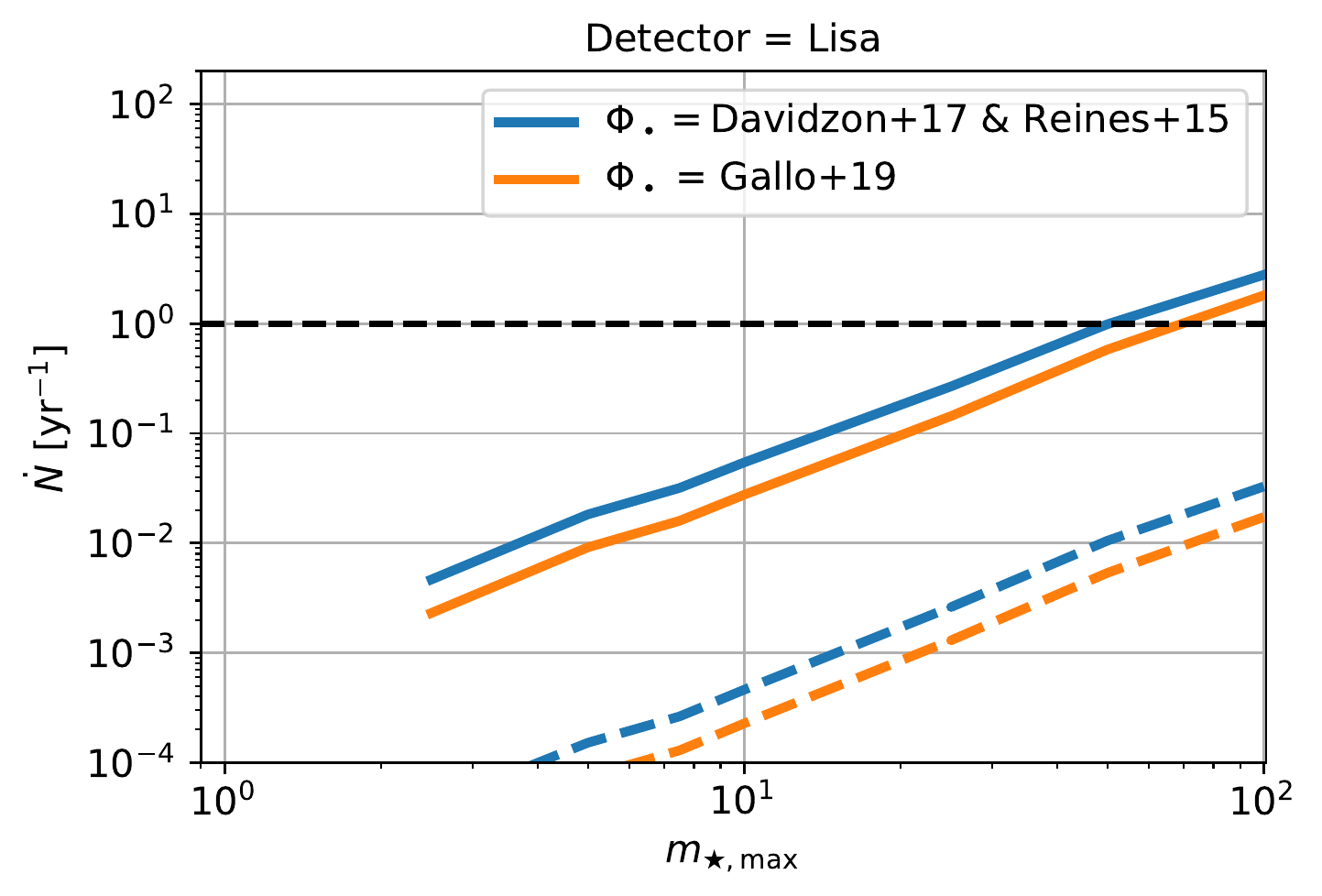} \hfill 
\includegraphics[width=\columnwidth]{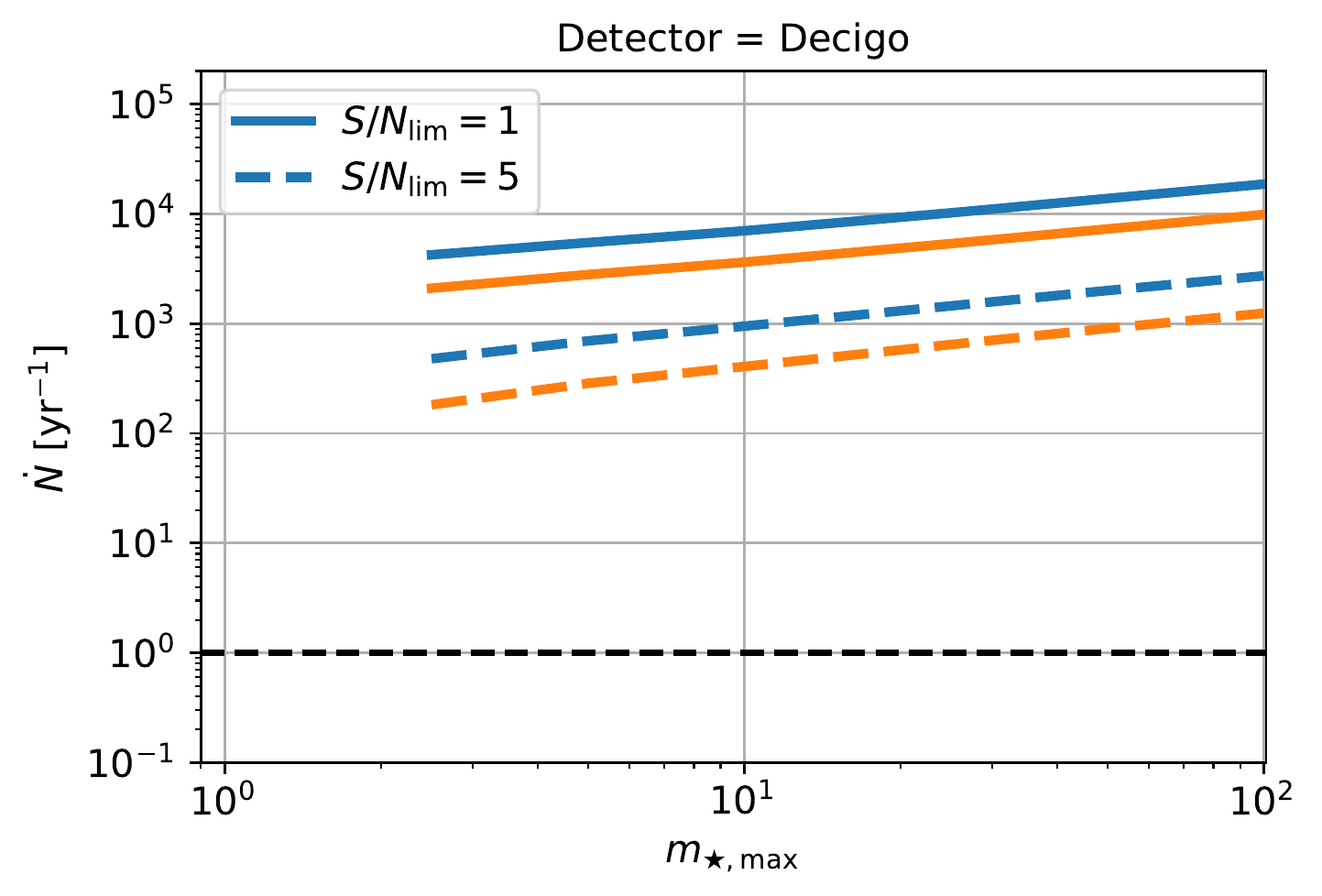}
\caption{Observable TDE rate with gravitational waves as a function of the maximum stellar mass in the Kroupa stellar mass function. We show the results for LISA (left) and Decigo (right), and in both cases for our two models for the BH mass function (color, see \S\ref{sec:BHMassFunction}) and ability to detect the signal from the noise (line style).}
\label{fig:NobsVSmstar}
\end{figure*}

The different models explored in this paper differ by the BH mass function and the maximum stellar mass in the Kroupa stellar mass function (see \S~\ref{sec:Caveats}). We show in Fig.~\ref{fig:NobsVSmstar} the observable rate of TDEs as a function of $m_{\star,\, \max}$, for the two BH mass functions (colors), for two $S/N_{\lim}$ (line style), and for LISA (left) and Decigo (right).

In all cases we note that there is still a strong dependency with our ability to detect events from the noise.

The case of LISA (left) is interesting because, if we are able to detect events with $S/N_{\lim} = 1$, there exists a set of models for which there will be up to few detections per year. This means that (non) observations will favour (rule out) these models. To be more precise, if we observe TDEs with LISA, this implies that typical stellar population around BHs is rather young with $m_{\star,\, \max} \gtrsim 60 \Msun$, independently of the BH mass function; and vice versa.

This information can then be used to better interpret the future observations of Decigo (right). Chosing again the example of $S/N_{\lim} = 1$, if, for instance, $\sim 10^4$ events are yearly observed, one cannot know if the underlying population is $\Phi_{\bullet,\, 1}$ with $m_{\star,\, \max}\sim20\Msun$ or $\Phi_{\bullet,\, 2}$ with $m_{\star,\, \max}\sim100\Msun$, but previous (non) observations with LISA can help in disentangling the two scenarii.

In summary, apart from the optimistic case $m_{\star,\, \max}\sim100\Msun$, it is quite unlikely that LISA will observe any TDEs during its 4 years mission. However, (non) observations will already constrain the typical stellar age around BHs, and will be useful to better understand future observations.

\section{Conclusions}
\label{sec:conclusions}

We determine the rates of possible observations of TDEs with future gravitational wave spacecrafts as well as their possible electromagnetic counterpart. To this purpose we develop a simple semi-analytical model combining standard gravitational wave and electromagnetic results (\S\ref{sec:DetectionOfSingleEvents}) and rates estimates (\S\ref{sec:TDErates}). We summarize our findings below:

\begin{itemize}
\item LISA could detect gravitational waves from extreme TDEs ($m_\star\sim100\Msun$ on an orbit which skims the Schwarzschild radius of a $10^6\Msun$ BH) up to $z_\mathrm{max}\sim0.1$ (see Fig.~\ref{fig:zmaxVSMBH}). We provide an analytical expression for $z_\mathrm{max}$ (Eq.~\eqref{eq:zmax}).
\item Under the assumptions that all TDEs produce prompt and luminous optical or X-ray emissions, then all these LISA gravitational detections should be detectable electromagnetically (see Fig.~\ref{fig:g_band}~-~\ref{fig:LISA_xray}).
\item The TDE rate of a monochromatic stellar population is about 5 times lower than that of a Kroupa stellar population (Fig.~\ref{fig:GammaVSmBH}). Since we remove events for which the star is swallowed whole (assuming a Schwarzschild BH), we find a smooth decay of the TDE rate with $M_\bullet$ starting at $\sim 10^7\Msun$ BH where $\sim0.1\Msun$ stars are being swallowed, and finishing at $\gtrsim 10^8\Msun$ where $>1\Msun$ stars are being swallowed (see Fig.~\ref{fig:mstarBHVSMBH}). 
\item This enhancement is in broad agreement with previous analytical results (Fig.~\ref{fig:GammaVSmstarmax}), although we derive in this paper more detailed rates, in particular regarding the dependency with the penetration factor $\beta$.  \item The TDE rate overall decreases with the stellar mass and the penetration factor, but its complex variations are depicted in Fig.~\ref{fig:Example_GammaVSbetaANDrp}, and will be discussed in more details in Wong et al. (in prep).
\item LISA should not detect any TDEs (Fig.~\ref{fig:ObsAll}), unless BHs are typically embedded by a young stellar population with $m_{\star,\, \max}~\gtrsim 60~\Msun$ which, in this situation, could lead up to few 10 events during the duration of the mission  (Fig.~\ref{fig:NobsVSmstar}). As such, the number of (non) detections will reveal the typical age surrounding BHs, with (non) detections if BHs are typically embedded in a young (old) stellar population.
\item The following generation of detectors (Alia, Bbo and Decigo) will be more sensitive and will be able to yearly detect thousands to millions of events (Fig.~\ref{fig:ObsAll}) at cosmological redshift (Fig.~\ref{fig:zmaxVSMBH}), allowing to probe the BH mass function (Fig.~\ref{fig:NobsVSmstar}).
\item \hugo{For each detectors and models, we obtain the distribution of parameters (Fig.~\ref{fig:PropObs}). The most probable BH mass corresponds to BHs for which the Keplerian frequency around the critical radius for direct plunge ($\kappa \times r_{\rm Sch}$) is the same as the optimal frequency of the detector (see Table~\ref{tab:fdet}). Assuming this most probable BH mass and some typical $\mstar \sim 1\Msun$, the most probable penetration factor corresponds to the maximum possible value (Eq.~\eqref{eq:beta_max}), and the most probable redshift corresponds to the maximum redshift for the detector (Eq.~\eqref{eq:zmax}).}
\end{itemize}

In order to end up with a finite paper, several assumptions have been made (Kroupa stellar mass function, one single choice of the $\Mbh-\sigma$ relation, non-spinning BHs etc...) which may affect the exacts rates obtained in this paper. As such, these predictions should be regarded as guidelines. It would be interesting to investigate in depth the detailed effects of other parameters in future studies.

\section*{Acknowledgments}
{We thank the anonymous referee carefully reading the first version of the paper and for providing useful comments.} HP acknowledge support from the Danish National Research Foundation (DNRF132) and the Hong Kong government (GRF grant HKU27305119, HKU17304821). MT and GL have received funding from the European Union’s Horizon 2020 research and innovation program under the Marie Sk\l{}odowska-Curie grant agreement NO 823823 (RISE DUSTBUSTERS project). TW and JLD are supported by the GRF grant from the Hong Kong government under HKU 27305119 and HKU 17304821.

HP work would not be possible without the exceptional following Python tools: \textsc{Astropy} \citep{astropy}, \textsc{iPython} \citep{iPython}, \textsc{Jupyter-Notebook} \citep{jupyter}, \textsc{Matplotlib} \citep{matplotlib}, \textsc{Numpy} \citep{numpy}, \textsc{Pandas} \citep{pandas}, \textsc{Scipy} \citep{scipy}, and \textsc{YT} \citep{yt}.

\section*{Data availability}
Scripts and data used for this paper are available upon request.

\appendix

\section{Sentivity of gravitational waves detector}
\label{sec:SensitivityOfGWDetectors}

\begin{figure}
\centering
\includegraphics[width=\columnwidth]{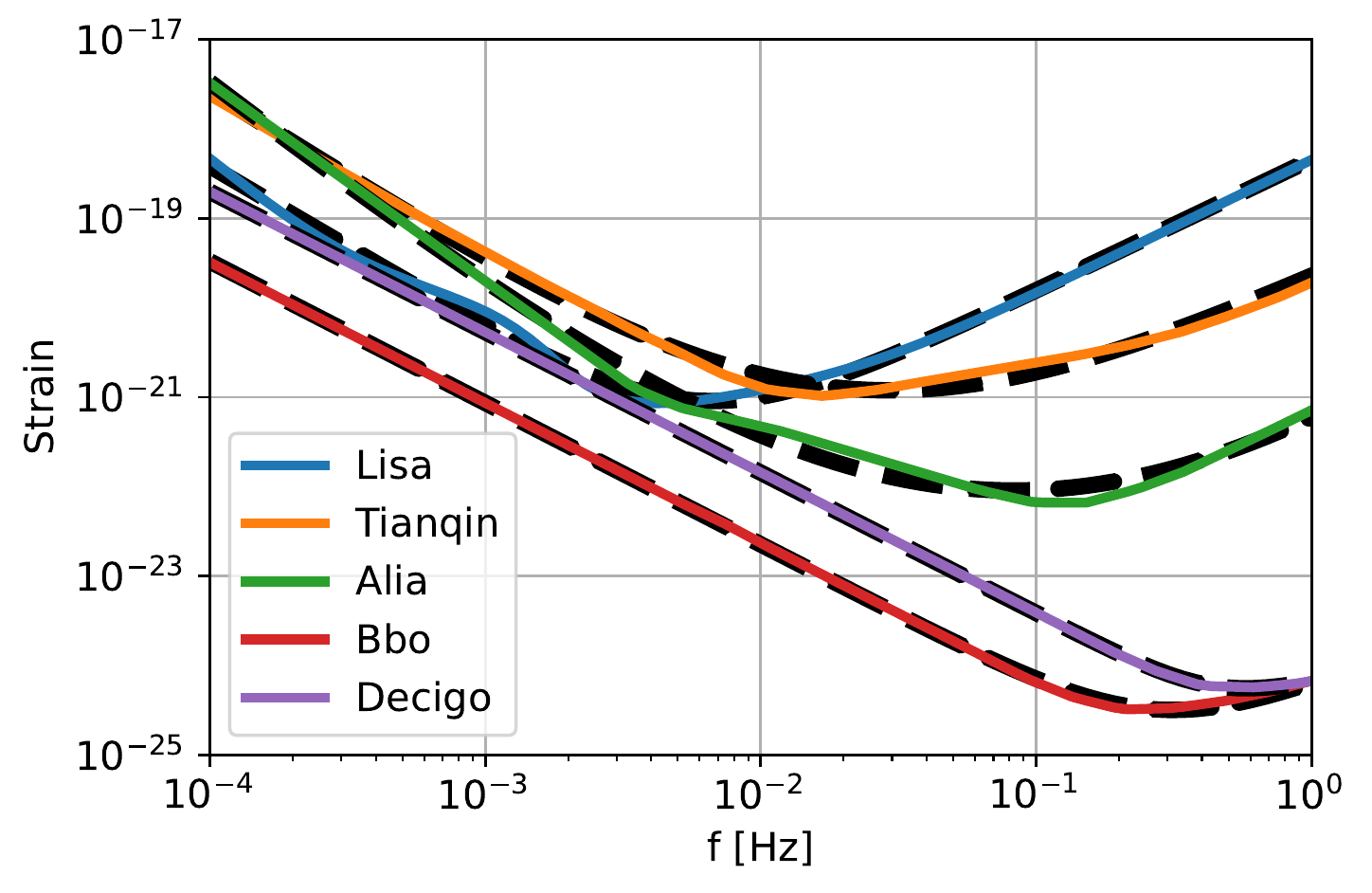}
\caption{Sensitivity curve as a function of frequency for several gravitational wave detectors (colors) {as well as the fit (black dashed lines) with Eq.~\eqref{eq:fit_strain}}.}
\label{fig:SensitivityDet}
\end{figure}

We download the sensitivity curves of several instruments on \hyperlink{http://gwplotter.com/}{http://gwplotter.com/} and report them in Fig.~\ref{fig:SensitivityDet}. {While we use the numerical value for this work, it can be useful to have a simple analytical expression to understand the results. Using a simple $\chi$-square regression in the log-log plane, we fit each curves with the following functional:
\begin{eqnarray}
h_\mathrm{det}(f) = h_\mathrm{opt} \left( \frac{f}{f_\mathrm{opt}} \right)^{-a} \left( \left( \frac{f}{f_\mathrm{opt}} \right)^{c} +1  \right)^{\frac{b + a}{c}} \, , \label{eq:fit_strain}
\end{eqnarray}
where $f_\mathrm{opt}$ and $h_\mathrm{opt}$ are respectively the optimal frequency and strain of the detector, $a$ and $b$ are the logarithmic slopes for $f \ll f_\mathrm{opt}$ and $f \gg f_\mathrm{opt}$. The fitting parameters can be found in Table~\ref{tab:fit_det} with the resulting curve shown with dashed black lines in Fig.~\ref{fig:SensitivityDet}.
}

\begin{table}
    \centering
    \begin{tabular}{|l|c|c|c|c|c|}
    \hline
         \textbf{Detector} & $f_\mathrm{opt}$ & $h_\mathrm{opt}$  & $a$ & $b$ & $c$ \\
                           &             Hz   & $10^{-21}$        &     &     &     \\
    \hline
        LISA    & $6\times 10^{-3}$& 0.2    &   1.8 &   1.5 & 1.7 \\
        Tianqin & $0.02$           &  7     &   2.0 &   1.4 & 0.8 \\
        Alia    & $0.08$           &  0.02  &   2.5 &   2.3 & 0.6 \\
        Bbo     & $0.3$            &  0.01  &   1.6 &   1.2 & 2.0 \\
        Decigo  & $0.4$            &  0.04  &   1.6 &   0.7 & 3.3 \\
         \hline
    \end{tabular}
    \caption{{Best fitting parameters of the strain of the different detectors with Eq.~\eqref{eq:fit_strain}.}}
    \label{tab:fit_det}
\end{table}

\section{Fokker-Planck equation with a stellar population}
\label{sec:FPEqWithAStellarPopulation}

In this Appendix we show how we can use the formalism developed for monochromatic stellar population to study a more complex population of stars. It is mostly a summary and combination of the works of \cite{Magorrian_99} and \cite{Strubbe_11}.

\subsection{Diffusion coefficients}

If the medium is homogeneous such that the stellar mass function ($\phi$) is independent of position and time, then we may write the density function $f_\star$ as:
\begin{eqnarray}
f_\star(\vv{r},\vv{v},m_\star) \d \vv{r} \d \vv{v} \d m_\star  = f(\vv{r},\vv{v})\d \vv{r} \d \vv{v} \times \phi(m_\star) \d m_\star\, .
\end{eqnarray}.

Following \S8.3 from \cite{BT_87}, assuming the background follow the density $f(\vv{r},\vv{v})$, we define $\psi(\vv{w},\Delta \vv{w}, m_\star, m_\mathrm{scat}) \d^6 \Delta \vv{w} \d t$ as the probability that a star of mass $m_\star$ is scattered from $\vv{w}$ to $\vv{w}+\Delta \vv{w}$ ($\vv{w}=(\vv{r},\vv{v})$) by a background star of mass $m_\mathrm{scat}$ during a transition time $\d t$. Since the background is made of a population of star with distribution $\phi$, the probability that a star of mass $m_\star$ is scattered from $\vv{w}$ to $\vv{w}+\Delta \vv{w}$ ($\vv{w}=(\vv{r},\vv{v})$) by the background during $\d t$ is:
\begin{eqnarray}
\psi_\star(\vv{w},\Delta \vv{w}, m_\star) \d^6 \Delta \vv{w}   \d t  = \\
\d^6 \Delta \vv{w}   \d t  \int \phi(m_\mathrm{scat}) \psi(\vv{w},\Delta \vv{w}, m_\star, m_\mathrm{scat}) \d m_\mathrm{scat} \, . \nonumber
\end{eqnarray}

This allows to classically write the evolution of $f_\star$ as a Fokker-Planck equation:
\begin{eqnarray}
\frac{\d f_\star}{\d t} &=& - \sum_{i=1}^3\frac{\partial}{\partial v_i}\left[f_\star(\vv{w}, m_\star) D_\star(\Delta v_i, m_\star) \right] + \nonumber \\
&& \frac{1}{2} \sum_{i,j=1}^3\frac{\partial^2}{\partial v_i\partial v_j}\left[f_\star(\vv{w}, m_\star) D_\star(\Delta v_i \Delta v_j, m_\star) \right] \, , \label{eq:FP_complex}
\end{eqnarray}
where we have defined the diffusion coefficient which, \textit{a priori}, depend on $m_\star$:
\begin{eqnarray}
D_\star(\Delta v_i ,m_\star) &=& \int \Delta v_i \psi_\star(\vv{w},\Delta \vv{w}, m_\star) \d^6\Delta\vv{w} \\
&=& \int \phi(m_\mathrm{scat}) \d m_\mathrm{scat} \Big[   \\ 
&& \int \Delta v_i \psi(\vv{w},\Delta \vv{w}, m_\star, m_\mathrm{scat}) \d^6\Delta\vv{w}\, \Big] \nonumber  \\
&=& \int \phi(m_\mathrm{scat}) D(\Delta v_i, m_\star, m_\mathrm{scat}) \d m_\mathrm{scat} \label{eq:Dstar_BT} \, ,
\end{eqnarray}
and $D(\Delta v_i, m_\star, m_\mathrm{scat})$ is the diffusion coefficient of test particle of mass $m_\star$ moving in a monochromatic population of star $m_\mathrm{scat}$ with density function $f(\vv{r},\vv{v}) $ \citep[see \S8.3 from ][]{BT_87}.

When written as this, Eq.~\eqref{eq:FP_complex} show that the evolution of a star with mass $m_{\star,1}$ differs from the evolution of a star with mass $m_{\star,2}$ because their diffusion coefficients differ. This means that to study the evolution of our system, one should study a set of coupled equations, with one Fokker-Planck equation for each mass.

\subsection{The particular case of TDEs}

This set of coupled Fokker-Planck equations can be greatly simplified in the case of TDEs. After changing variables from $\vv{w}$ to (minus the specific) energy ($E=-v^2/2+\Psi(r)$, $\Psi$ is the potential) and  $R=J^2/J^2_c(E)$ ($J=|\vv{r} \wedge \vv{v}|$ is the angular momentum and $J_c(E)$ is the circular angular momentum at a given energy), a Fokker-Planck equation for $f_\star$ can still be written. Neglecting diffusion in $E$ as it relaxes on a longer timescale than $R$ \citep{Lightman_77, Cohn_78, Merrit_book}, this yields:
\begin{eqnarray}
\frac{\d f_\star}{\d t} &=& \frac{\partial}{\partial R}\Bigl[ -f_\star D_\star(\Delta R, m_\star) +\\
&& \frac{1}{2} \frac{\partial}{\partial R} [f_\star D_\star((\Delta R)^2, m_\star)] \Bigr] \, .
\end{eqnarray}
Here again, $D_\star(., m_\star)$ are the diffusion coefficients and \textit{a priori} depend in $m_\star$. 

The diffusion coefficient of interest can be computed \citep{Lightman_77,Magorrian_99, Stone_16a, BarOr_16}:
\begin{eqnarray}
D_\star(\Delta R, m_\star) &=& \frac{r^2 D_\star((\Delta v_\perp)^2, m_\star)}{J^2_c} \\
D_\star((\Delta R)^2, m_\star) &=& 2RD_\star(\Delta R, m_\star)\, ,
\end{eqnarray}
where $D_\star( (\Delta v_\perp)^2, m_\star)$ is the diffusion coefficient in $(\Delta v_\perp)^2$, corresponding to the orthogonal component of the variation of the velocity. Using Eq.~\eqref{eq:Dstar_BT} and that $D( (\Delta v_\perp)^2, m_\star, m_{\rm scat})$ is \citep[Eq.~(8A-22) of ][]{BT_87}:
\begin{eqnarray}
D( (\Delta v_\perp)^2, m_\star, m_\mathrm{scat}) = \frac{32 \pi^2 \G^2 m^2_\mathrm{scat} \ln\Lambda}{3 v}\ \times \\
\left( \int_0^v (3 \tilde{v}^2 - \frac{\tilde{v}^4}{v^2}) f(\vv{r}, \tilde{v}) \d \tilde{v} + 2v \int_v^\infty \tilde{v} f(\vv{r}, \tilde{v}) \d \tilde{v} \right) \nonumber \, ,
\end{eqnarray}
we have the remarkable result that $D_\star( (\Delta v_\perp)^2, m_\star)$, hence $D_\star(\Delta R, m_\star)$ and $D_\star((\Delta R)^2, m_\star)$ which are the two diffusion coefficients of interest for us, are \textit{independent}\footnote{There is actually a small dependency through the Coulomb logarithm factor $\ln \Lambda$, but this dependency is logarithmic and usually neglected.} of $m_\star$. This means that, under our assumptions, the evolution of a test particle is independent of its mass, and everything happens as if the particle was moving in a background composed by stars of mass $\langle m^2_\star \rangle^{1/2}=\left(\int m^2_{\rm scat}\phi(m_{\rm scat})\d m_{\rm scat}\right)^{1/2}$, that is $D_\star(\Delta R)=D(\Delta R, m_\mathrm{scat}=\langle m^2_\star \rangle^{1/2})=\mu$ (note that we have now dropped the $m_\star$ dependency).

All this allows to write a Fokker-Planck equation for $f$:
\begin{eqnarray}
\frac{\d f}{\d t} &=& \frac{\partial}{\partial R}\Bigl[ -f D_\star(\Delta R) +\\
&& \frac{1}{2} \frac{\partial}{\partial R} [f D_\star((\Delta R)^2)] \Bigr] \\
&=& \mu \frac{\partial }{\partial R}\left( R \frac{\partial f}{\partial R} \right)\, ,
\end{eqnarray}
where we recognize the starting point of estimates of TDE rates \citep[\eg][]{Merrit_book}.

In conclusion, for our purpose and under approximation, the evolution of a test mass in a medium composed by a stellar population is the same as a test mass in a medium composed single type of stars with mass $\langle m^2_\star \rangle^{1/2}$.

\section{BH mass function}

\begin{figure}
\includegraphics[width=\columnwidth]{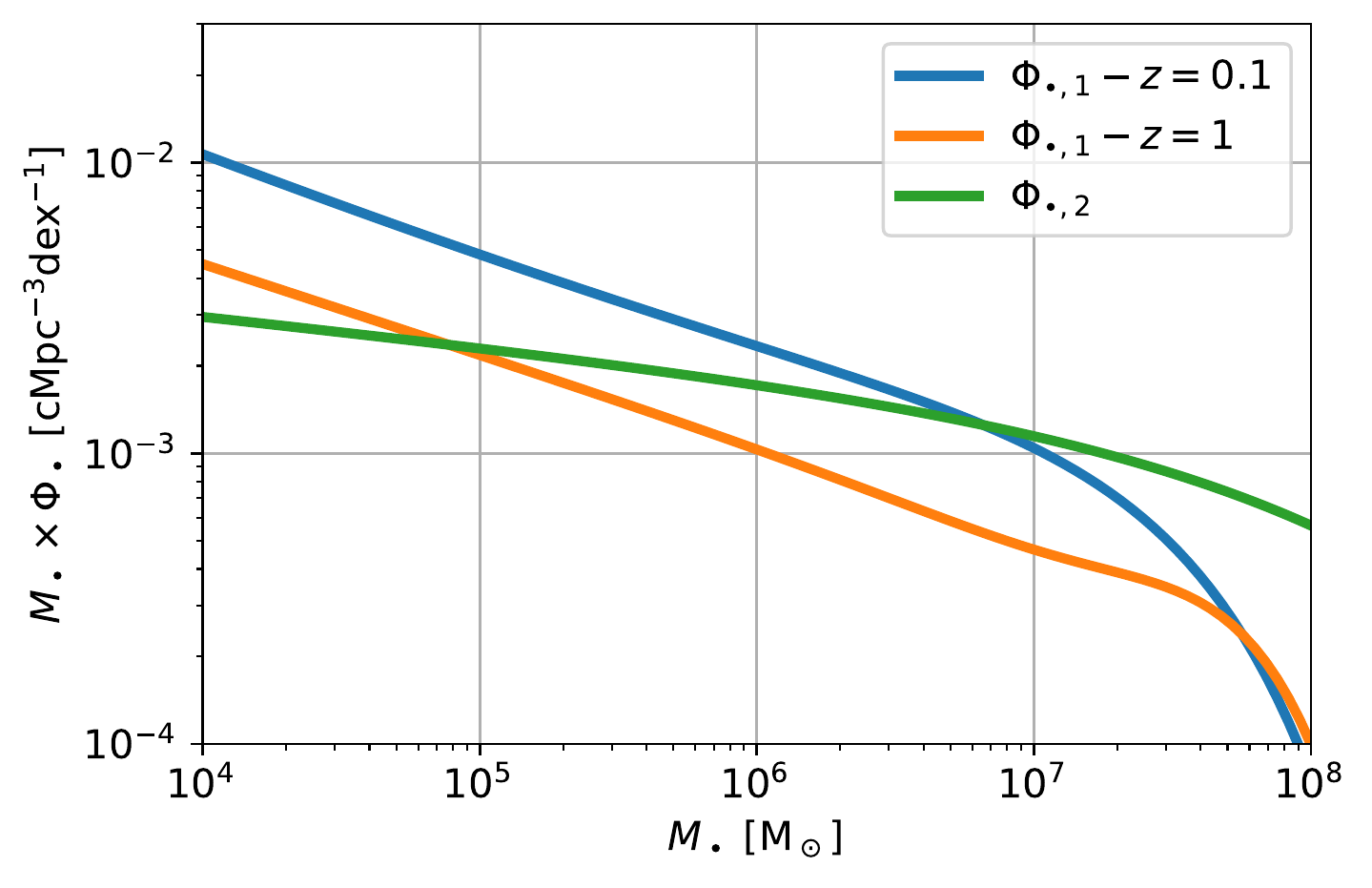}
\caption{The BH mass functions used in this work: $\Phi_{\bullet,1}$, which results from the combination of {\protect \cite{Davidzon_17}} and {\protect \cite{Reines_15}}, is shown at two different redshifts; and $\Phi_{\bullet,2}$ which is directly taken from {\protect \cite{Gallo_19}}.}
\label{fig:BHMF}
\end{figure}

We report in Fig.~\ref{fig:BHMF} the two BH mass functions used in this work. This simple plot emphasizes our current poor knowledge about BH population, even at low redshift.

\bsp	\label{lastpage}

\begin{thebibliography}{}
\makeatletter
\relax
\def\mn@urlcharsother{\let\do\@makeother \do\$\do\&\do\#\do\^\do\_\do\%\do\~}
\def\mn@doi{\begingroup\mn@urlcharsother \@ifnextchar [ {\mn@doi@}
  {\mn@doi@[]}}
\def\mn@doi@[#1]#2{\def\@tempa{#1}\ifx\@tempa\@empty \href
  {http://dx.doi.org/#2} {doi:#2}\else \href {http://dx.doi.org/#2} {#1}\fi
  \endgroup}
\def\mn@eprint#1#2{\mn@eprint@#1:#2::\@nil}
\def\mn@eprint@arXiv#1{\href {http://arxiv.org/abs/#1} {{\tt arXiv:#1}}}
\def\mn@eprint@dblp#1{\href {http://dblp.uni-trier.de/rec/bibtex/#1.xml}
  {dblp:#1}}
\def\mn@eprint@#1:#2:#3:#4\@nil{\def\@tempa {#1}\def\@tempb {#2}\def\@tempc
  {#3}\ifx \@tempc \@empty \let \@tempc \@tempb \let \@tempb \@tempa \fi \ifx
  \@tempb \@empty \def\@tempb {arXiv}\fi \@ifundefined
  {mn@eprint@\@tempb}{\@tempb:\@tempc}{\expandafter \expandafter \csname
  mn@eprint@\@tempb\endcsname \expandafter{\@tempc}}}

\bibitem[\protect\citeauthoryear{{Abbott} et~al.,}{{Abbott}
  et~al.}{2017}]{LIGO_17b}
{Abbott} B.~P.,  et~al., 2017, \mn@doi [ApJ] {10.3847/2041-8213/aa91c9}, \href
  {https://ui.adsabs.harvard.edu/abs/2017ApJ...848L..12A} {848, L12}

\bibitem[\protect\citeauthoryear{{Amaro-Seoane} et~al.,}{{Amaro-Seoane}
  et~al.}{2017}]{LISA_Proposal}
{Amaro-Seoane} P.,  et~al., 2017, arXiv:1702.00786, \href
  {https://ui.adsabs.harvard.edu/abs/2017arXiv170200786A} {}

\bibitem[\protect\citeauthoryear{{Astropy Collaboration} et~al.,}{{Astropy
  Collaboration} et~al.}{2013}]{astropy}
{Astropy Collaboration} et~al., 2013, \mn@doi [\aap]
  {10.1051/0004-6361/201322068}, \href
  {http://adsabs.harvard.edu/abs/2013A%26A...558A..33A} {558, A33}

\bibitem[\protect\citeauthoryear{{Baker} et~al.,}{{Baker}
  et~al.}{2019}]{Alia_proposal}
{Baker} J.,  et~al., 2019, arXiv e-prints, \href
  {https://ui.adsabs.harvard.edu/abs/2019arXiv190706482B} {p. arXiv:1907.06482}

\bibitem[\protect\citeauthoryear{{Bar-Or} \& {Alexander}}{{Bar-Or} \&
  {Alexander}}{2016}]{BarOr_16}
{Bar-Or} B.,  {Alexander} T.,  2016, \mn@doi [ApJ]
  {10.3847/0004-637X/820/2/129}, \href
  {https://ui.adsabs.harvard.edu/abs/2016ApJ...820..129B} {820, 129}

\bibitem[\protect\citeauthoryear{{Barausse}, {Dvorkin}, {Tremmel}, {Volonteri}
  \& {Bonetti}}{{Barausse} et~al.}{2020}]{2020ApJ...904...16B}
{Barausse} E.,  {Dvorkin} I.,  {Tremmel} M.,  {Volonteri} M.,   {Bonetti} M.,
  2020, \mn@doi [\apj] {10.3847/1538-4357/abba7f}, \href
  {https://ui.adsabs.harvard.edu/abs/2020ApJ...904...16B} {904, 16}

\bibitem[\protect\citeauthoryear{Binney \& Tremaine}{Binney \&
  Tremaine}{1987}]{BT_87}
Binney J.,  Tremaine S.,  1987, Galactic Dynamics, first edn.
Princeton Series in Astrophysics, Princeton University Press

\bibitem[\protect\citeauthoryear{{Bonnerot} \& {Stone}}{{Bonnerot} \&
  {Stone}}{2021}]{Bonnerot_21}
{Bonnerot} C.,  {Stone} N.~C.,  2021, \mn@doi [\ssr]
  {10.1007/s11214-020-00789-1}, \href
  {https://ui.adsabs.harvard.edu/abs/2021SSRv..217...16B} {217, 16}

\bibitem[\protect\citeauthoryear{{Chabrier}}{{Chabrier}}{2003}]{2003PASP..115..763C}
{Chabrier} G.,  2003, \mn@doi [\pasp] {10.1086/376392}, \href
  {https://ui.adsabs.harvard.edu/abs/2003PASP..115..763C} {115, 763}

\bibitem[\protect\citeauthoryear{{Chen}, {Yu}  \& {Lu}}{{Chen}
  et~al.}{2020}]{2020ApJ...900..191C}
{Chen} Y.,  {Yu} Q.,   {Lu} Y.,  2020, \mn@doi [\apj]
  {10.3847/1538-4357/aba950}, \href
  {https://ui.adsabs.harvard.edu/abs/2020ApJ...900..191C} {900, 191}

\bibitem[\protect\citeauthoryear{{Choi}, {Dotter}, {Conroy}, {Cantiello},
  {Paxton}  \& {Johnson}}{{Choi} et~al.}{2016}]{Choi_16}
{Choi} J.,  {Dotter} A.,  {Conroy} C.,  {Cantiello} M.,  {Paxton} B.,
  {Johnson} B.~D.,  2016, \mn@doi [ApJ] {10.3847/0004-637X/823/2/102}, \href
  {https://ui.adsabs.harvard.edu/abs/2016ApJ...823..102C} {823, 102}

\bibitem[\protect\citeauthoryear{{Cohn} \& {Kulsrud}}{{Cohn} \&
  {Kulsrud}}{1978}]{Cohn_78}
{Cohn} H.,  {Kulsrud} R.~M.,  1978, \mn@doi [ApJ] {10.1086/156685}, \href
  {https://ui.adsabs.harvard.edu/abs/1978ApJ...226.1087C} {226, 1087}

\bibitem[\protect\citeauthoryear{{Colpi} \& {Sesana}}{{Colpi} \&
  {Sesana}}{2017}]{Colpi_17}
{Colpi} M.,  {Sesana} A.,  2017, {Gravitational Wave Sources in the Era of
  Multi-Band Gravitational Wave Astronomy}.
G. Auger and E. Plagnol

\bibitem[\protect\citeauthoryear{{Dai}, {McKinney}, {Roth}, {Ramirez-Ruiz}  \&
  {Miller}}{{Dai} et~al.}{2018}]{Dai_18}
{Dai} L.,  {McKinney} J.~C.,  {Roth} N.,  {Ramirez-Ruiz} E.,   {Miller} M.~C.,
  2018, \mn@doi [ApJ] {10.3847/2041-8213/aab429}, \href
  {http://adsabs.harvard.edu/abs/2018ApJ...859L..20D} {859, L20}

\bibitem[\protect\citeauthoryear{{Dai}, {Lodato}  \& {Cheng}}{{Dai}
  et~al.}{2021}]{Dai_21}
{Dai} J.~L.,  {Lodato} G.,   {Cheng} R.,  2021, \mn@doi [\ssr]
  {10.1007/s11214-020-00747-x}, \href
  {https://ui.adsabs.harvard.edu/abs/2021SSRv..217...12D} {217, 12}

\bibitem[\protect\citeauthoryear{{Davidzon} et~al.,}{{Davidzon}
  et~al.}{2017}]{Davidzon_17}
{Davidzon} I.,  et~al., 2017, \mn@doi [AAP] {10.1051/0004-6361/201730419},
  \href {https://ui.adsabs.harvard.edu/abs/2017A&A...605A..70D} {605, A70}

\bibitem[\protect\citeauthoryear{{Evans} \& {Kochanek}}{{Evans} \&
  {Kochanek}}{1989}]{Evans_89}
{Evans} C.~R.,  {Kochanek} C.~S.,  1989, \mn@doi [ApJ] {10.1086/185567}, \href
  {https://ui.adsabs.harvard.edu/abs/1989ApJ...346L..13E} {346, L13}

\bibitem[\protect\citeauthoryear{{Gallo} \& {Sesana}}{{Gallo} \&
  {Sesana}}{2019}]{Gallo_19}
{Gallo} E.,  {Sesana} A.,  2019, \mn@doi [ApJ] {10.3847/2041-8213/ab40c6},
  \href {https://ui.adsabs.harvard.edu/abs/2019ApJ...883L..18G} {883, L18}

\bibitem[\protect\citeauthoryear{{Goicovic}, {Springel}, {Ohlmann}  \&
  {Pakmor}}{{Goicovic} et~al.}{2019}]{2019MNRAS.487..981G}
{Goicovic} F.~G.,  {Springel} V.,  {Ohlmann} S.~T.,   {Pakmor} R.,  2019,
  \mn@doi [\mnras] {10.1093/mnras/stz1368}, \href
  {https://ui.adsabs.harvard.edu/abs/2019MNRAS.487..981G} {487, 981}

\bibitem[\protect\citeauthoryear{{Golightly}, {Coughlin}  \&
  {Nixon}}{{Golightly} et~al.}{2019a}]{golightly19}
{Golightly} E. C.~A.,  {Coughlin} E.~R.,   {Nixon} C.~J.,  2019a, \mn@doi
  [\apj] {10.3847/1538-4357/aafd2f}, \href
  {https://ui.adsabs.harvard.edu/abs/2019ApJ...872..163G} {872, 163}

\bibitem[\protect\citeauthoryear{{Golightly}, {Nixon}  \&
  {Coughlin}}{{Golightly} et~al.}{2019b}]{2019ApJ...882L..26G}
{Golightly} E.~C.~A.,  {Nixon} C.~J.,   {Coughlin} E.~R.,  2019b, \mn@doi
  [\apjl] {10.3847/2041-8213/ab380d}, \href
  {https://ui.adsabs.harvard.edu/abs/2019ApJ...882L..26G} {882, L26}

\bibitem[\protect\citeauthoryear{{Greene}, {Strader}  \& {Ho}}{{Greene}
  et~al.}{2019}]{Greene_19}
{Greene} J.~E.,  {Strader} J.,   {Ho} L.~C.,  2019, arXiv e-prints, \href
  {https://ui.adsabs.harvard.edu/abs/2019arXiv191109678G} {p. arXiv:1911.09678}

\bibitem[\protect\citeauthoryear{{Guillochon} \& {Ramirez-Ruiz}}{{Guillochon}
  \& {Ramirez-Ruiz}}{2013}]{Guillochon_13}
{Guillochon} J.,  {Ramirez-Ruiz} E.,  2013, \mn@doi [ApJ]
  {10.1088/0004-637X/767/1/25}, \href
  {http://adsabs.harvard.edu/abs/2013ApJ...767...25G} {767, 25}

\bibitem[\protect\citeauthoryear{{Guillochon}, {Ramirez-Ruiz}, {Rosswog}  \&
  {Kasen}}{{Guillochon} et~al.}{2009}]{2009ApJ...705..844G}
{Guillochon} J.,  {Ramirez-Ruiz} E.,  {Rosswog} S.,   {Kasen} D.,  2009,
  \mn@doi [\apj] {10.1088/0004-637X/705/1/844}, \href
  {https://ui.adsabs.harvard.edu/abs/2009ApJ...705..844G} {705, 844}

\bibitem[\protect\citeauthoryear{Harris et~al.,}{Harris et~al.}{2020}]{numpy}
Harris C.~R.,  et~al., 2020, \mn@doi [Nature] {10.1038/s41586-020-2649-2}, 585,
  357

\bibitem[\protect\citeauthoryear{{Harry}, {Fritschel}, {Shaddock}, {Folkner}
  \& {Phinney}}{{Harry} et~al.}{2006}]{BBO_proposal}
{Harry} G.~M.,  {Fritschel} P.,  {Shaddock} D.~A.,  {Folkner} W.,   {Phinney}
  E.~S.,  2006, \mn@doi [Classical and Quantum Gravity]
  {10.1088/0264-9381/23/15/008}, \href
  {https://ui.adsabs.harvard.edu/abs/2006CQGra..23.4887H} {23, 4887}

\bibitem[\protect\citeauthoryear{{Hills}}{{Hills}}{1975}]{Hills_75}
{Hills} J.~G.,  1975, \mn@doi [Nature] {10.1038/254295a0}, \href
  {http://adsabs.harvard.edu/abs/1975Natur.254..295H} {254, 295}

\bibitem[\protect\citeauthoryear{{Hung} et~al.,}{{Hung} et~al.}{2017}]{Hung_17}
{Hung} T.,  et~al., 2017, \mn@doi [ApJ] {10.3847/1538-4357/aa7337}, \href
  {https://ui.adsabs.harvard.edu/abs/2017ApJ...842...29H} {842, 29}

\bibitem[\protect\citeauthoryear{Hunter}{Hunter}{2007}]{matplotlib}
Hunter J.~D.,  2007, \mn@doi [Computing in Science \& Engineering]
  {10.1109/MCSE.2007.55}, 9, 90

\bibitem[\protect\citeauthoryear{{Ivezi{\'c}} et~al.,}{{Ivezi{\'c}}
  et~al.}{2019}]{Ivezic_19}
{Ivezi{\'c}} {\v{Z}}.,  et~al., 2019, \apj, 873, 111

\bibitem[\protect\citeauthoryear{{Jones} et~al.,}{{Jones}
  et~al.}{2020}]{YSE_paper}
{Jones} D.~O.,  et~al., 2020, arXiv e-prints, \href
  {https://ui.adsabs.harvard.edu/abs/2020arXiv201009724J} {p. arXiv:2010.09724}

\bibitem[\protect\citeauthoryear{{Kesden}}{{Kesden}}{2012}]{Kesden_12}
{Kesden} M.,  2012, \mn@doi [Physical Review] {10.1103/PhysRevD.85.024037},
  \href {https://ui.adsabs.harvard.edu/abs/2012PhRvD..85b4037K} {85, 024037}

\bibitem[\protect\citeauthoryear{{Kippenhahn} \& {Weigert}}{{Kippenhahn} \&
  {Weigert}}{1990}]{Kippenhahn_90}
{Kippenhahn} R.,  {Weigert} A.,  1990, {Stellar Structure and Evolution}

\bibitem[\protect\citeauthoryear{Kluyver et~al.,}{Kluyver
  et~al.}{2016}]{jupyter}
Kluyver T.,  et~al., 2016, in Loizides F.,  Schmidt B.,  eds, Positioning and
  Power in Academic Publishing: Players, Agents and Agendas. pp 87 -- 90

\bibitem[\protect\citeauthoryear{{Kobayashi}, {Laguna}, {Phinney}  \&
  {M{\'e}sz{\'a}ros}}{{Kobayashi} et~al.}{2004}]{Kobayashi_04}
{Kobayashi} S.,  {Laguna} P.,  {Phinney} E.~S.,   {M{\'e}sz{\'a}ros} P.,  2004,
  \mn@doi [ApJ] {10.1086/424684}, \href
  {https://ui.adsabs.harvard.edu/abs/2004ApJ...615..855K} {615, 855}

\bibitem[\protect\citeauthoryear{{Kochanek}}{{Kochanek}}{2016a}]{Kochanek_16}
{Kochanek} C.~S.,  2016a, \mn@doi [MNRAS] {10.1093/mnras/stw267}, \href
  {https://ui.adsabs.harvard.edu/abs/2016MNRAS.458..127K} {458, 127}

\bibitem[\protect\citeauthoryear{{Kochanek}}{{Kochanek}}{2016b}]{Kochanek_16b}
{Kochanek} C.~S.,  2016b, \mn@doi [MNRAS] {10.1093/mnras/stw1290}, \href
  {https://ui.adsabs.harvard.edu/abs/2016MNRAS.461..371K} {461, 371}

\bibitem[\protect\citeauthoryear{{Kormendy} \& {Ho}}{{Kormendy} \&
  {Ho}}{2013}]{Kormendy_13}
{Kormendy} J.,  {Ho} L.~C.,  2013, \mn@doi [ARAA]
  {10.1146/annurev-astro-082708-101811}, \href
  {http://adsabs.harvard.edu/abs/2013ARA%26A..51..511K} {51, 511}

\bibitem[\protect\citeauthoryear{{Krolik}, {Piran}  \& {Ryu}}{{Krolik}
  et~al.}{2020}]{2020ApJ...904...68K}
{Krolik} J.,  {Piran} T.,   {Ryu} T.,  2020, \mn@doi [\apj]
  {10.3847/1538-4357/abc0f6}, \href
  {https://ui.adsabs.harvard.edu/abs/2020ApJ...904...68K} {904, 68}

\bibitem[\protect\citeauthoryear{{Kroupa}}{{Kroupa}}{2001}]{Kroupa_01}
{Kroupa} P.,  2001, \mn@doi [MNRAS] {10.1046/j.1365-8711.2001.04022.x}, \href
  {https://ui.adsabs.harvard.edu/abs/2001MNRAS.322..231K} {322, 231}

\bibitem[\protect\citeauthoryear{{Lacy}, {Townes}  \& {Hollenbach}}{{Lacy}
  et~al.}{1982}]{Lacy_82}
{Lacy} J.~H.,  {Townes} C.~H.,   {Hollenbach} D.~J.,  1982, \mn@doi [ApJ]
  {10.1086/160402}, \href
  {https://ui.adsabs.harvard.edu/abs/1982ApJ...262..120L} {262, 120}

\bibitem[\protect\citeauthoryear{{Lauer} et~al.,}{{Lauer}
  et~al.}{2007}]{Lauer_07}
{Lauer} T.~R.,  et~al., 2007, \mn@doi [ApJ] {10.1086/519229}, \href
  {http://adsabs.harvard.edu/abs/2007ApJ...664..226L} {664, 226}

\bibitem[\protect\citeauthoryear{{Law-Smith}, {Coulter}, {Guillochon},
  {Mockler}  \& {Ramirez-Ruiz}}{{Law-Smith} et~al.}{2020}]{LawSmith_20}
{Law-Smith} J. A.~P.,  {Coulter} D.~A.,  {Guillochon} J.,  {Mockler} B.,
  {Ramirez-Ruiz} E.,  2020, arXiv e-prints, \href
  {https://ui.adsabs.harvard.edu/abs/2020arXiv200710996L} {p. arXiv:2007.10996}

\bibitem[\protect\citeauthoryear{{Lightman} \& {Shapiro}}{{Lightman} \&
  {Shapiro}}{1977}]{Lightman_77}
{Lightman} A.~P.,  {Shapiro} S.~L.,  1977, \mn@doi [ApJ] {10.1086/154925},
  \href {https://ui.adsabs.harvard.edu/#abs/1977ApJ...211..244L} {211, 244}

\bibitem[\protect\citeauthoryear{{Liptai}, {Price}, {Mandel}  \&
  {Lodato}}{{Liptai} et~al.}{2019}]{liptai19}
{Liptai} D.,  {Price} D.~J.,  {Mandel} I.,   {Lodato} G.,  2019, arXiv
  e-prints, \href {https://ui.adsabs.harvard.edu/abs/2019arXiv191010154L} {p.
  arXiv:1910.10154}

\bibitem[\protect\citeauthoryear{{Lodato}}{{Lodato}}{2012}]{lodato12}
{Lodato} G.,  2012, in European Physical Journal Web of Conferences. p. 01001
  (\mn@eprint {arXiv} {1211.6109}), \mn@doi{10.1051/epjconf/20123901001}

\bibitem[\protect\citeauthoryear{{Lodato} \& {Rossi}}{{Lodato} \&
  {Rossi}}{2011}]{Lodato_11}
{Lodato} G.,  {Rossi} E.~M.,  2011, \mnras, 410, 359

\bibitem[\protect\citeauthoryear{{Lodato}, {King}  \& {Pringle}}{{Lodato}
  et~al.}{2009}]{LKP09}
{Lodato} G.,  {King} A.~R.,   {Pringle} J.~E.,  2009, \mn@doi [\mnras]
  {10.1111/j.1365-2966.2008.14049.x}, \href
  {https://ui.adsabs.harvard.edu/abs/2009MNRAS.392..332L} {392, 332}

\bibitem[\protect\citeauthoryear{{Lodato}, {Cheng}, {Bonnerot}  \&
  {Dai}}{{Lodato} et~al.}{2020}]{lodato20}
{Lodato} G.,  {Cheng} R.~M.,  {Bonnerot} C.,   {Dai} J.~L.,  2020, \mn@doi
  [\ssr] {10.1007/s11214-020-00697-4}, \href
  {https://ui.adsabs.harvard.edu/abs/2020SSRv..216...63L} {216, 63}

\bibitem[\protect\citeauthoryear{{Luo} et~al.,}{{Luo}
  et~al.}{2016}]{Tianqin_proposal}
{Luo} J.,  et~al., 2016, \mn@doi [Classical and Quantum Gravity]
  {10.1088/0264-9381/33/3/035010}, \href
  {https://ui.adsabs.harvard.edu/abs/2016CQGra..33c5010L} {33, 035010}

\bibitem[\protect\citeauthoryear{Maggiore}{Maggiore}{2008}]{Maggiore_08}
Maggiore M.,  2008, Gravitational waves. Vol 1, Theory and Experiments.
Oxford University Press (OUP)

\bibitem[\protect\citeauthoryear{{Magorrian} \& {Tremaine}}{{Magorrian} \&
  {Tremaine}}{1999}]{Magorrian_99}
{Magorrian} J.,  {Tremaine} S.,  1999, \mn@doi [MNRAS]
  {10.1046/j.1365-8711.1999.02853.x}, \href
  {https://ui.adsabs.harvard.edu/abs/1999MNRAS.309..447M} {309, 447}

\bibitem[\protect\citeauthoryear{{Merloni} et~al.,}{{Merloni}
  et~al.}{2012}]{Merloni_12}
{Merloni} A.,  et~al., 2012, arXiv e-prints, \href
  {https://ui.adsabs.harvard.edu/abs/2012arXiv1209.3114M} {p. arXiv:1209.3114}

\bibitem[\protect\citeauthoryear{{Merritt}}{{Merritt}}{2013}]{Merrit_book}
{Merritt} D.,  2013, Dynamics and Evolution of Galactic Nuclei.
Princeton University Press

\bibitem[\protect\citeauthoryear{{Merritt} \& {Ferrarese}}{{Merritt} \&
  {Ferrarese}}{2001}]{2001ASPC..249..335M}
{Merritt} D.,  {Ferrarese} L.,  2001, in {Knapen} J.~H.,  {Beckman} J.~E.,
  {Shlosman} I.,   {Mahoney} T.~J.,  eds,  Astronomical Society of the Pacific
  Conference Series Vol. 249, The Central Kiloparsec of Starbursts and AGN: The
  La Palma Connection. p.~335 (\mn@eprint {arXiv} {astro-ph/0107134})

\bibitem[\protect\citeauthoryear{{Mockler}, {Guillochon}  \&
  {Ramirez-Ruiz}}{{Mockler} et~al.}{2019}]{Mockler_19}
{Mockler} B.,  {Guillochon} J.,   {Ramirez-Ruiz} E.,  2019, \mn@doi [ApJ]
  {10.3847/1538-4357/ab010f}, \href
  {https://ui.adsabs.harvard.edu/abs/2019ApJ...872..151M} {872, 151}

\bibitem[\protect\citeauthoryear{{Nandra} et~al.,}{{Nandra}
  et~al.}{2013}]{Nandra_13}
{Nandra} K.,  et~al., 2013, arXiv e-prints, p. arXiv:1306.2307

\bibitem[\protect\citeauthoryear{{Pechetti}, {Seth}, {Neumayer}, {Georgiev},
  {Kacharov}  \& {den Brok}}{{Pechetti} et~al.}{2019}]{Pechetti_19}
{Pechetti} R.,  {Seth} A.,  {Neumayer} N.,  {Georgiev} I.,  {Kacharov} N.,
  {den Brok} M.,  2019, arXiv e-prints, \href
  {https://ui.adsabs.harvard.edu/abs/2019arXiv191109686P} {p. arXiv:1911.09686}

\bibitem[\protect\citeauthoryear{P\'erez \& Granger}{P\'erez \&
  Granger}{2007}]{iPython}
P\'erez F.,  Granger B.~E.,  2007, \mn@doi [Computing in Science and
  Engineering] {10.1109/MCSE.2007.53}, 9, 21

\bibitem[\protect\citeauthoryear{{Pestoni}, {Bortolas}, {Capelo}  \&
  {Mayer}}{{Pestoni} et~al.}{2021}]{2021MNRAS.500.4628P}
{Pestoni} B.,  {Bortolas} E.,  {Capelo} P.~R.,   {Mayer} L.,  2021, \mn@doi
  [\mnras] {10.1093/mnras/staa3496}, \href
  {https://ui.adsabs.harvard.edu/abs/2021MNRAS.500.4628P} {500, 4628}

\bibitem[\protect\citeauthoryear{{Pfister}, {Lupi}, {Capelo}, {Volonteri},
  {Bellovary}  \& {Dotti}}{{Pfister} et~al.}{2017}]{Pfister_17}
{Pfister} H.,  {Lupi} A.,  {Capelo} P.~R.,  {Volonteri} M.,  {Bellovary} J.~M.,
    {Dotti} M.,  2017, \mn@doi [MNRAS] {10.1093/mnras/stx1853}, \href
  {https://ui.adsabs.harvard.edu/abs/2017MNRAS.471.3646P} {471, 3646}

\bibitem[\protect\citeauthoryear{{Pfister}, {Volonteri}, {Dubois}, {Dotti}  \&
  {Colpi}}{{Pfister} et~al.}{2019a}]{Pfister_19a}
{Pfister} H.,  {Volonteri} M.,  {Dubois} Y.,  {Dotti} M.,   {Colpi} M.,  2019a,
  \mn@doi [MNRAS] {10.1093/mnras/stz822}, \href
  {https://ui.adsabs.harvard.edu/abs/2019MNRAS.486..101P} {486, 101}

\bibitem[\protect\citeauthoryear{{Pfister}, {Bar-Or}, {Volonteri}, {Dubois}  \&
  {Capelo}}{{Pfister} et~al.}{2019b}]{Pfister_19b}
{Pfister} H.,  {Bar-Or} B.,  {Volonteri} M.,  {Dubois} Y.,   {Capelo} P.~R.,
  2019b, \mn@doi [MNRAS] {10.1093/mnrasl/slz091}, \href
  {https://ui.adsabs.harvard.edu/abs/2019MNRAS.488L..29P} {488, L29}

\bibitem[\protect\citeauthoryear{{Pfister}, {Volonteri}, {Lixin Dai}  \&
  {Colpi}}{{Pfister} et~al.}{2020a}]{Pfister_20b}
{Pfister} H.,  {Volonteri} M.,  {Lixin Dai} J.,   {Colpi} M.,  2020a, arXiv
  e-prints, \href {https://ui.adsabs.harvard.edu/abs/2020arXiv200308133P} {p.
  arXiv:2003.08133}

\bibitem[\protect\citeauthoryear{{Pfister}, {Dai}, {Volonteri}, {Auchettl},
  {Trebitsch}  \& {Ramirez-Ruiz}}{{Pfister} et~al.}{2020b}]{Pfister_20c}
{Pfister} H.,  {Dai} J.,  {Volonteri} M.,  {Auchettl} K.,  {Trebitsch} M.,
  {Ramirez-Ruiz} E.,  2020b, arXiv e-prints, \href
  {https://ui.adsabs.harvard.edu/abs/2020arXiv200606565P} {p. arXiv:2006.06565}

\bibitem[\protect\citeauthoryear{{Piran}, {Svirski}, {Krolik}, {Cheng}  \&
  {Shiokawa}}{{Piran} et~al.}{2015}]{Piran_15}
{Piran} T.,  {Svirski} G.,  {Krolik} J.,  {Cheng} R.~M.,   {Shiokawa} H.,
  2015, \mn@doi [ApJ] {10.1088/0004-637X/806/2/164}, \href
  {https://ui.adsabs.harvard.edu/abs/2015ApJ...806..164P} {806, 164}

\bibitem[\protect\citeauthoryear{{Planck Collaboration} et~al.,}{{Planck
  Collaboration} et~al.}{2016}]{Planck_15}
{Planck Collaboration} et~al., 2016, \mn@doi [AAP]
  {10.1051/0004-6361/201525830}, \href
  {http://adsabs.harvard.edu/abs/2016A%26A...594A..13P} {594, A13}

\bibitem[\protect\citeauthoryear{{Rees}}{{Rees}}{1988}]{Rees_88}
{Rees} M.~J.,  1988, \mn@doi [Nature] {10.1038/333523a0}, \href
  {http://adsabs.harvard.edu/abs/1988Natur.333..523R} {333, 523}

\bibitem[\protect\citeauthoryear{{Reines} \& {Volonteri}}{{Reines} \&
  {Volonteri}}{2015}]{Reines_15}
{Reines} A.~E.,  {Volonteri} M.,  2015, \mn@doi [ApJ]
  {10.1088/0004-637X/813/2/82}, \href
  {http://adsabs.harvard.edu/abs/2015ApJ...813...82R} {813, 82}

\bibitem[\protect\citeauthoryear{{Roth} \& {Kasen}}{{Roth} \&
  {Kasen}}{2018}]{Roth_18}
{Roth} N.,  {Kasen} D.,  2018, \apj, 855, 54

\bibitem[\protect\citeauthoryear{{Roth}, {van Velzen}, {Cenko}  \&
  {Mushotzky}}{{Roth} et~al.}{2020a}]{Roth_20}
{Roth} N.,  {van Velzen} S.,  {Cenko} S.~B.,   {Mushotzky} R.~F.,  2020a, arXiv
  e-prints, \href {https://ui.adsabs.harvard.edu/abs/2020arXiv200811231R} {p.
  arXiv:2008.11231}

\bibitem[\protect\citeauthoryear{{Roth}, {Rossi}, {Krolik}, {Piran}, {Mockler}
  \& {Kasen}}{{Roth} et~al.}{2020b}]{Roth_20b}
{Roth} N.,  {Rossi} E.~M.,  {Krolik} J.,  {Piran} T.,  {Mockler} B.,   {Kasen}
  D.,  2020b, \mn@doi [\ssr] {10.1007/s11214-020-00735-1}, \href
  {https://ui.adsabs.harvard.edu/abs/2020SSRv..216..114R} {216, 114}

\bibitem[\protect\citeauthoryear{{Ryu}, {Krolik}  \& {Piran}}{{Ryu}
  et~al.}{2020a}]{Ryu_20b}
{Ryu} T.,  {Krolik} J.,   {Piran} T.,  2020a, \mn@doi [ApJ]
  {10.3847/1538-4357/abbf4d}, \href
  {https://ui.adsabs.harvard.edu/abs/2020ApJ...904...73R} {904, 73}

\bibitem[\protect\citeauthoryear{{Ryu}, {Krolik}, {Piran}  \& {Noble}}{{Ryu}
  et~al.}{2020b}]{Ryu_20}
{Ryu} T.,  {Krolik} J.,  {Piran} T.,   {Noble} S.~C.,  2020b, \mn@doi [ApJ]
  {10.3847/1538-4357/abb3cf}, \href
  {https://ui.adsabs.harvard.edu/abs/2020ApJ...904...98R} {904, 98}

\bibitem[\protect\citeauthoryear{{Ryu}, {Krolik}, {Piran}  \& {Noble}}{{Ryu}
  et~al.}{2020c}]{2020ApJ...904...99R}
{Ryu} T.,  {Krolik} J.,  {Piran} T.,   {Noble} S.~C.,  2020c, \mn@doi [\apj]
  {10.3847/1538-4357/abb3cd}, \href
  {https://ui.adsabs.harvard.edu/abs/2020ApJ...904...99R} {904, 99}

\bibitem[\protect\citeauthoryear{{Ryu}, {Krolik}, {Piran}  \& {Noble}}{{Ryu}
  et~al.}{2020d}]{2020ApJ...904..100R}
{Ryu} T.,  {Krolik} J.,  {Piran} T.,   {Noble} S.~C.,  2020d, \mn@doi [\apj]
  {10.3847/1538-4357/abb3ce}, \href
  {https://ui.adsabs.harvard.edu/abs/2020ApJ...904..100R} {904, 100}

\bibitem[\protect\citeauthoryear{{Sacchi} \& {Lodato}}{{Sacchi} \&
  {Lodato}}{2019}]{sacchi19}
{Sacchi} A.,  {Lodato} G.,  2019, \mn@doi [\mnras] {10.1093/mnras/stz981},
  \href {https://ui.adsabs.harvard.edu/abs/2019MNRAS.486.1833S} {486, 1833}

\bibitem[\protect\citeauthoryear{{Salpeter}}{{Salpeter}}{1955}]{1955ApJ...121..161S}
{Salpeter} E.~E.,  1955, \mn@doi [\apj] {10.1086/145971}, \href
  {https://ui.adsabs.harvard.edu/abs/1955ApJ...121..161S} {121, 161}

\bibitem[\protect\citeauthoryear{{S{\'a}nchez-Janssen}
  et~al.,}{{S{\'a}nchez-Janssen} et~al.}{2019}]{SanchezJanssen_19}
{S{\'a}nchez-Janssen} R.,  et~al., 2019, \mn@doi [ApJ]
  {10.3847/1538-4357/aaf4fd}, \href
  {https://ui.adsabs.harvard.edu/abs/2019ApJ...878...18S} {878, 18}

\bibitem[\protect\citeauthoryear{{Sato} et~al.,}{{Sato}
  et~al.}{2009}]{Decigo_proposal}
{Sato} S.,  et~al., 2009, in Journal of Physics Conference Series. p. 012040,
  \mn@doi{10.1088/1742-6596/154/1/012040}

\bibitem[\protect\citeauthoryear{{Saxton}, {Komossa}, {Auchettl}  \&
  {Jonker}}{{Saxton} et~al.}{2020}]{Saxton_20}
{Saxton} R.,  {Komossa} S.,  {Auchettl} K.,   {Jonker} P.~G.,  2020, \ssr, 216,
  85

\bibitem[\protect\citeauthoryear{{Shiokawa}, {Krolik}, {Cheng}, {Piran}  \&
  {Noble}}{{Shiokawa} et~al.}{2015}]{Shiokawa_15}
{Shiokawa} H.,  {Krolik} J.~H.,  {Cheng} R.~M.,  {Piran} T.,   {Noble} S.~C.,
  2015, \apj, 804, 85

\bibitem[\protect\citeauthoryear{{Spitzer} \& {Hart}}{{Spitzer} \&
  {Hart}}{1971}]{Spitzer_71}
{Spitzer} Lyman J.,  {Hart} M.~H.,  1971, \mn@doi [ApJ] {10.1086/150855}, \href
  {https://ui.adsabs.harvard.edu/abs/1971ApJ...164..399S} {164, 399}

\bibitem[\protect\citeauthoryear{{Stone} \& {Metzger}}{{Stone} \&
  {Metzger}}{2016}]{Stone_16a}
{Stone} N.~C.,  {Metzger} B.~D.,  2016, \mn@doi [MNRAS]
  {10.1093/mnras/stv2281}, \href
  {http://adsabs.harvard.edu/abs/2016MNRAS.455..859S} {455, 859}

\bibitem[\protect\citeauthoryear{{Stone}, {Sari}  \& {Loeb}}{{Stone}
  et~al.}{2013}]{Stone_13}
{Stone} N.,  {Sari} R.,   {Loeb} A.,  2013, \mn@doi [MNRAS]
  {10.1093/mnras/stt1270}, \href
  {https://ui.adsabs.harvard.edu/abs/2013MNRAS.435.1809S} {435, 1809}

\bibitem[\protect\citeauthoryear{{Stone}, {Kesden}, {Cheng}  \& {van
  Velzen}}{{Stone} et~al.}{2019}]{Stone_19}
{Stone} N.~C.,  {Kesden} M.,  {Cheng} R.~M.,   {van Velzen} S.,  2019, \mn@doi
  [General Relativity and Gravitation] {10.1007/s10714-019-2510-9}, \href
  {https://ui.adsabs.harvard.edu/abs/2019GReGr..51...30S} {51, 30}

\bibitem[\protect\citeauthoryear{{Stone}, {Vasiliev}, {Kesden}, {Rossi},
  {Perets}  \& {Amaro-Seoane}}{{Stone} et~al.}{2020}]{Stone_20}
{Stone} N.~C.,  {Vasiliev} E.,  {Kesden} M.,  {Rossi} E.~M.,  {Perets} H.~B.,
  {Amaro-Seoane} P.,  2020, \mn@doi [SSR] {10.1007/s11214-020-00651-4}, \href
  {https://ui.adsabs.harvard.edu/abs/2020SSRv..216...35S} {216, 35}

\bibitem[\protect\citeauthoryear{{Strubbe}}{{Strubbe}}{2011}]{Strubbe_11}
{Strubbe} L.~E.,  2011, PhD thesis, University of California, Berkeley

\bibitem[\protect\citeauthoryear{{The Lynx Team}}{{The Lynx
  Team}}{2018}]{lynx_18}
{The Lynx Team} 2018, arXiv e-prints, p. arXiv:1809.09642

\bibitem[\protect\citeauthoryear{{Toscani}, {Lodato}  \& {Nealon}}{{Toscani}
  et~al.}{2019}]{Toscani_19}
{Toscani} M.,  {Lodato} G.,   {Nealon} R.,  2019, \mn@doi [MNRAS]
  {10.1093/mnras/stz2201}, \href
  {https://ui.adsabs.harvard.edu/abs/2019MNRAS.489..699T} {489, 699}

\bibitem[\protect\citeauthoryear{{Toscani}, {Rossi}  \& {Lodato}}{{Toscani}
  et~al.}{2020}]{Toscani_20}
{Toscani} M.,  {Rossi} E.~M.,   {Lodato} G.,  2020, \mn@doi [MNRAS]
  {10.1093/mnras/staa2290}, \href
  {https://ui.adsabs.harvard.edu/abs/2020MNRAS.498..507T} {498, 507}

\bibitem[\protect\citeauthoryear{{Tremmel}, {Governato}, {Volonteri}  \&
  {Quinn}}{{Tremmel} et~al.}{2015}]{Tremmel_15}
{Tremmel} M.,  {Governato} F.,  {Volonteri} M.,   {Quinn} T.~R.,  2015, \mn@doi
  [MNRAS] {10.1093/mnras/stv1060}, \href
  {http://adsabs.harvard.edu/abs/2015MNRAS.451.1868T} {451, 1868}

\bibitem[\protect\citeauthoryear{{Turk}, {Smith}, {Oishi}, {Skory}, {Skillman},
  {Abel}  \& {Norman}}{{Turk} et~al.}{2011}]{yt}
{Turk} M.~J.,  {Smith} B.~D.,  {Oishi} J.~S.,  {Skory} S.,  {Skillman} S.~W.,
  {Abel} T.,   {Norman} M.~L.,  2011, \mn@doi [\apjs]
  {10.1088/0067-0049/192/1/9}, \href
  {https://ui.adsabs.harvard.edu/abs/2011ApJS..192....9T} {192, 9}

\bibitem[\protect\citeauthoryear{{Ulmer}}{{Ulmer}}{1999}]{Ulmer_99}
{Ulmer} A.,  1999, \mn@doi [ApJ] {10.1086/306909}, \href
  {https://ui.adsabs.harvard.edu/abs/1999ApJ...514..180U} {514, 180}

\bibitem[\protect\citeauthoryear{{Vasiliev}}{{Vasiliev}}{2017}]{Vasiliev_17}
{Vasiliev} E.,  2017, \mn@doi [ApJ] {10.3847/1538-4357/aa8cc8}, \href
  {https://ui.adsabs.harvard.edu/#abs/2017ApJ...848...10V} {848, 10}

\bibitem[\protect\citeauthoryear{{Virtanen} et~al.,}{{Virtanen}
  et~al.}{2020}]{scipy}
{Virtanen} P.,  et~al., 2020, \mn@doi [Nature Methods]
  {10.1038/s41592-019-0686-2}, \href
  {https://ui.adsabs.harvard.edu/abs/2020NatMe..17..261V} {17, 261}

\bibitem[\protect\citeauthoryear{{Volonteri}, {Haardt}  \& {Madau}}{{Volonteri}
  et~al.}{2003}]{Volonteri_03}
{Volonteri} M.,  {Haardt} F.,   {Madau} P.,  2003, \mn@doi [ApJ]
  {10.1086/344675}, \href {http://adsabs.harvard.edu/abs/2003ApJ...582..559V}
  {582, 559}

\bibitem[\protect\citeauthoryear{{Wang} \& {Merritt}}{{Wang} \&
  {Merritt}}{2004}]{Wang_04}
{Wang} J.,  {Merritt} D.,  2004, \mn@doi [ApJ] {10.1086/379767}, \href
  {http://adsabs.harvard.edu/abs/2004ApJ...600..149W} {600, 149}

\bibitem[\protect\citeauthoryear{{Wevers}, {van Velzen}, {Jonker}, {Stone},
  {Hung}, {Onori}, {Gezari}  \& {Blagorodnova}}{{Wevers}
  et~al.}{2017}]{Wevers_17}
{Wevers} T.,  {van Velzen} S.,  {Jonker} P.~G.,  {Stone} N.~C.,  {Hung} T.,
  {Onori} F.,  {Gezari} S.,   {Blagorodnova} N.,  2017, \mnras, 471, 1694

\bibitem[\protect\citeauthoryear{pandas~development team}{pandas~development
  team}{2020}]{pandas}
pandas~development team T.,  2020, pandas-dev/pandas: Pandas,
  \mn@doi{10.5281/zenodo.3509134}, \url
  {https://doi.org/10.5281/zenodo.3509134}

\bibitem[\protect\citeauthoryear{{van Velzen}}{{van
  Velzen}}{2018}]{VanVelzen_18}
{van Velzen} S.,  2018, \mn@doi [ApJ] {10.3847/1538-4357/aa998e}, \href
  {https://ui.adsabs.harvard.edu/abs/2018ApJ...852...72V} {852, 72}

\bibitem[\protect\citeauthoryear{{van Velzen} et~al.,}{{van Velzen}
  et~al.}{2016}]{2016Sci...351...62V}
{van Velzen} S.,  et~al., 2016, \mn@doi [Science] {10.1126/science.aad1182},
  \href {https://ui.adsabs.harvard.edu/abs/2016Sci...351...62V} {351, 62}

\bibitem[\protect\citeauthoryear{{van Velzen} et~al.,}{{van Velzen}
  et~al.}{2020a}]{VanVelzen_20}
{van Velzen} S.,  et~al., 2020a, arXiv e-prints, \href
  {https://ui.adsabs.harvard.edu/abs/2020arXiv200101409V} {p. arXiv:2001.01409}

\bibitem[\protect\citeauthoryear{{van Velzen}, {Holoien}, {Onori}, {Hung}  \&
  {Arcavi}}{{van Velzen} et~al.}{2020b}]{2020SSRv..216..124V}
{van Velzen} S.,  {Holoien} T. W.~S.,  {Onori} F.,  {Hung} T.,   {Arcavi} I.,
  2020b, \mn@doi [\ssr] {10.1007/s11214-020-00753-z}, \href
  {https://ui.adsabs.harvard.edu/abs/2020SSRv..216..124V} {216, 124}

\makeatother
\end{thebibliography}
\end{document}